\documentclass[12pt]{article}
\usepackage{epsfig,amsfonts,amssymb}
\usepackage{hyperref}
\usepackage {cite}
\usepackage{cite}

\def\ZZZ{{\hbox{ Z\kern-1.6mm Z}}}
\def\RRR{{\hbox{ R\kern-2.4mm R}}}
\def\CCC{{\hbox{ C\kern-2.5mm C}}}
\def\zzz{{\hbox{z\kern-1mm z}}}

\newcommand{\qeq}{{\hbox{=\kern-2.3mm ? \kern.5mm }}}
\renewcommand{\qeq}{=}

\newcommand{\eps}{\epsilon}

\newcommand{\BB}{{\cal B}}
\newcommand{\DD}{{\cal D}}

\newcommand{\BBB}{{\bf B}}
\newcommand{\II}{{\cal I}}

\newcommand{\GG}{{\cal G}}

\newcommand{\JJ}{{\cal J}}
\newcommand{\HH}{{\cal H}}

\newcommand{\CC}{{\cal C}}

\newcommand{\OO}{{\cal O}}

\newcommand{\PP}{{\cal P}}

\newcommand{\LL}{{\cal L}}

\newcommand{\XX}{{\cal X}}

\newcommand{\wt}{\widetilde}

\newcommand{\RR}{{\cal R}}

\newcommand{\be}{\begin{equation}}
\newcommand{\ee}{\end{equation}}
\newcommand{\ben}{\begin{eqnarray}\displaystyle}
\newcommand{\een}{\end{eqnarray}}

\newcommand{\refb}[1]{(\ref{#1})}
\newcommand{\p}{\partial}
\newcommand{\sectiono}[1]{\section{#1}\setcounter{equation}{0}}

\def\one{{\hbox{ 1\kern-.8mm l}}}
\def\zero{{\hbox{ 0\kern-1.5mm 0}}}

\newcommand{\bea}[1]{\begin{eqnarray}\label{#1} }
\newcommand{\eea}{\end{eqnarray}}

\newcommand{\eqref}{\refb}

\usepackage{bm}
\usepackage[table]{xcolor}

\def\Imm{{\rm Im\, }}

\newcommand{\xtau}{u}

\newcommand{\cl}{\rm cl}

\def\figgraph{

\def\JPicScale{1}
\ifx\JPicScale\undefined\def\JPicScale{1}\fi
\unitlength \JPicScale mm
\begin{picture}(130,65)(0,0)
\linethickness{0.3mm}
\multiput(10,60)(0.12,-0.12){83}{\line(1,0){0.12}}
\linethickness{0.3mm}
\multiput(10,40)(0.12,0.12){83}{\line(1,0){0.12}}
\linethickness{0.3mm}
\put(20,50){\line(1,0){10}}
\linethickness{0.3mm}
\multiput(30,50)(0.12,0.12){83}{\line(1,0){0.12}}
\linethickness{0.3mm}
\put(30,50){\line(1,0){10}}
\linethickness{0.3mm}
\multiput(30,50)(0.12,-0.12){83}{\line(1,0){0.12}}
\linethickness{0.3mm}
\multiput(50,60)(0.12,-0.12){83}{\line(1,0){0.12}}
\linethickness{0.3mm}
\multiput(50,40)(0.12,0.12){83}{\line(1,0){0.12}}
\linethickness{0.3mm}
\put(50,50){\line(1,0){10}}
\linethickness{0.3mm}
\put(60,50){\line(1,0){10}}
\linethickness{0.3mm}
\multiput(70,50)(0.12,0.12){83}{\line(1,0){0.12}}
\linethickness{0.3mm}
\multiput(70,50)(0.12,-0.12){83}{\line(1,0){0.12}}
\linethickness{0.3mm}
\multiput(90,60)(0.12,-0.12){83}{\line(1,0){0.12}}
\linethickness{0.3mm}
\multiput(90,40)(0.12,0.12){83}{\line(1,0){0.12}}
\linethickness{0.3mm}
\put(100,50){\line(1,0){10}}
\linethickness{0.3mm}
\put(110,50){\line(1,0){10}}
\linethickness{0.3mm}
\multiput(120,50)(0.12,0.12){83}{\line(1,0){0.12}}
\linethickness{0.3mm}
\multiput(120,50)(0.12,-0.12){83}{\line(1,0){0.12}}
\linethickness{0.3mm}
\put(110,40){\line(0,1){10}}
\put(10,63){\makebox(0,0)[cc]{3}}

\put(10,36){\makebox(0,0)[cc]{4}}

\put(40,62){\makebox(0,0)[cc]{1}}

\put(40,47){\makebox(0,0)[cc]{2}}

\put(40,37){\makebox(0,0)[cc]{5}}

\put(50,63){\makebox(0,0)[cc]{3}}

\put(48,50){\makebox(0,0)[cc]{4}}

\put(50,36){\makebox(0,0)[cc]{1}}

\put(80,63){\makebox(0,0)[cc]{2}}

\put(80,36){\makebox(0,0)[cc]{5}}

\put(90,63){\makebox(0,0)[cc]{3}}

\put(90,36){\makebox(0,0)[cc]{4}}

\put(110,36){\makebox(0,0)[cc]{1}}

\put(130,63){\makebox(0,0)[cc]{2}}

\put(130,36){\makebox(0,0)[cc]{5}}

\put(25,47){\makebox(0,0)[cc]{$a$}}

\put(65,47){\makebox(0,0)[cc]{$b$}}

\put(105,47){\makebox(0,0)[cc]{$a$}}

\put(115,47){\makebox(0,0)[cc]{$b$}}

\end{picture}

}

\def\figgraphrep{

\def\JPicScale{1}
\ifx\JPicScale\undefined\def\JPicScale{1}\fi
\unitlength \JPicScale mm
\begin{picture}(130,65)(0,0)
\linethickness{0.3mm}
\multiput(10,60)(0.12,-0.12){83}{\line(1,0){0.12}}
\linethickness{0.3mm}
\multiput(10,40)(0.12,0.12){83}{\line(1,0){0.12}}
\linethickness{0.3mm}
\put(20,50){\line(1,0){10}}
\linethickness{0.3mm}
\multiput(30,50)(0.12,0.12){83}{\line(1,0){0.12}}
\linethickness{0.3mm}
\put(30,50){\line(1,0){10}}
\linethickness{0.3mm}
\multiput(30,50)(0.12,-0.12){83}{\line(1,0){0.12}}
\linethickness{0.3mm}
\multiput(50,60)(0.12,-0.12){83}{\line(1,0){0.12}}
\linethickness{0.3mm}
\multiput(50,40)(0.12,0.12){83}{\line(1,0){0.12}}
\linethickness{0.3mm}
\put(50,50){\line(1,0){10}}
\linethickness{0.3mm}
\put(60,50){\line(1,0){10}}
\linethickness{0.3mm}
\multiput(70,50)(0.12,0.12){83}{\line(1,0){0.12}}
\linethickness{0.3mm}
\multiput(70,50)(0.12,-0.12){83}{\line(1,0){0.12}}
\linethickness{0.3mm}
\multiput(90,60)(0.12,-0.12){83}{\line(1,0){0.12}}
\linethickness{0.3mm}
\multiput(90,40)(0.12,0.12){83}{\line(1,0){0.12}}
\linethickness{0.3mm}
\put(100,50){\line(1,0){10}}
\linethickness{0.3mm}
\put(110,50){\line(1,0){10}}
\linethickness{0.3mm}
\multiput(120,50)(0.12,0.12){83}{\line(1,0){0.12}}
\linethickness{0.3mm}
\multiput(120,50)(0.12,-0.12){83}{\line(1,0){0.12}}
\linethickness{0.3mm}
\put(110,40){\line(0,1){10}}
\put(10,63){\makebox(0,0)[cc]{3}}

\put(10,36){\makebox(0,0)[cc]{4}}

\put(40,62){\makebox(0,0)[cc]{1}}

\put(40,47){\makebox(0,0)[cc]{2}}

\put(40,37){\makebox(0,0)[cc]{5}}

\put(50,63){\makebox(0,0)[cc]{3}}

\put(48,50){\makebox(0,0)[cc]{4}}

\put(50,36){\makebox(0,0)[cc]{1}}

\put(80,63){\makebox(0,0)[cc]{2}}

\put(80,36){\makebox(0,0)[cc]{5}}

\put(90,63){\makebox(0,0)[cc]{3}}

\put(90,36){\makebox(0,0)[cc]{4}}

\put(110,36){\makebox(0,0)[cc]{1}}

\put(130,63){\makebox(0,0)[cc]{2}}

\put(130,36){\makebox(0,0)[cc]{5}}

\put(25,30){\makebox(0,0)[cc]{(i)}}

\put(65,30){\makebox(0,0)[cc]{(ii)}}

\put(110,30){\makebox(0,0)[cc]{(iii)}}

\put(25,47){\makebox(0,0)[cc]{$a$}}

\put(65,47){\makebox(0,0)[cc]{$b$}}

\put(105,47){\makebox(0,0)[cc]{$a$}}

\put(115,47){\makebox(0,0)[cc]{$b$}}

\put(20,50){\circle{5}}

\put(31,50){\circle{5}}

\put(60,50){\circle{5}}

\put(70,50){\circle{5}}

\put(120,50){\circle{5}}

\put(100,50){\circle{5}}

\put(110,50){\circle{5}}

\end{picture}

}

\begin{document}

\vskip 12pt

\baselineskip 24pt

\begin{center}

% {\Large \bf Analyticity and Crossing Symmetry of Loop Amplitudes  in Superstring Field Theory}

%{\Large \bf Applications of Classical String Field Theory}

{\Large \bf String Field Theory as World-sheet UV Regulator
%Applications of Classical String Field Theory: Regulating World-sheet Ultraviolet Divergences
% Using Classical String Field Theory
}

\end{center}

\vskip .6cm
\medskip

\vspace*{4.0ex}

\baselineskip=18pt

\begin{center}

{\large 
\rm Ashoke Sen}

\end{center}

\vspace*{4.0ex}

\centerline{ \it  Harish-Chandra Research Institute, HBNI,}
\centerline{\it Chhatnag Road, Jhusi,
Allahabad 211019, India}

\vspace*{1.0ex}
\centerline{\small E-mail: sen@hri.res.in}

\vspace*{5.0ex}

\baselineskip=18pt

\centerline{\bf Abstract} \bigskip

Even at tree level, the first quantized string theory suffers from apparent short distance singularities 
associated with collision of vertex operators 
that prevent us from straightforward numerical computation of various quantities.
Examples include string theory S-matrix for generic external momenta and
computation of the spectrum of string theory under a marginal deformation of the world-sheet theory. 
The former requires us to define the S-matrix via analytic continuation or  as limits of
contour integrals in complexified moduli space, while the latter requires us
to use an ultraviolet cut-off at intermediate steps.
In contrast, string field theory does not suffer from such divergences. 
In this paper we show how string field theory can be used to generate an explicit algorithm 
for computing tree level amplitudes
in any string theory that does not suffer from any short distance divergence from integration over the 
world-sheet variables. We also
use string field theory to compute second order mass shift
of string states under a marginal deformation without having to use any cut-off at intermediate steps.
We carry out the analysis in a broad class of string field theories, thereby making
it manifest that the final results are independent of the extra data that go into the formulation of 
string field theory. We also comment on 
the generalization of this analysis to higher genus amplitudes.

\vfill \eject

\baselineskip=18pt

\tableofcontents

\sectiono{Introduction and summary} \label{s1}

String field theory was originally formulated with the hope of getting non-perturbative results in string theory.
With the exception of some non-perturbative classical solutions in open string field theory\cite{0511286}, 
this hope has not been
realized so far. However string field theory can play another useful role -- in getting a better description of
perturbative string theory.  Indeed, 
superstring field theory has been useful for giving a definition of perturbative string amplitudes free from all
divergences other than the ones expected on physical grounds, {\it e.g.} the usual infrared divergences in
four space-time dimensions. It has  also been useful 
for proving general properties of the perturbative S-matrix {\it e.g.}
unitarity, analyticity etc. A  review of these developments can be found in \cite{1703.06410}.

The goal of this paper will be to use string field theory to remove some of the inconvenient features
of {\it tree level} string theory -- divergences associated with short distance singularities on the world-sheet.
We analyze two classes of problems -- computation of on-shell amplitudes and studying the effect of marginal
deformation on the world-sheet (super-) conformal field theory (CFT) on the spectrum of string theory.
Since at tree level, theories with tachyons {\it e.g.} 
bosonic string theory and superstring theory formulated
around certain non-supersymmetric backgrounds, also give sensible results, our analysis will also be
valid for the corresponding string field theories. For this reason we shall use the
phrase `string field theory'
instead of `superstring field theory', with the understanding that superstring field theory should 
be considered as
a special case of string field theory.
We shall see that all the apparent short distance divergences on the world-sheet
arise from wrong treatment of $(L_0+\bar L_0)^{-1}$ where
$L_n,\bar L_n$ are the usual Virasoro generators of matter ghost CFT and, for
superstring theory, also from wrong treatment of picture changing operators (PCOs)\cite{FMS}.

There are many versions of string field theory -- for our analysis we shall work with a broad class of string field theories
whose interaction vertices are constructed in such a way that the equality of the S-matrix computed from string field theory
and first quantized string theory is manifest. This includes for example the original open bosonic string field theory of
Witten\cite{wittensft}, its generalization to closed bosonic string field theory in \cite{saadi,kugo,sonoda,
9206084,9705241} 
and closed and open superstring
field theories described {\it e.g.} in \cite{9202087,1312.2948,1403.0940,1508.05387,1602.02582,1602.02583}, 
but does not include some versions of string 
field theory where this
equality is not manifest {\it e.g.} those described in \cite{9503099,0109100,0406212,0409018,1508.00366}.
Although our analysis can be applied to both open and closed string
theories, for definiteness we shall focus on closed string theories. For open string theories, the $L_0+\bar L_0$
factors in the subsequent discussion 
will be replaced by $L_0$ and there will be no analog of the projection to $L_0=\bar L_0$ states.

%\begin{enumerate}
%\item
Let us first consider tree level scattering amplitudes in string theory.
The usual definition of these amplitudes is in terms of integrals of certain correlation functions of 
vertex operators in a 
CFT, but this often suffers from divergences from regions of integration where the 
locations of the vertex operators
come close to each other. A simple example of this is the Koba-Nielsen 
formula for multi-tachyon amplitudes in bosonic string theory. Usually we avoid this problem by defining these
integrals via analytic continuation, but this prevents us from directly evaluating the integrals numerically when
an analytic formula is absent. Refs.\cite{0510064,1307.5124,1706.08527} 
suggest alternative approaches by regarding the integrals
as (limits of) integrals
in the complexified moduli space -- some aspects of this will be discussed later.

In string field theory the tree level S-matrix is evaluated by summing over Feynman diagrams. 
Part of the contribution to the S-matrix with
$N$ external legs comes from the diagram involving an $N$-point interaction
vertex to which the external legs connect,
without any internal propagator.\footnote{We shall use the phrase vertex operator to denote a local operator of the
CFT on the world-sheet and the phrase interaction vertex to describe a term 
in the action of string
field theory. These two terms should not be confused.}
This diagram gives the same world-sheet integral as the first quantized
string theory, except that the integration
is over a restricted region, avoiding regions where two or more vertex operators come close to each other.
These missing contributions are given by Feynman diagrams with one or more internal propagators and
interaction vertices with less number of external legs. Formally the contribution from the latter diagrams 
also can be represented as the same world-sheet integral as the first quantized
string theory, but the integration runs over regions where one or more vertex operators come close, 
complementing the region associated with the the $N$-string interaction vertex.  However this correspondence is
only formal. To understand the difference between the formal expression and the actual contribution from the
Feynman diagram of string field theory, we first note that in Siegel gauge\cite{siegel}
the propagator of the string field takes
the form 
\be \label{e1}
2\, b_0 \, \bar b_0 \, (L_0+\bar L_0)^{-1} \, \delta_{L_0, \bar L_0} =
{1\over\pi} \, b_0 \, \bar b_0 \, (L_0+\bar L_0)^{-1} \, \int_0^{2\pi} d\theta e^{-i\theta(L_0-\bar L_0)}\, ,
\ee
with some additional numerator factors for the Ramond sector states in superstring theories. Here $b_0,\bar b_0$ denote
the $b$, $\bar b$ ghost zero modes and $L_n$, $\bar L_n$ denote the total Virasoro generators.  In order to
express the contribution from the Feynman diagrams 
in the usual form of the string amplitude in the first quantized formalism, we need to express \refb{e1}
as
\ben \label{e2}
&& {1\over \pi} 
\, b_0 \, \bar b_0 \, \int_0^\infty \, ds \, e^{-s(L_0+\bar L_0)} \, \int_0^{2\pi} d\theta\, e^{-i \theta (L_0-\bar L_0)}
= -{1\over 2\pi i} \,  b_0 \, \bar b_0 \, \int_{|q|\le 1} d^2 q \, q^{L_0-1} \bar q^{\bar L_0-1} \, , \nonumber \\
&& q\equiv e^{-s-i\theta}, \quad d^2 q \equiv dq\wedge d\bar q\, .
\een
We use the convention that $ds\wedge d\theta$ describes positive integration measure.
The collection of the variables $(s,\theta)$ for different propagators, together with the integration variables
that enter in the definition of the interaction vertices, become the coordinates of the vertex operators on the
world-sheet after some change of variables.

Note however that the equality of \refb{e1} and \refb{e2} holds only when both sides act on states with positive
$L_0+\bar L_0$ eigenvalues. Even tachyon free superstring field theories have off-shell internal
states with negative $L_0+\bar L_0$ eigenvalues since
they
carry Lorentzian space-time momentum. Acting on such states \refb{e1} is perfectly well-defined,
but \refb{e2} is divergent. It is these divergences that show up as apparent divergences in the integration over
the locations of the vertex operators (punctures) in the first quantized formalism.

String field theory suggests different (but equivalent) ways of resolving this problem. The first approach will be to
directly use the expressions for amplitudes given as sum over Feynman diagrams without using \refb{e2}. It is possible to formulate string field theory in a way that the sum over intermediate states converges rapidly
even though the sum runs over infinite number of states\cite{1703.06410}. 
The second approach will be to  rewrite \refb{e1} by:
\ben \label{e3.5}
&& {1\over \pi} \, b_0 \, \bar b_0 \, \left[\int_0^\Lambda \, ds \, e^{-s(L_0+\bar L_0)} + (L_0+\bar L_0)^{-1} 
e^{-\Lambda (L_0+\bar L_0)} \right]\, \int_0^{2\pi} d\theta \, e^{-i \theta (L_0-\bar L_0)}\nonumber \\
&=& -{1\over 2\pi i} \,  b_0 \, \bar b_0 \, \left[\int_{e^{-\Lambda}\le |q|\le 1} d^2 q \, q^{L_0-1} \bar q^{\bar L_0-1}
- {1\over \bar L_0} \int_{|q|=e^{-\Lambda}} dq \, q^{L_0-1} \bar q^{\bar L_0}
\right] \, ,
\een
where $\Lambda$ is a large number and in the last term the integration over $q$ runs in the anti-clockwise 
direction. In the second expression in \refb{e3.5} we have used the fact that in the second term the integration over
$q$ projects to $L_0=\bar L_0$ states, allowing us to replace the $L_0+\bar L_0$ factor in the denominator by
$2\bar L_0$. The total  expression is  independent of $\Lambda$ and is equal to \refb{e1} both for positive and 
negative values of
$L_0+\bar L_0$. However the advantage of this representation is that
for large $\Lambda$ the contribution from the first term can be represented as  the usual string amplitude
with integration range over the world-sheet variables having certain excluded domains corresponding to the
$|q|<e^{-\Lambda}$ regions, while the second term inside the square bracket receives appreciable
contribution only from finite number of states with 
$L_0+\bar L_0\le 0$.\footnote{As discussed in the last paragraph of \S\ref{slocal}, 
for on-shell amplitudes the organization of various terms based on 
the representation \refb{e3.5} can be reinterpreted as the one based on the 
representation \refb{e1} by redefining the interactions
vertices of string field theory.
This can be achieved by absorbing the first term inside the square bracket in \refb{e3.5} 
and the $e^{-\Lambda (L_0+\bar L_0)}$
factor from the second term
into the definition of the interaction vertices, leaving only the propagator factor proportional to 
$(L_0+\bar L_0)^{-1}$. In the string field theory literature, this operation is known 
as `adding stubs'\cite{9301097}. \label{fo11}}
A third representation of the propagator \refb{e1} is:
\be  \label{e2alt}
{1\over \pi} \, b_0 \, \bar b_0 \, \left[\int_0^\Lambda \, ds \, e^{-s(L_0+\bar L_0)} 
+ \int_\Lambda^{\Lambda+i\infty} \, ds \, e^{-s(L_0+\bar L_0-i\eps)}  
\right]\, \int_0^{2\pi} d\theta \, e^{-i \theta (L_0-\bar L_0)}\, ,
\ee
where $\eps$ is a small positive parameter which we eventually take to 0. This 
representation of the propagator was used in \cite{1307.5124} to get finite results for string amplitudes.
This is equivalent to the Feynman $i\eps$
prescription, but applied to tree level diagrams. In order to use \refb{e2alt} for
numerical evaluation of the amplitude, we must work with a finite $\eps$ since for $\eps=0$ the integrand
is oscillatory and not damped. The actual answer is then obtained by evaluating this expression for various
values of $\eps$ and then extrapolating the result to $\eps\to 0$.

Of course in all these approaches, we have a genuine divergence when 
an intermediate state has exactly vanishing $L_0+\bar L_0$. This 
represents the physical poles of the tree level S-matrix, and happens when we
choose the external momenta such that an intermediate state goes on-shell. 

%\item

In 
the 
computation of on-shell amplitudes in
superstring theory,
the apparent divergence arising from intermediate states with negative
$(L_0+\bar L_0)$ eigenvalue can some time be enhanced due to the use of vertex operators of wrong
picture number. 
String field theory gives a definite prescription for the choice of locations of PCOs
when two or more vertex
operators come close. This in particular guarantees that only $-1$ and $-3/2$ picture states propagate as
intermediate states. 
If we use the wrong picture number,  {\it e.g.} by converting some of the NS sector vertex operators
into zero picture, states in 0 picture propagate in the intermediate state when a pair
of zero picture vertex operators approach each other. 
This sector has states with negative conformal weights leading to
apparent divergences which are not present in the correct treatment of string field theory.

In this paper we show that using the representation \refb{e3.5} of the propagator (equivalently 
representation  \refb{e1}
following footnote
\ref{fo11}) and the fact that the change in the locations of
PCOs results in addition of total derivative terms to the integrand, one can arrive at the following prescription for
computing (super-)string tree amplitudes that avoids all divergences:
\begin{enumerate}
\item Let us suppose that  for an $(n+3)$-point 
amplitude of an arbitrary set of vertex operators in an arbitrary (compactified) string theory, 
$\II^{(0)}(\sigma_1,\cdots ,\sigma_n)$ is the integrand obtained by computing
the relevant correlation functions of $(n+3)$ vertex operators on the sphere, with three of the vertex operators
at fixed positions and the $n$ vertex operators inserted at $\sigma_1,\cdots,\sigma_n$. 
Naively the amplitude
is given by the integral of the $2n$-form $\II^{(0)}$ over the whole moduli space parametrized by $\sigma_i$'s,
but this integral diverges from the regions where the vertex operators come close, or equivalently,
the sphere degenerates. 
\item The basic degenerations are single
degenerations where 
the sphere degenerates into a pair of spheres, one carrying $p$ external 
punctures and an `internal' puncture and the
other carrying $(n+3-p)$ external punctures and an internal puncture,
with $p\ge 2$, $(n+3-p)\ge 2$. The two spheres are sewed to each other by cutting out small disks around the
internal punctures and gluing the boundaries of the disks.
We label all such degenerations by the label $s$
-- different values of $s$ correspond to different ways of dividing $(n+3)$ punctures into $p$ punctures and
$(n+3-p)$ punctures. 
Near such a degeneration we can construct from the
$\{\sigma_i\}$'s a new coordinate system, containing a single complex 
coordinate $\xtau_s$ that vanishes at the degeneration and a set of complex coordinates 
$m_{(s)}$ labelling the moduli of the pair of punctured spheres into which the original sphere degenerates.
\item Multiple degenerations correspond to the region of the moduli space where the original sphere degenerates
into several spheres.
A degeneration into $(k+1)$ spheres is characterized uniquely by a set of $k$ unordered 
labels $s_1,\cdots, s_k$ where each $s_i$ corresponds to one of the labels characterizing single degenerations.
Near such degenerations we can construct from the $\{\sigma_i\}$'s a set of complex
coordinates $\xtau_{s_1},\cdots , \xtau_{s_k}$ that vanish at the degeneration and another
set of $n-k$ complex coordinates $m_{(s_1,\cdots, s_k)}$, labelling the moduli of the punctured spheres to which
the original sphere degenerates.
\item We now denote by $\DD_s$ an open tubular neighborhood 
of the subspace $\xtau_s=0$ and define $\CC^{(0)}$ to be the region
of the moduli space that excludes $\DD_s$ for all $s$.\footnote{For example in a local patch we could take
$|\xtau_s|<\eps$ for some small number $\eps$
as the definition of $\DD_s$. The choice of $\xtau_s$'s is of 
course not unique, 
but the final result is independent
of this choice. Furthermore the coordinates
$\{\xtau_{s_1},\cdots, \xtau_{s_k}\}$ may need to be defined differently in different local patches of the
space spanned by $m_{(s_1,\cdots, s_k)}$. 
In this case on the overlap one needs to choose appropriate interpolation in the definition of $\DD_s$ between
the two patches.
An explicit example elaborating the choice of the coordinates
$\{\xtau_s\}$, $\{m_{(s_1,\cdots, s_k)}\}$ and the definition of $\DD_s$ 
for a five punctured sphere has been described in appendix 
\ref{sb}.}
Therefore by construction $\CC^{(0)}$ excludes all degenerations.
We also denote by $\CC^{(1)}_s$ 
the intersection $\p\DD_s\cap \CC^{(0)}$, forming a component of the boundary of $\CC^{(0)}$,
and by
$\CC^{(k)}_{s_1\cdots s_k}$ the codimension $k$ 
intersection $\CC^{(1)}_{s_1}\cap \CC^{(1)}_{s_2}\cap\cdots\cap \CC^{(1)}_{s_k}$.
The orientation of
$\CC^{(k)}_{s_1\cdots s_k}$ is fixed by the equation
\be
\p \CC^{(k)}_{s_1\cdots s_k} = -\sum_s \CC^{(k+1)}_{s_1\cdots s_k s}\, ,
\ee
where the sum over $s$ runs over all labels for which $\CC^{(k+1)}_{s_1\cdots s_k s}$  exists.
\item Then the full amplitude
(for generic external momenta) is given by
\be \label{etotalIintro}
I = \sum_{k=0}^n  \sum_{\{s_1,\cdots ,s_k\}\atop s_1<s_2<\cdots < s_k}
\int_{\CC^{(k)}_{s_1\cdots s_k}} \II^{(k)}_{s_1\cdots s_k}\, ,
\ee
where the $(2n-k)$ forms $\II^{(k)}_{s_1\cdots s_k}$, defined in a neighborhood of 
$\CC^{(k)}_{s_1\cdots s_k}$, are obtained by solving the equations
\be \label{edikeqintro}
d \, \II^{(k)}_{s_1\cdots s_k} = \II^{(k-1)}_{s_1\cdots s_{k-1}} - \II^{(k-1)}_{s_1 \cdots  s_{k-2} s_k} + \cdots
+(-1)^{k-1} \II^{(k-1)}_{s_2 \cdots  s_k} \, .
\ee
Beginning with the known expression for $\II^{(0)}$,
we could solve for $\II^{(k)}_{s_1\cdots s_k}$ iteratively in $k$ by expanding both sides of \refb{edikeqintro} in a
power series in $\xtau_{s_1},\cdots,\xtau_{s_k}$ and their complex conjugates and matching the
coefficients. 
\item The solution to \refb{edikeqintro} is not unique but the final result is not
affected by this ambiguity, as long as we work with generic external momenta so that the exponents
in the power series expansion in $\xtau_s,\bar\xtau_s$ are not integers.
Similarly the choice of $\CC^{(1)}_s$'s and therefore of the $\CC^{(k)}_{s_1\cdots s_k}$'s
depend on the choice of the tubular neighborhoods $\DD_s$ which are not unique. However the expression
\refb{etotalIintro} can be shown to be invariant under deformations of the $\CC^{(1)}_s$'s.
\end{enumerate}
Note that for $k=0$ term in \refb{etotalIintro}, the integrand is the same as the original integrand but the 
integration runs over the region $\CC^{(0)}$ that avoids all degenerations. The missing regions are compensated
for by the remaining integrals involving $k\ge 1$ terms. Also note that the final formula for the amplitude encoded
in \refb{etotalIintro} and \refb{edikeqintro} does not require any detailed knowledge of string field theory even though
we use string field theory to arrive at these formul\ae.

We can regard $\int_{\CC^{(k)}_{s_1\cdots s_k}} \II^{(k)}_{s_1\cdots s_k}$ as the result of using
 the second term in \refb{e3.5} -- or equivalently \refb{e2alt} -- 
 for the propagators labelled by $s_1,\cdots, s_k$ if we identify the variable 
 $\xtau_s$ with the variable $q$ for the $s$-th propagator and the tubular neighborhood $\DD_s$ as the
 region $|\xtau_s| < e^{-\Lambda}$.
As we shall see, this choice of $\{u_s\}$ and $\{\DD_s\}$ 
is not strictly necessary. Nevertheless this
 makes the connection to the $i\eps$ prescription of \cite{1307.5124} clear:
 $\int_{\CC^{(k)}_{s_1\cdots s_k}} \II^{(k)}_{s_1\cdots s_k}$ represents part of the contribution where the integrals
 over $\ln|\xtau_s|$ runs parallel to the imaginary axis for $s=s_1,\cdots, s_k$. We note however that
the prescription given in \refb{etotalIintro} works equally well for complex external momenta, while the $i\eps$
prescription will typically require different choices of the contour of integration in the complexified moduli
space for different complex external momenta to ensure convergence.

The procedure described above works even if the original integrand $\II^{(0)}$ is computed with wrong choice
of PCO locations, {\it e.g.} by taking the PCO locations to coincide with some of the vertex operators. This adds
a total derivative to $\II^{(0)}$ and could introduce additional divergences when the vertex operators collide.
However the addition of such terms to $\II^{(0)}$ also requires us to add corrections to $\II^{(1)}_s$ when we
solve \refb{edikeqintro}, and these cancel in \refb{etotalIintro}. Therefore while implementing this procedure, one
does not need to be careful about the choice of PCO locations.

As already mentioned, during the analysis described above we do not assume any particular choice of string 
background or any specific form of the
external states. Our analysis holds for generic external states in a generic (compactified) string theory. 

%\item 
The second application of classical string field theory that
we discuss is in the study of marginal deformations of the world-sheet CFT describing the
background space-time.
String field theory is formulated around some specific classical solution, corresponding to a CFT 
of matter and ghost fields on the world-sheet. If this CFT has marginal deformations, then
one can in principle compute the spectrum and correlations functions of all the operators in the deformed theory
using conformal perturbation theory. This in turn will determine the spectrum of physical states and the
S-matrix of string theory formulated in the deformed background, in terms of the spectrum and correlation
functions of the original CFT. In practice this requires carefully regulating the ultraviolet divergences
on the world-sheet since the marginal operator needs to be integrated on the world-sheet together with other vertex
operators and we have to regulate the divergences when the locations of the operators come close to each other.
In particular at higher order in perturbation theory when there are several insertions of the marginal operator we need to
carefully remove all the divergences, while making sure not to  remove the finite pieces. 

Now given the deformed CFT, one can formulate a string field theory around this deformed CFT. 
It is known that this new string field theory is related to the string field theory formulated around the original
CFT 
by a field redefinition that includes a shift\cite{back1,back2,1711.08468}. 
This means that the original string field  theory has a classical solution
that describes string field theory around the deformed 
background.\footnote{For open string field theory, such classical
solutions have been constructed analytically\cite{0701248,0701249,0704.2222,1009.6185,1406.3021}.}
Furthermore the spectrum and S-matrix of the string
field theory around the deformed CFT can be computed by expanding the original string field theory action around the
classical solution describing the deformed background.
Since sting field  theory has no divergences, this procedure also does not suffer from any 
divergences. In other worlds string field theory automatically provides 
an ultraviolet regulator for the world-sheet theory.
A fully systematic procedure for constructing the classical solution in string field theory 
to any given order in the expansion
in terms of the deformation parameters, and computing the spectrum and S-matrix of string theory in the deformed 
background, can be found in \cite{1411.7478,1703.06410}. Its application to a particular class
of examples has also 
been described in \cite{1508.02481} (see also \cite{mukherji,9201040} 
for construction of the 
solution).
The apparent ultraviolet divergences in the conformal perturbation theory can be
traced either to wrong use of Schwinger parametrization as in \refb{e2}, or applying $(L_0+\bar L_0)^{-1}$ on states
that have zero $(L_0+\bar L_0)$ eigenvalue. The first case is dealt with by using \refb{e1} or \refb{e3.5}, 
while the second
case requires a more elaborate treatment as explained in sections 3.3 and 4.1 of 
\cite{1411.7478}.

We apply this method to the special case of string compactification on a circle, and consider the marginal
deformation to be the one that changes the radius of the circle. The mass of a state carrying momentum 
$n/R$ along the circle gets shifted under this deformation. We find that both for the bosonic and the heterotic string
theory the mass$^2$ shift to second order in the deformation parameter $\mu$ is given by
\be\label{emassshift}
\Delta m^2 = -n^2 R^{-2} \left(\mu - {\mu^2\over 2}\right)\, .
\ee
Since the leading order expression for $m^2$ has an additive term given by $n^2 R^{-2}$, \refb{emassshift}
corresponds to a scaling of $R^{-2}$ by $(1-\mu +\mu^2/2)$. This is consistent since the marginal deformation
is expected to change the radius. If $\Delta m^2$ had not been proportional to $n^2$ 
then this would not be possible.
Indeed at the intermediate stages of the calculation there are other terms proportional to $n^4R^{-4}$ and 
$n^0R^0$,
but they cancel at the end. Nowhere at any stage of the calculation we need an ultraviolet regulator on the
world-sheet. Furthermore the method we use is completely general and can be used to compute the shift in the
spectrum under a general marginal deformation where the answer may not be {\it a priori} known, {\it e.g.} shift in
the masses of heavy string states under a blowup of the orbifold singularity.

%\item 
Another possible application of string field theory is in the study of Ramond-Ramond 
background\cite{1811.00032}.
Once a string field theory is formulated around a given world-sheet CFT, 
one may have a family of consistent classical solutions that involve switching on Ramond-Ramond background.
In this case one can use the same
procedure used for marginal deformations to systematically construct the classical solution in string field theory to
any order in deformation parameter, and compute the spectrum and S-matrix of the theory around the deformed
background\cite{1811.00032}. 
As in the case of marginal deformations, this procedure never requires an ultraviolet regulator on the
world-sheet if we 
apply the formalism of string field theory systematically.

%\end{enumerate}

As is well known, formulation of string field theory requires us to specify certain data -- the choice of local coordinates
at the punctures of the Riemann surface and locations of the PCOs. The results we have quoted are independent
of the data. Indeed, one of the goals in our analysis will be to manipulate the various expressions in such a way that
even though at the intermediate stages of the calculation the result depends on these additional data (which
we leave unspecified), this dependence cancels at the end.  This is manifest in our expressions \refb{etotalIintro}
and \refb{edikeqintro}. One finds for example that different choice of local coordinate system at the punctures lead
to different choices of the subspaces $\CC^{(1)}_s$ (and hence also $\CC^{(k)}_{s_1\cdots s_k}$), but
it can be shown that
\refb{etotalIintro} is invariant under deformations of the $\CC^{(1)}_s$'s.\footnote{For 
example if we use hyperbolic metric to introduce the local coordinate 
system as in \cite{1703.10563,1706.07366,1708.04977}
then the subspaces 
$\CC^{(1)}_s$'s will be obtained by setting to some small number $\ell_0$ 
the length of the closed geodesic that shrinks to a point at the $s$-th degeneration of the punctured sphere.
Similar regularization was used in \cite{1611.08003} 
for computing tree level string amplitudes with large number of external
states, but only the $\int_{\CC^{(0)}}\II^{(0)}$ contribution was analyzed in
\cite{1611.08003}.
}
Similarly different choice of PCO locations
can be shown to induce an additive total derivative term in the expression for $\II^{(0)}$, but \refb{etotalIintro} 
can be shown to remain unchanged under addition of such a term. In essence, such an additive
term in $\II^{(0)}$ forces us to also add certain terms in $\II^{(1)}_s$ in order to satisfy \refb{edikeqintro}. These
extra terms cancel the effect of additive term in $\II^{(0)}$ when we evaluate \refb{etotalIintro}.

Similarly we see that \refb{emassshift} does not depend on the choice of local coordinate systems or PCO locations.
At intermediate stages of the analysis various expressions we get do depend on the additional data, but this dependence
cancels at the end. We must note however that the total independence of the result \refb{emassshift}
of the additional data is
accidental since different choice of local coordinates and PCO locations lead to string field theories that are related
by field redefinition and we expect that under such a field redefinition the deformation parameter $\mu$ -- which is the
component of a field -- will also get redefined. Therefore at higher order in the expansion in powers of $\mu$, 
we do expect
the result to depend on the additional data involved in the construction of string field theory, but this dependence 
should be removable by a redefinition of $\mu$.

The analysis leading to \refb{emassshift} is
close to the spirit of the analysis in \cite{1811.00032} in that the latter paper computed
the shift in mass under deformation involving RR background to second order in the deformation parameter $\mu$. 
There
is however one important technical difference between the two analysis. 
In the case of RR background, the leading contribution to the mass
shift appears at second order in $\mu$. Therefore for computing the mass shift, 
the vertex operators of the states whose mass shift is being
calculated could be taken to be dimension zero primaries. For the
deformation we consider in this paper, 
the first correction to the mass already appears at order $\mu$, and therefore to compute the
mass at second order in $\mu$ we need to use vertex operators whose dimensions differ from zero by order $\mu$.
This introduces non-trivial dependence on the choice of local coordinate system at intermediate steps of the
calculation even though the dependence cancels at the end. This difference is similar to the difference between the
calculation of one loop mass renormalization and two loop mass renormalization. The former requires computing
the torus two point function of a pair of vertex operators that satisfy tree level on-shell condition. However the latter
requires use of vertex operators that take into account one loop mass renormalization, and therefore do not
satisfy the tree level on-shell condition.

The rest of the paper is organized as follows. In \S\ref{s2} we review some conventions we use to
compute correlation functions in the world-sheet CFT
and briefly review some aspects of 
string field theory that we need for our analysis.
In \S\ref{s3} we discuss how choice of local coordinate system and PCO locations affect the results for Feynman
diagrams in string field theory. In \S\ref{s5} we make use of the representation \refb{e3.5} of the propagator to
give a manifestly finite expression for the four tachyon amplitude in bosonic string theory. In \S\ref{shigher} we use
\refb{e3.5} to give a general algorithm for getting manifestly finite expressions for general amplitudes in bosonic
string theory involving arbitrary number of external states carrying arbitrary quantum numbers. 
Our analysis in this section leads to eqs.\refb{etotalIintro}, \refb{edikeqintro}.
In \S\ref{ssuper} we use superstring field theory to arrive at 
the same formul\ae\ \refb{etotalIintro}, \refb{edikeqintro}, giving manifestly 
finite expressions for the superstring tree amplitudes. In particular we show that \refb{etotalIintro}, \refb{edikeqintro}
give the correct amplitude even if $\II^{(0)}$ is computed using  wrong choice of PCO locations.
In \S\ref{s6} we apply string field theory to study the effect of marginal deformations in 
bosonic and heterotic
string theory and arrive at \refb{emassshift} in both theories. 
We conclude in \S\ref{sdiss} with some comments on higher genus amplitudes.
Appendix \ref{sb} contains examples of the choice of the coordinate $\{\xtau_s\}$, $\{m_{(s_1\cdots s_k)}\}$ for
a five punctured sphere. Appendix \ref{sa} contains some technical results needed
to complete the analysis in \S\ref{shetb}.

\sectiono{Conventions} \label{s2}

In this section we shall briefly review some aspects of the world-sheet theory and the string field theory
that we shall need for our analysis . More details can be found in \cite{1703.06410}.
We begin
by describing the normalization conventions for the vacuum of the world-sheet (super-)conformal field theory
of the matter ghost system.
For bosonic string theory the ghost system has $b,c,\bar b$ and $\bar c$ ghosts with the usual mode expansion.
We normalize the SL(2,C) invariant vacuum of the bosonic string as
\be \label{ebosenorm}
\langle 0| c_{-1} \bar c_{-1} c_0 \bar c_0 c_1 \bar c_1 |0\rangle = -1\, ,
\ee
up to a factor given by the overall volume of space-time which eventually generates the momentum
conserving delta function in a correlator. We shall not write this factor explicitly.
For heterotic string theory, besides the $b,c,\bar b, \bar c$ ghost fields we also have
the $\beta,\gamma$ ghosts, related to the $\xi,\eta,\phi$ system via the relations
\be
\beta =\p\xi e^{-\phi}, \quad \gamma=\eta\, e^\phi\, .
\ee
We choose the normalization of the vacuum of the heterotic string such that 
\be \label{ehetnorm}
\langle 0| c_{-1} \bar c_{-1} c_0 \bar c_0 c_1 \bar c_1 e^{-2\phi(z)} |0\rangle = 1\, .
\ee
For type II string theories we also have anti-holomorphic $\bar\beta,\bar\gamma$ system and the normalization
of the vacuum will be chosen as
\be\label{eiinorm}
\langle 0| c_{-1} \bar c_{-1} c_0 \bar c_0 c_1 \bar c_1 e^{-2\phi(z)}  e^{-2\bar\phi(\bar z)} |0\rangle = -1\, .
\ee

We shall denote by $X^\mu$ the world-sheet fields corresponding to non-compact space time coordinates,
and, for the analysis in \S\ref{s6},
 by $Y$ the world-sheet field corresponding to a compact space direction of radius $R$. 
We also denote by
$\psi^\mu$ and $\chi$ their holomorphic 
superpartners on the world-sheet -- in type II theories we also have anti-holomorphic fields
$\bar\psi^\mu$ and
$\bar \chi$. Their operator product expansions have the form:
\ben \label{ematterope}
&&
\p X^\mu(z) \p X^\nu(w) = -{\eta^{\mu\nu}\over 2 (z-w)^2}+\cdots, \quad \psi^\mu (z) \psi^\nu(w) = -{\eta^{\mu\nu}\over 2 (z-w)}+\cdots\, ,
\nonumber \\
&& \p Y(z) \p Y(w) = -{1\over 2 (z-w)^2}+\cdots, \quad
\chi (z) \chi(w) = -{1\over 2 (z-w)}+\cdots\, ,
\nonumber \\
&& \bar\p X^\mu(\bar z) \bar\p X^\nu(\bar w) 
= -{\eta^{\mu\nu}\over 2 (\bar z-\bar w)^2}
+\cdots, \quad
\bar \psi^\mu (\bar z) \bar\psi^\nu (\bar w) 
= -{\eta^{\mu\nu}\over 2 (\bar z-\bar w)}
+\cdots, \nonumber \\
&& \bar \p Y(\bar z) \bar\p Y(\bar w) = -{1\over 2 (\bar z-\bar w)^2}+\cdots, \quad
\bar\chi (\bar z) \bar\chi(\bar w) = -{1\over 2 (\bar z-\bar w)}+\cdots\, ,
\een
where $\cdots$ denote less singular terms whose knowledge will not be
needed for our analysis. There may also be additional component of the matter CFT
describing other compact directions that 
will not be relevant for our analysis.
The operator product expansion of the ghost fields take the form
\ben \label{eghostope}
&& c(z) b(w) =(z-w)^{-1}+\cdots, \quad \xi(z)\eta(w) = (z-w)^{-1}+\cdots, \nonumber \\
&& \bar c(\bar z) \bar b(\bar w) = (\bar z-\bar w)^{-1}
+\cdots, \quad \bar\xi(\bar z)\bar\eta(\bar w) = (\bar z-\bar w)^{-1}+\cdots, \nonumber \\
&& e^{q_1\phi(z)} e^{q_2\phi(w)} = (z-w)^{-q_1q_2} e^{(q_1+q_2)\phi(w)}+ \cdots \, , \quad
\p \phi (z)\,  \p\phi(w) = -{1\over (z-w)^2} +\cdots, \nonumber \\
&&
e^{q_1\bar\phi(\bar z)} e^{q_2\bar\phi(\bar w)} 
= (\bar z-\bar w)^{-q_1q_2} e^{(q_1+q_2)\bar\phi(\bar w)}+ \cdots \, , \quad 
 \bar\p \bar\phi (\bar z)\,  \bar\p\bar\phi(\bar w) = -{1\over ( \bar z-\bar w)^2} +\cdots \, ,
\een
where $\cdots$ denote less singular terms. 

In the heterotic string theory we have holomorphic PCO given by
\be \label{epicture}
\XX(z) = \{Q_B, \xi(z)\} = c \, \partial \xi + 
e^\phi T_F - {1\over 4} \p \eta \, e^{2\phi} \, b
- {1\over 4} \p\left(\eta \, e^{2\phi} \, b\right)\, ,
\ee
where $Q_B$ is the BRST charge:
\ben \label{ebrs1}
Q_B &=& \ointop dz \jmath_B(z) + \ointop d\bar z \bar \jmath_B(\bar z)\, , \nonumber \\
\bar \jmath_B(\bar z) &=& \bar c(\bar z) \bar T_m(\bar z)
+\bar b(\bar z) \bar c(\bar z) \bar\p \bar c(\bar z)\, , \nonumber \\
\jmath_B(z) &=& c(z) (T_{m}(z) + T_{\beta,\gamma}(z) )+ \gamma (z) T_F(z) 
+ b(z) c(z) \p c(z) 
-{1\over 4} \gamma(z)^2 b(z)\, .
\een
$T_m$, $\bar T_m$ denote  components of the matter stress tensor and $T_{\beta,\gamma}$ is the stress
tensor of the $\beta,\gamma$ system.
$\ointop_z$ includes the $1/2\pi i$ factor for holomorphic integral and
$-1/2\pi i$ factor for the anti-holomorphic integral.
$T_F$ is the super-stress tensor of the matter SCFT, given by
\be \label{edefstress}
T_F(z) = - \psi_\mu \p X^\mu -\chi \p Y + (T_F)_{int}\, . \nonumber \\
\ee
Here $(T_F)_{int}$ denotes the contribution from the additional compact target space directions
other than the $Y$-$\chi$ system. In the bosonic string theory the contribution from the $\beta,\gamma$
system will be absent, while in type II theory there will be additional contribution involving 
$\bar\beta,\bar\gamma$. In type II theory we also have the anti-holomorphic 
PCO, obtained by replacing the holomorphic fields by anti-holomorphic fields in \refb{epicture}.

In the bosonic string theory the physical unintegrated
vertex operators take the form $c\bar c V$ where $V$ is a dimension
(1,1) primary in the matter CFT. From this one can construct the 
integrated vertex operator:
\be
\left(-\ointop_z dw b(w)\right) \, \left(-\ointop_z d\bar w \bar b(\bar w)\right)  
\, c(z) \bar c(\bar z) V (z,\bar z) = - V(z, \bar z)\, ,
\ee
where $\ointop_z$ denotes a contour around $z$.

In the heterotic string theory the unintegrated $-1$ picture NS sector vertex operator takes the form
\be \label{eunint}
c\bar c e^{-\phi} V \, ,
\ee
where $V$ is a dimension $(1,1/2)$ superconformal primary in the matter SCFT. The unintegrated zero picture
vertex operator takes the form
\be \label{eunint0}
\lim_{w\to z} \XX(w) \, c\bar c\, e^{-\phi} V(z,\bar z) = c \bar c \, W(z,\bar z) -{1\over 4} \eta \, \bar c\, e^\phi V(z,\bar z)\, ,
\ee
where $W$ is a dimension (1,1) matter sector vertex operator defined via
\be \label{edefW}
W(z,\bar z) = -\lim_{w\to z} (w-z) T_F (w) V(z,\bar z)\, .
\ee
The integrated $-1$ picture vertex operator takes the form
\be \label{eintm1}
\left(-\ointop_z dw b(w)\right) \, \left(-\ointop_z d\bar w \bar b(\bar w)\right)  c\bar c e^{-\phi} V(z, \bar z)
= - e^{-\phi} V(z,\bar z)\, ,
\ee
and
the integrated 0 picture vertex operator is given by
\be\label{eint0}
\left(-\ointop_z dw b(w)\right) \, \left(-\ointop_z d\bar w \bar b(\bar w)\right) 
 \left[c \bar c \, W(z,\bar z) -{1\over 4} \eta \, \bar c\, e^\phi V(z,\bar z)\right] = -W(z,\bar z) \, .
 \ee
In type II string theory we can similarly define integrated and unintegrated 
NSNS sector vertex operators carrying
picture numbers $(-1,-1)$, $(0,-1)$, $(-1,0)$ and $(0,0)$. We shall not write down the explicit
form of the Ramond sector vertex operators since they will not be needed for our analysis.

Tree level
$(n+3)$-point amplitude of vertex operators $c\bar c V_i$ for $1\le i\le (n+3)$ in bosonic string
theory is given by converting $n$ of them to integrated vertex operators and
integrating the resulting correlation function over the locations of the
integrated vertex operators:
\be \label{eampcon}
A=\left(-{1\over 2\pi i}\right)^n \int \prod_{i=1}^n d\sigma_i \wedge d\bar\sigma_i \,
 \left\langle c\bar c V_{n+1}(y_{n+1}) \, c\bar c V_{n+2} (y_{n+2}) \, c\bar c V_{n+3}(y_{n+3}) 
 \, \prod_{i=1}^n (-V_i(\sigma_i)) \right\rangle\, .
\ee
Here $y_{n+1}$, $y_{n+2}$ and $y_{n+3}$ represent arbitrary points in the complex plane.
The $(-1/2\pi i)^n$ factor is a normalization factor that appears in the definition of
the interaction vertices of string field theory and is related to the $-1/2\pi i$
factor in \refb{e2}.
$A$ given in \refb{eampcon} can be regarded as a contribution to the $T$-matrix, related to the 
$S$-matrix by $S=1+i\, T$.

The results for the tree level $(n+3)$ point function of NS sector states 
in the heterotic string theory and NSNS sector states in type II string 
theories are similar, except that the
vertex operators are taken in the $-1$ picture and  
the amplitude has insertion of $(n+1)$
of PCOs in the heterotic theory and  $(n+1)$ holomorphic PCO's and $(n+1)$
pair of anti-holomorphic PCO's in type II theory. The PCO locations are
arbitrary when the vertex operators are well separated from each other but
need to satisfy certain relations when two or more vertex operators approach
each other. These rules are induced from superstring field theory and essentially
tell us that when $n$ vertex operators approach each other, we must also have $(n-1)$
PCOs approaching them so that the picture numbers of all the operators add up to
$-1$. If there are R sector vertex operators present, then they are taken in the
$-1/2$ picture. The number of PCO's need to be adjusted so that the total picture number
of all vertex operators and PCOs add up to $-2$ in the heterotic string theory and
$(-2,-2)$ in the type II string theory.

We shall now briefly review some aspects of (super-)string field theory. In bosonic string
theory we denote by $\HH$ the Hilbert space of matter-ghost CFT satisfying the conditions:
\be \label{econdproj}
|\phi\rangle \in \HH \quad \hbox{if $b_0^-|\phi\rangle  = 0$, $L_0^-|\phi\rangle = 0$}, \quad
b_0^\pm \equiv b_0\pm \bar b_0, \quad L_0^\pm \equiv L_0\pm \bar L_0, \quad 
c_0^\pm \equiv {1\over 2} (c_0\pm\bar c_0)\, .
\ee
For NS sector of heterotic string theory and NSNS sector of type II string theory we have similar
constraints except that we also require the states in $\HH$ to carry picture number $-1$ in
heterotic string theory and picture number $(-1,-1)$ in type II string theory. In all the theories the
string field $|\Psi\rangle$ is an arbitrary  element of $\HH$. We shall denote by $\Psi$ the
corresponding vertex operator in the CFT.

The classical action of bosonic string field theory and the NS sector fields in superstring field theory 
takes the form:\footnote{For notational simplicity we have dropped the string coupling
constant $g_s$ from this expression. It appears  in the action via an overall multiplicative factor $g_s^{-2}$.}
\be \label{esftaction}
S = {1\over 2} \langle\Psi| c_0^- Q_B |\Psi\rangle + \sum_{N=3}^\infty {1\over N!} \{\Psi^N\}\, ,
\ee
where for $|A_i\rangle\in\HH$, 
$\{A_1\cdots A_n\}$ is a multilinear function of the $A_i$'s obtained by first constructing the sphere correlation
functions of certain ghost operators and PCOs
and the vertex operators $A_i$ in specified coordinate system, 
and then integrating the result over an appropriate subspace of the moduli space
that excludes all the singular regions where two or more vertex operators come close. The
precise choice of these subspaces, or the coordinate system in which the vertex operators are
inserted, or the PCO locations are not fixed completely but are subject to stringent constraints, and different
choices lead to different string field theories which are related by field redefinition.
Note that in this definition 
we do not require the $A_i$'s to be BRST invariant.

We also define $[A_2\cdots A_n]\in \HH$ such that
\be \label{edefsquare}
\langle A_1|c_0^- | [A_2\cdots A_N]\rangle = \{A_1\cdots A_n\} \quad \hbox{for $|A_i\rangle\in \HH$}\, .
\ee
In terms of this, the equations of motion of the string field $|\Psi\rangle$, derived from the action 
\refb{esftaction}, takes the form:
\be\label{eeom}
Q_B|\Psi\rangle + \sum_{N=2}^\infty {1\over N!} [\Psi^N] = 0\, .
\ee

Inclusion of the Ramond (R) sector states requires additional structure. In heterotic string theory
we introduce
a pair of string fields $\Psi_R$ and $\wt\Psi_R$ belonging to  $-1/2$ and $-3/2$ picture number sectors
satisfying \refb{econdproj}. 
In superstring theory $\Psi_R$ will include fields in the $(-1, -1/2)$, $(-1/2, -1)$ and
$(-1/2,-1/2)$ sectors while 
$\wt\Psi_R$ will include fields in the $(-1, -3/2)$, $(-3/2, -1)$ and
$(-3/2,-3/2)$ sectors.
The kinetic 
term takes the form
\be 
-{1\over 2} \langle\wt\Psi_R| c_0^- Q_B \GG|\wt\Psi_R\rangle + \langle\wt\Psi_R| c_0^- Q_B |\Psi_R\rangle\, .
\ee
In heterotic string theory $\GG$ is the zero mode of the PCO. In type II string theory $\GG$ is the
zero mode of the holomorphic (anti-holomorphic) PCOs in the NSR (RNS) sectors
and product of zero modes of 
holomorphic and anti-holomorphic PCOs in the RR sector. The interaction term involving
the NS and R sector fields have the form
\be
\sum_{N=3}^\infty {1\over N!} \{(\Psi+\Psi_R)^N\}\, .
\ee
In particular the field $\wt\Psi_R$ does not enter the interaction term. Therefore the Feynman rules 
only require the $\Psi-\Psi$ propagator. In Siegel gauge this has the form similar to \refb{e1} except
for an extra factor of $\GG$:
\be\label{erprop}
2\, b_0 \, \bar b_0 \, (L_0+\bar L_0)^{-1} \, \GG\, \delta_{L_0, \bar L_0}\, .
\ee
The relevant equations of motion in the Ramond sector, obtained by taking a linear combination of the
equations of motion of $\Psi_R$ and $\wt\Psi_R$ is given by
\be
Q_B|\Psi_R\rangle + \sum_{N=2}^\infty {1\over N!} \GG\, [(\Psi+\Psi_R)^N] = 0\, .
\ee
The other linear combination satisfies free field equations of motion and decouples from the theory. Of course we must also replace $\Psi$ by $\Psi+\Psi_R$ in the second term
in \refb{eeom}.

\sectiono{Local coordinates and picture changing operators} \label{s3}

As mentioned earlier, for the construction of a string field theory action, we need to define the interaction
vertices involving off-shell string fields. For $n$-point interaction vertex this requires identifying certain 
subspaces of the moduli space of a sphere with $n$ punctures, and a choice of local coordinates at
the punctures, satisfying certain consistency conditions. For superstring theory we also need to 
specify choice of the locations of the picture changing operators. In this section we shall describe 
some general criteria for choosing the 
local coordinate system at the punctures of a three punctured sphere involved
in the definition of 3-string interaction vertex and
the corresponding choice of a subspace of the moduli space of a sphere with 4 punctures, involved
in the definition of the 4-point interaction vertex. We shall also discuss the 
general criteria for the choice of locations of the
PCOs. However we shall refrain from committing ourselves to
any particular choice of local coordinates or PCO locations since one of our goals will be to 
demonstrate that
the final result is independent of these choices.

\subsection{Local coordinate system} \label{slocal}

On a sphere with three punctures, we can use SL(2,C) transformation to choose the locations of the
punctures to be at
\be \label{e2.1}
z_1=0, \qquad z_2=\infty, \qquad z_3= 1\, .
\ee
We shall choose the local coordinates $w_i$ at these punctures to be related to the global coordinate
$z$ by
\be \label{e2.2}
w_1 = \lambda\, h_1(z), \qquad w_2 = \lambda \,  h_2(z), 
\qquad w_3 = \lambda \, h_3(z), 
\ee
up to arbitrary phases.  
$h_i(z)$ satisfies $h_i(z_i)=0$ for $1\le i\le 3$, so that $w_i$ vanishes at $z=z_i$.
$\lambda$ is an
arbitrary positive real number that could have been included in the definitions of $h_i(z)$, but we have displayed it
explicitly since at the end we shall try to simplify our analysis by taking the large $\lambda$ limit. 
$h_i(z)$ is an analytic map between an open neighborhood around  $z=z_i$  and the unit disk $|w_i|<1$, but
may have singularities outside this domain. We require the images of the $|w_i|\le 1$ regions in the $z$-plane to
be non-overlapping -- this can be achieved by taking $\lambda$ sufficiently large.
Furthermore,
up to overall phases, this choice of local coordinates should be invariant under permutation of the punctures.
For example, since the transformation $z\to 1/z$ exchanges the punctures 1 and 2 and leaves $z_3$ 
invariant, we must have
\be \label{e2.2a}
h_1(1/z) = h_2(z), \quad h_3(1/z)=h_3(z)\, ,
\ee
up to phases. Similarly we should have
\be \label{e2.2b}
h_1(1-z) = h_3(z), \quad h_2(1-z)=h_2(z)\, ,
\ee
and 
\be
h_1(z/(z-1))= h_1(z), \quad h_2(z/(z-1))=h_3(z)\, ,
\ee
up to phases.\footnote{A particular choice of local coordinates satisfying these relations is given by\cite{vecc}
$h_1(z) = z /(z-2)$, $h_2(z) = 1/(2 z -1)$ and $h_3(z)= (1-z) /(1+z)$, but we shall proceed without committing
ourselves to any particular choice. \label{fo5}}

Let us now consider another sphere with three punctures, carrying global coordinate $z'$ and the punctures
situated at
\be \label{e2.3}
z_1'=0, \qquad z_2'=\infty, \qquad z_3'= 1\, ,
\ee
carrying local coordinates
\be \label{e2.4}
w_1' = \lambda\, h_1(z'), \qquad w_2' = \lambda \,  h_2(z'), 
\qquad w_3' = \lambda \, h_3(z') \, ,
\ee
around the three punctures. We can construct a two parameter family of 
spheres with four punctures by sewing the two spheres at their
third punctures via the relation
\be \label{e2.5}
w_3 \, w_3' =  e^{-s-i\theta}\equiv q, \quad 0\le s<\infty, \quad 0\le\theta<2\pi\, .
\ee
This family of four punctured 
spheres is what we shall obtain from the s-channel Feynman diagram of string field theory.
Using \refb{e2.2} and \refb{e2.4} we get the relation between the global coordinates $z$ and $z'$ on the two
spheres:
\be \label{e2.6}
\lambda^2 \, h_3(z) \, h_3(z')= q\, , 
\ee
inside open neighborhoods of $z=z_3$ and $z'=z'_3$. Since the glued Riemann surface is a sphere 
with four punctures, we can introduce a global coordinate $y$ on the four punctured sphere. If the
$h_i$'s had been SL(2,C) transformations as in footnote \ref{fo5} then $y$ could be taken to be either $z$ or
$z'$ or related to these by an SL(2,C) transformation, but in general the relation between $y$ and the
original coordinates $z,z'$ is more complicated. By an SL(2,C) transformation we can ensure that in the
$y$ plane the original punctures at $z=0$ and $\infty$ are located at $y=0$ and $\infty$, and 
the puncture at $z'=\infty$ is located at $y=1$. We shall denote by $\sigma$ the location of the puncture at
$z'=0$ in the $y$ plane. It takes the form of a function of $q/\lambda^2$ due to \refb{e2.6}. Therefore in
the $y$ plane the four punctures are located at:
\be \label{emap1}
y_1=0, \quad y_2=\infty, \quad y_3=1, \quad y_4=\sigma = g(q/\lambda^2)\, ,
\ee
for some function $g$. $y_4=\sigma$ is a holomorphic function of $q$ for $|q|<1$, 
but we shall not make use of this
information in an essential way.
We also define $v_1$, $v_2$, $v_3$ and $v_4$ to be the original local coordinates $w_1$, $w_2$, 
$w_1'$ and $w_2'$,
but now expressed as function of $y$. These take the form
\be \label{elocals}
v_i = \lambda \, \hat h_i(y;q/\lambda^2)\, ,
\ee
for some functions $\hat h_i$. The overall multiplicative factors of $\lambda$ comes from the
multiplicative factors of $\lambda$ in the definitions of $w_1$, $w_2$, $w_1'$ and $w_2'$ while the
dependence on $q/\lambda^2$ enters through \refb{e2.6}.

It follows from \refb{e2.6} that as $q\to 0$, the four punctured sphere degenerates into 
two spheres, with the original punctures at $z=0$ and $z=\infty$ on one sphere and the
original punctures at $z'=0$ and $z'=\infty$ on the other sphere. In the $y$ plane this will correspond
to the punctures at $\sigma$ and $1$ coming together. 
To find the behavior of $g(q/\lambda^2)$ for small $q$, 
we
note that for small $q/\lambda^2$ we can take $z,z'$ close to 1 and express the relation 
\refb{e2.6} as
\be \label{eapprox}
\lambda^2 \, (h_3'(1))^2 \, (z-1) (z'-1) = q\, .
\ee
Since in this case the local coordinates are related to global coordinates $z$ and $z'$ via S(2,C) transformation,
we can identify $z$ (and $z'$) with $y$ up to SL(2,C) transformation. Now under the identification \refb{eapprox}
the puncture at $z'=\infty$ is mapped to $z=1$ and the puncture $z'=0$ is mapped to
$z=1 - q\, \lambda^{-2} (h_3'(1))^{-2}$. Therefore we can make the identification $y=z$ and
$y_4\equiv \sigma = 1 - q\, \lambda^{-2} (h_3'(1))^{-2}$. Comparing this with \refb{emap1} we get
\be \label{e3.13x}
g(q/\lambda^2) \simeq 1 - q\, \lambda^{-2} (h_3'(1))^{-2} + \OO(q^2/\lambda^4)\, .
\ee
Therefore we have
\be \label{egder}
g(0) =1, \quad g'(0) = - (h_3'(1))^{-2}\, .
\ee

The region $|q| \le 1$ corresponds to a
neighborhood of the point 1 in the $\sigma$ plane.
In string field theory, if we use the local coordinates given in \refb{e2.2} to define the 3-point interaction vertex,
then the family of four punctured spheres \refb{emap1} 
corresponding to  $|q|\le 1$, and the local coordinates
given in \refb{elocals}, describe the contribution to the four point Green's function due to `s-channel diagrams'. In this
the external states represented by the
vertex operators inserted at the punctures at $y_1$ and $y_2$ merge to form an intermediate state
which then splits into the states represented by the vertex operators inserted at the punctures $y_3$ and
$y_4$. 

Contribution from the $u$-channel diagram is obtained by exchanging 1 and 3. Denoting the global coordinate
on the plane by $\tilde y$, we get, by exchanging 1 and 3 in \refb{emap1},
\be \label{e2.15}
\tilde y_1=1, \quad \tilde y_2=\infty, \quad \tilde y_3=0, \quad \tilde y_4=g(q/\lambda^2)\, ,
\ee
and the local coordinates at the punctures are
\be
v_1 = \lambda\, \hat h_3(\tilde y;q/\lambda^2)\, ,
\quad v_2 = \lambda\, \hat h_2(\tilde y;q/\lambda^2)\, ,
\quad  v_3 = \lambda \,  \hat h_1(\tilde y;q/\lambda^2)\, ,\quad
 v_4 =\lambda\, \hat h_4(\tilde y;q/\lambda^2)\, . \nonumber \\
\ee
We now make a change of variable
\be \label{eyty}
y = 1 - \tilde y\, ,
\ee
so that in the $y$ coordinate system the locations of the punctures are given by
\be \label{e2.18}
y_1=0, \quad y_2=\infty, \quad y_3=1, \quad y_4 = \sigma = 1 - g(q/\lambda^2)\, ,
\ee
and the local coordinates around the punctures are given by,
\be \label{euchannel}
v_1 = \lambda\, \hat h_3(1-y;q/\lambda^2)\, ,
\quad v_2 = \lambda\, \hat h_2(1-y;q/\lambda^2)\, ,
\quad  v_3 = \lambda \,  \hat h_1(1-y;q/\lambda^2)\, ,\quad
 v_4 =\lambda\, \hat h_4(1-y;q/\lambda^2)\, . \nonumber \\
\ee

Contribution from the $t$-channel diagram is given by exchanging 2 and 3 in the $s$-channel contribution. 
Denoting the global coordinate
on the plane by $\hat y$, we get
\be
\hat y_1=0, \quad \hat y_2=1, \quad \hat y_3=\infty, \quad \hat y_4=g(q/\lambda^2)\, ,
\ee
\be
v_1 = \lambda\, \hat h_1(\hat y;q/\lambda^2)\, ,
\quad v_2 = \lambda\, \hat h_3(\hat y;q/\lambda^2)\, ,
\quad  v_3 = \lambda \,  \hat h_2(\hat y;q/\lambda^2)\, ,\quad
 v_4 =\lambda\, \hat h_4(\hat y;q/\lambda^2)\, . \nonumber \\
\ee
We make a change of variables
\be \label{eyhy}
y = {\hat y \over \hat y -1}, \quad \hat y = {y\over y-1}\, .
\ee
In this coordinate system the punctures are located at
\be \label{e2.23}
y_1 = 0, \quad y_2=\infty, \quad y_3=1, \quad y_4 =\sigma
= {g(q/\lambda^2) \over g(q/\lambda^2) -1}\, ,
\ee
and the local coordinates are given by
\ben \label{e2.24}
&& v_1 = \lambda\, \hat h_1(y/(y-1);q/\lambda^2)\, ,
\quad v_2 = \lambda\, \hat h_3(y/(y-1);q/\lambda^2)\, , \nonumber \\ &&
 v_3 = \lambda \,  \hat h_2(y/(y-1);q/\lambda^2)\, ,\quad
 v_4 =\lambda\, \hat h_4(y/(y-1);q/\lambda^2)\, . 
\een

Using \refb{e3.13x} we see that 
in the $y_4=\sigma$ plane, the region $|q|\le 1$ in \refb{emap1}, \refb{e2.18} and 
\refb{e2.23} map respectively 
to some regions around 1, 0 and $\infty$ which we shall denote by $\RR_s$, $\RR_u$ and
$\RR_t$. Integrals over these regions describe the contributions from $s$, $u$ and 
$t$-channel Feynman 
diagrams. The rest of the region in the $\sigma$ plane, which we shall call $\RR$, must 
come from the elementary 
four point interaction vertex.

In order to define the four point interaction vertex for off-shell string states, we need 
to look for local coordinates of the form
\be
v_i =\lambda\, \tilde h_i(y; \sigma), \quad 1\le i\le 4, \quad \sigma\in \RR,
\ee
satisfying the following conditions:
\begin{enumerate}
\item $v_i$ must vanish as $y\to y_i$:
\be
\tilde h_i(y_i; \sigma)=0\, .
\ee
\item $\tilde h_i(y;\sigma)$ must match the results \refb{elocals}, \refb{euchannel} and
\refb{e2.24} of $s$, $u$ and $t$ channel diagrams when 
$\sigma$ takes value at the
common boundary of $\RR$ and $\RR_s$, $\RR$ and $\RR_u$ and $\RR$ and $\RR_t$ 
respectively, corresponding to
setting $q=e^{-i\theta}$ in \refb{e2.15}, \refb{e2.18} and \refb{e2.23}.
\item $v_i(\sigma)$ must obey permutation symmetry. This means the following. Let us suppose that we 
exchange the punctures $k$ and $\ell$ for some fixed $k$ and $\ell$. 
This exchanges $y_k$ with $y_\ell$ and also $v_k$ with $v_\ell$. Now by an SL(2,C) transformation we can 
bring the locations of the punctures 1, 2 and 3 to their original values 0, $\infty$ and 1, but in that process
$y_4=\sigma$ will typically map to a different point $\sigma'$ and the local coordinates $v_i$, 
expressed in terms of the new
complex coordinate, which we shall still denote by $y$, will change to $v_i'$. The requirement that we would
like to impose is that $v_i'$ should agree with $v_i$ at $\sigma'$. This will guarantee that interaction vertex is
invariant under permutation of the external states.
\end{enumerate}
While it will be nice to have an interaction vertex of this type, the third condition
is not strictly necessary. Given a choice of local
coordinates that satisfy conditions 1 and 2 but not 3, we can define the interaction vertex by taking average of
the result computed from this choice of local coordinates and the images of these local coordinates under
permutation. 

Similar procedure can be followed for the construction of the local coordinates needed for higher point
interaction vertices of string field theory.

We shall end this section with a few comments on the parameter $\lambda$.
\begin{enumerate}
\item While we expect the on-shell amplitudes to be independent of the parameter $\lambda$ (and the functions
$h_i$), the
off-shell amplitudes depend on $\lambda$. Since $dw_i/dz$ computed from \refb{e2.2} is proportional to
$\lambda$, the off-shell amplitude of a mode with $L_0+\bar L_0$ eigenvalue $h$ is proportional to
$|dw_i/dz|^{-h} \sim \lambda^{-h}$. Therefore for large $\lambda$ the contribution from the modes with large
$L_0+\bar L_0$ eigenvalue is highly
suppressed. This suppression factor in turn makes the sum over massive states in the intermediate state
finite even though the number of such modes has a Hagedorn growth.
\item As can be seen from \refb{emap1}, \refb{e2.18} and \refb{e2.23}, for large $\lambda$ the $s$, $u$ and $t$
channel diagrams cover small regions of the $\sigma\equiv y_4$ plane around 1,  0 and $\infty$ respectively, 
given by the $|q|\le 1$ regions in \refb{emap1}, \refb{e2.18} and \refb{e2.23}. Therefore most of the integration 
over $\sigma=y_4$ is generated by the four point interaction vertex. 
This is related to the fact that the regions $|w_i|\le 1$ in \refb{e2.2} cover small regions in the $z$ plane.
\item Note that $q$ always appears in the combination $q/\lambda^2$. This means in particular that if
we change the cut-off $\Lambda$ in \refb{e3.5} and simultaneously change $\lambda$ so that $\lambda^2 
e^{\Lambda}$ remains fixed, then the cut-off $|q|=e^{-s}\ge e^{-\Lambda}$
remains unchanged in the $\sigma=y_4$ plane. Using this  we can set $\Lambda=0$ by scaling $\lambda$
by $e^{\Lambda/2}$.  In this case the first term in \refb{e3.5} vanishes and the second term becomes 
\refb{e1}. Therefore use of \refb{e3.5} to deal with the negative $(L_0+\bar L_0)$ states is equivalent to the use
of \refb{e1}, with a rescaled value of $\lambda$.
\end{enumerate}

\subsection{Locations of picture changing operators} \label{s4}

For constructing the interaction vertices of superstring theories, we also need to make a choice of PCO
locations. 
We shall now briefly describe the procedure. For simplicity we focus on the heterotic
string theory. For type II string theories we have to repeat the procedure for the left and the right-moving
sectors of the world-sheet theory separately.

We begin with the cubic interaction vertex. There are two types of vertices: R-R-NS and NS-NS-NS. 
For the R-R-NS vertex the total picture number of the vertex operators 
is $-2$ and we do not need to add any PCO. Therefore
the vertex is the same as that in bosonic string theory.
For the NS-NS-NS vertex,
the picture numbers of the three NS-sector vertex operators add up to $-3$ and we
need to insert one PCO. 
The location of the PCO needs to be invariant under permutation symmetry, i.e.\ under
the SL(2,C) transformations that exchange the positions 0, 1 and $\infty$ of the vertex operators. One can 
either try to choose the location to be invariant under this symmetry group, or pick an arbitrary location and
average over all its images. It is easy to verify that there is no single location that is invariant under 
the subgroup of SL(2,C) generating arbitrary permutation of 0, 1 and $\infty$. 
The brute force way will be to choose an arbitrary point $p_1$ and its images $p_2,\cdots p_6$ under
the symmetry generated by $z\to 1-z$ and $z\to z/(z-1)$ and take the PCO insertion to be
\be \label{eavpco}
{1\over 6} \sum_{i=1}^6 \XX(p_i)\, .
\ee
A more economical procedure will be 
to
pick a pair of points, each of which is  invariant under the subgroup that generates cyclic permutation of
0, 1 and $\infty$, and the pair gets exchanged under a $Z_2$ transformations that exchange 0 and 1, leaving
$\infty$ fixed. These pair of points are:
\be \label{ep1p2}
p_1 = {1\over 2}  + i {\sqrt 3\over 2}, \quad p_2 = {1\over 2}  - i {\sqrt 3\over 2}\, .
\ee
We can now insert $(\XX(p_1)+\XX(p_2))/2$ to get a symmetric
3-point interaction vertex. In the following we shall denote the PCO location for a 3-point
interaction vertex by $p$, keeping in mind that $p$ may stand for averages over several 
locations.

Let us now consider the contribution to the 4-point amplitudes associated with $s$, $t$ and $u$ channel
diagrams. First consider the case of four R-sector external states. In this case the intermediate state
in each channel is an NS sector state. No PCO is needed at any stage of this
calculation, reducing the analysis to that of bosonic string field theory. 

Next consider the case of four NS sector external states. In the $s$-channel diagrams the PCOs 
are located at 
\be
z=p, \quad z'=p\, ,
\ee
where $p$ corresponds to average of several insertions as in \refb{eavpco}. Their
images $W_1$ and $W_2$ 
in the $y$ plane may be found using the known map between $z$, $z'$ and $y$. 
The information that will be useful to us later is that in the small $q$ or large $\lambda$
limit, one of the PCO's (say $W_1$) is finite distance away from the punctures $1$ and $\sigma$ 
while the other
one (say $W_2$) is close to the punctures 1 and $\sigma$. 
For $t$ and
$u$ channel diagrams the locations of the PCOs are related to 
$W_1, W_2$ by the transformations $y\to y/(y-1)$ and $y\to (1-y)$ respectively.

While constructing the 4-point interaction vertex of four NS sector states, one can choose the local
coordinates at the punctures for $\sigma\in \RR$
as in the case of bosonic string theory, but we also need to fix the
location of two
PCOs as a function of the modulus $\sigma$. The PCO locations $(W_1, W_2)$ on $\p\RR_s$,
$\p\RR_u$ and $\p\RR_t$ induced from the s, u and t-channel diagrams  
provide appropriate boundary
condition that the PCO locations must satisfy. They also must respect the permutation symmetry.
This can be achieved by brute force by choosing one configuration satisfying the boundary condition and
then averaging over all its images under permutation. However, one may be able to reduce the number of
configurations to average over by making judicious choice of the initial configuration so that it is
invariant under a subgroup of the permutation group. Other than these constraints the PCO locations on the
four punctured sphere can be chosen arbitrarily for $\sigma\in\RR$. If necessary ({\it e.g.} to avoid collision
with each other) we can even allow the PCO locations to jump discontinuously across codimension one
subspaces inside $\RR$ and add correction terms\cite{1408.0571,1504.00609}, so there is 
no obstruction to choosing the PCOs
satisfying the desired conditions.

The case of two NS and two R vertex operators can be analyzed similarly. In this case in the channel
where an R sector state propagates as intermediate state, the PCO insertion comes from the factor of
$\GG$ in the R sector propagator \refb{erprop}. Therefore the PCO insertion involves
an average not over discrete number of possibilities but a continuous set of possibilities.
As before, for $\sigma\in\RR$ the
PCO locations can be chosen arbitrarily subject to the boundary conditions 
on $\p\RR_s$, $\p\RR_t$ and $\p\RR_u$ and the symmetry requirement
(which in this case corresponds to the exchange of the two NS punctures and (independently) the two R punctures.

\sectiono{Warm up with four point amplitude} \label{s5}

We shall now show how bosonic string field theory can be used to compute the four point amplitude over
the full range of external momenta without any need for analytic continuation. 
In \S\ref{s5.1} we shall outline the general procedure for determining the boundary term that arises
when we use \refb{e3.5} to deal with the contribution from states with negative $L_0+\bar L_0$ 
eigenvalue. In \S\ref{s5.2} we shall use this to express the four tachyon amplitude as 
an integral over the world-sheet
that does not have any divergence.

\subsection{A concise description of the boundary terms} \label{s5.1}

Let us consider either the s, t or u-channel
Feynman diagram of string field theory four point amplitude. Each of these diagrams has a single
propagator.  
Using the representation \refb{e3.5} of the propagator we can express the contribution to the
diagram from the first term
on the right hand side of \refb{e3.5} as a two dimensional integral over the complex variable $q$, with the
integrand given by
\be \label{e5x.1}
\II^{(0)} = dq\wedge d\bar q \sum_i A_i q^{-1+\gamma_i} \bar q^{-1+\delta_i}\, ,
\ee
where $\gamma_i$ and $\delta_i$ are the $L_0$ and $\bar L_0$ eigenvalues of 
the $i$-th internal state and
the constants $A_i$ are given by the products of the vertices of the Feynman diagram.
We shall work with generic external momenta so that  
there are no terms with integer $\gamma_i,\delta_i$. 
On the other hand the contribution to the Feynman diagram from the second term on the right hand side of
\refb{e3.5} may be represented as a boundary integral of the form:
\be \label{e5x.2}
\BB = \int_{|q|=e^{-\Lambda}}  \II^{(1)}\, ,
\ee
\be \label{ei1def} 
\II^{(1)} = - dq\,  \sum_i A_i \, \delta_i^{-1} \, q^{-1+\gamma_i} \bar q^{\delta_i}\, .
\ee
Note that $\II^{(1)}$ satisfies
\be \label{ei1i0}
d \, \II^{(1)} = \II^{(0)}\, .
\ee
The choice \refb{ei1def} of $\II^{(1)}$ is not unique since we could have taken this to be proportional to
$d\bar q$ or a linear combination of $dq$ and $d\bar q$ which all satisfy \refb{ei1i0}. This ambiguity corresponds
to the freedom of adding exact forms to $\II^{(1)}$.

By the standard procedure in string field theory illustrated in \S\ref{slocal}
one can convert the integration over the parameters
$q,\bar q$ associated with the propagators to the integration over the standard world-sheet variables
$\sigma,\bar\sigma$ denoting the  world-sheet coordinate of one of the vertex operators, keeping
fixed the positions of the other
three  vertex operators.
In the convention of \S\ref{slocal} the region $|q|\le 1$ gets mapped to the regions $\RR_s$, $\RR_t$ and
$\RR_u$ in the $\sigma$-plane for the s, t and u-channel diagrams. Let us suppose that the region
$|q|\le e^{-\Lambda}$ gets mapped to the regions $\wt\RR_s$, $\wt\RR_t$ and
$\wt\RR_u$ for these diagrams. Then the boundary integrals \refb{e5x.2} will run over $\p\wt\RR_s$, $\p\wt\RR_t$ and 
$\p\wt\RR_u$. On the other hand the bulk integral, after combining the contributions from the s, t and u-channel
Feynman diagrams and the diagram involving the four point interaction vertex, will run over the full complex 
$\sigma$-plane except the excluded regions $\wt\RR_s$, $\wt\RR_t$ and $\wt\RR_u$.

\refb{ei1i0} gives a simple way of determining 
$\II^{(1)}$ from the original integrand $\II^{(0)}$ without knowing the relation between $q$ and
$\sigma$.
Let us suppose that $\sigma$ approaches $\sigma_0$ as $q$ approaches 0. 
Here $\sigma_0$ is the location of one of the other vertex operators.
Then
near $q=0$ the relation between $\sigma$ and $q$ takes the form
\be \label{e5x.2.5}
q = f(\sigma) = \sum_{n\ge 1} a_n (\sigma-\sigma_0)^n\, .
\ee
In terms of the variables $\sigma$ and $\bar\sigma$ we can express \refb{e5x.1} 
as
\be \label{e5x.5}
\II^{(0)} = d\sigma\wedge d\bar\sigma \,
\sum_j B_j (\sigma-\sigma_0)^{-1+\alpha_j} (\bar\sigma-\bar\sigma_0)^{-1+\beta_j} \, ,
\ee
where $B_j$ are new coefficients and $\alpha_j,\beta_j$ take values that differ from $\gamma_j,\delta_j$
at most by integers. We can now solve \refb{ei1i0} to get
\be\label{e5x.6}
\II^{(1)} = -d\sigma \, \sum_j B_j \beta_j^{-1} 
(\sigma-\sigma_0)^{-1+\alpha_j} (\bar\sigma-\bar\sigma_0)^{\beta_j} \quad \hbox{near $\sigma=\sigma_0$}\, ,
\ee
up to closed differential forms. Now one can show that a closed differential
form with an expansion in non-integer powers of $(\sigma-\sigma_0)$ and 
$(\bar\sigma-\bar\sigma_0)$ is also exact, with an expansion
involving non-integer powers of $(\sigma-\sigma_0)$ and 
$(\bar\sigma-\bar\sigma_0)$. 
Therefore $\II^{(1)}$ given in \refb{e5x.6} differs from the one in \refb{ei1def} at most by an exact form and 
its integral over the cycle $|q|=e^{-\Lambda}$ 
gives the same result as that of \refb{ei1def}. When $\sigma_0=\infty$, one has to
use a slight variant of this procedure, with $\II^{(0)}$ expanded as $d\sigma\wedge d\bar\sigma \sum_j B_j 
\sigma^{-1-\alpha_j}\bar\sigma^{-1-\beta_j}$ and $\II^{(1)}$ given by $d\sigma \sum_j B_j \, \beta_j^{-1}
\sigma^{-1-\alpha_j}\bar\sigma^{-\beta_j}$.

In this procedure the only part that requires the knowledge of the relation between $q$ and $\sigma$ is the
curve $\CC=\p\wt\RR_u$, $\p\wt\RR_s$ or $\p\wt\RR_u$ --  
the image of the curve $|q|=e^{-\Lambda}$ in the $\sigma$ plane. However we shall
now show that the sum of the bulk and the boundary terms is independent of the form of $\CC$. For this
let us recall that the bulk integration runs over  
the region outside the curve $\CC$ in the $\sigma$ plane. 
Therefore if we change $\CC$ to $\CC'$, the bulk 
integration region will change by a region $\DD$ with $\p\DD=\CC-\CC'$. The net change
in the bulk and the boundary integral is given by
\be
\int_\DD \II^{(0)}
+ \int_{\CC'-\CC} \II^{(1)} =  \int_\DD  d\II^{(1)}
- \int_{\p\DD} \II^{(1)} = 0\, .
\ee
This shows that even the knowledge of $\CC$ is not needed and \refb{e5x.6} gives us complete information about
the boundary term in terms of the bulk integrand \refb{e5x.5}. This in turn shows that in order to
evaluate the sum of the  boundary terms and the bulk integral that defines the amplitude,
we do not need any knowledge
of the local coordinates used in defining the interaction vertices of string field theory.

While for computing amplitudes with generic external momenta, we can assume that $\gamma_j,\delta_j$ and
hence $\alpha_j, \beta_j$ are not integers, in \S\ref{s6}, where we analyze the effect of marginal
deformations, we shall encounter situations where $\alpha_j,\beta_j$ are integers. In this case there may be
an additional contribution to the boundary term involving the integral of a one form that is closed but not exact:
\be 
- \int_\CC d\sigma \,  K\, (\sigma - \sigma_0)^{-1}\, .
\ee 
Here $K$ is a constant. 
As we shall see in \S\ref{s6},
possible additional contributions from such terms can be calculated by 
working directly with \refb{e5x.1}.

\subsection{Four tachyon amplitude} \label{s5.2}

We shall now apply the procedure described in \S\ref{s5.1} to give an expression for the four tachyon amplitude
as a world-sheet integral that does not have any divergence.
First let us describe the conventional world-sheet computation in the $\alpha'=1$ unit. 
Let us denote by
$k_1$, $k_2$, $k_3$ and  $k_4$ the momenta of the tachyons, all counted as positive if ingoing, and 
satisfying the on-shell condition
\be 
k_i^2 = -m^2=4,
\ee  
where $m^2$ is the mass$^2$ of the tachyon. Then according to \refb{eampcon}, 
the amplitude takes the form
\be 
A= {1\over 2 \pi i} 
\int d^2 \sigma \, \langle \, c \, \bar c\, e^{i k_1.X}(y_1,\bar y_1) \,  \, c \, \bar c\, e^{i k_2.X}(y_2,\bar y_2) 
\, \,  c \, \bar c\, e^{i k_3.X}(y_3,\bar y_3) \, \, e^{i k_4.X}(\sigma , \bar \sigma )\rangle
\ee
where $y_1$, $y_2$, $y_3$ are arbitrary fixed points, which we can take to be 0,  $\infty$ and 1
respectively, and 
\be
d^2\sigma \equiv d\sigma\wedge d\bar\sigma\, .
\ee
After evaluating the correlation function using \refb{ebosenorm} the result takes the form:
\be \label{eamp}
A=-{1\over 2\pi i} \int d^2 \sigma  \, |\sigma |^{k_1.k_4} |\sigma -1|^{k_3.k_4} \, ,
\ee
up to overall momentum conserving delta function.
Defining 
\be \label{et2}
s= - (k_3+k_4)^2 = -8 - 2 k_3.k_4, \quad t= - (k_2+k_4)^2 = -8 - 2 k_2.k_4, 
\quad u= - (k_1+k_4)^2 = -8 - 2 k_1.k_4, 
\ee
we can express \refb{eamp} as
\be \label{edivexp}
A=-{1\over 2\pi i} \int d^2 \sigma  \, |\sigma |^{-u/2-4} |\sigma -1|^{-s/2 - 4} = -
{1\over 2\pi i} \int d^2 \sigma  \, |\sigma |^{-(u-m^2)/2-2} 
|\sigma -1|^{-(s-m^2)/2 -2} \, ,
\ee
This integral has divergences from the region $\sigma \to 0$, $\sigma \to 1$ and $\sigma \to\infty$ if, respectively,
\be \label{et1.5}
u \ge m^2, \quad s\ge m^2, \quad t\ge m^2\, .
\ee
Therefore the divergences appear whenever $s$, $t$ or $u$ exceeds the threshold of
production of a tachyonic particle in the intermediate state. 
Precisely in these domains the intermediate tachyon state propagating in the $s$, $t$ and $u$-channel respectively 
has negative $L_0+\bar L_0$ eigenvalue and the right hand side of \refb{e2} diverges.
Conventionally one first defines this
integral in the region where it converges and then goes out of this region via analytic continuation.
This leads to the result:
\be\label{evir}
A= {\Gamma(-1 - s/4) \Gamma(-1 - t/4) \Gamma(-1 - u/4) \over
\Gamma(2 + s/4) \Gamma(2 + t/4) \Gamma(2 + u/4)} \, .
\ee

Our goal will be to modify \refb{edivexp} suitably by drawing insights from string field theory so that 
we can evaluate the integral without having to invoke analytic continuation. For generic $s$, $t$ and $u$
we can invoke the results of \S\ref{s5.1} to arrive at such an expression. We cut out small regions $\wt\RR_s$,
$\wt \RR_u$ and $\wt\RR_t$ around 1, 0 and $\infty$ respectively, and denote by $\wt\RR$ the left over region 
in the complex plane.
Then we can express the amplitude as
\be \label{eprelim}
A = -{1\over 2\pi i} \, \int_{\wt\RR} d^2 \sigma  \, |\sigma |^{-u/2-4} |\sigma -1|^{-s/2 - 4} +
\BB_s +\BB_t + \BB_u\, ,
\ee
where the last three terms are boundary terms which can be evaluated from the 
bulk integrand using \refb{e5x.6}
as follows.
Suppose that near $\sigma=0$
\be \label{edefiu}
-{1\over 2\pi i} \, |\sigma |^{-u/2-4} |\sigma -1|^{-s/2 - 4} 
=  \sum_i \, C_i \, \sigma ^{-1+\alpha_i} \bar \sigma ^{-1+\beta_i}\, ,
\ee
where $\alpha_i,\beta_i$ are given by $-u/4-1$ plus non-negative integers and $C_i$ are constants.
Then \refb{e5x.6} gives
\be \label{euboundary}
\BB_u=-\int_{\p\wt\RR_u} \sum_i C_i \, {1\over \beta_i}\, 
d\sigma \, \sigma^{\alpha_i-1} \, \bar\sigma^{\beta_i}\, .
\ee
Similarly if we denote
the power series expansion of the integrand in the region around $\sigma=1$ as
\be \label{eexpand}
-{1\over 2\pi i} \, |\sigma|^{-u/2-4} |\sigma -1|^{-s/2 - 4} = \sum_{i} C'_{i} (\sigma -1)^{\alpha'_i-1} 
(\bar \sigma -1)^{\beta'_i-1}\, ,
\ee
then the required boundary terms on $\p\wt\RR_s$ can be expressed as 
\be  \label{esboundary}
\BB_s =  
-\int_{\p\wt \RR_s} d\sigma  \sum_{i} C'_{i} {1\over \beta_i'} (\sigma -1)^{\alpha_i'-1} 
 (\bar \sigma -1)^{\beta_i'}\, .
\ee
Finally if we expand the integrand  in a power series
expansion in $|\sigma |^{-1}$ around $\sigma =\infty$:
\be \label{eexpandc}
-{1\over 2\pi i} \, |\sigma |^{-u/2-4} |\sigma -1|^{-s/2 - 4} = \sum_{i} C''_{i} \sigma ^{-\alpha''_i-1} \bar \sigma ^{-\beta''_i-1}\, ,
\ee
then the required boundary terms on $\p\wt\RR_t$ are given by
\be \label{etboundary}
\BB_t= \int_{\p\wt \RR_t} d \sigma  \sum_{i} C''_{i} {1\over \beta''_i} \sigma ^{-\alpha''_i-1} 
 \bar \sigma ^{-\beta''_i}\, .
\ee

The expressions for $\BB_u$, $\BB_s$ and $\BB_t$ given above involve infinite sum over states.
However the sum converges and there is no difficulty in numerical evaluation of this expression.
We have tested \refb{eprelim} by numerically evaluating this and comparing this with the exact result 
\refb{evir} in the regime in $(s,t,u)$ space 
where the original integral
\refb{edivexp} diverges. By taking the
contours around 0 and 1 to be sufficiently small and the contour around $\infty$ to be sufficiently large, we can
ensure that only finite number of terms with negative $\alpha_i+\beta_i$, $\alpha'_i+\beta'_i$ and
$\alpha''_i+\beta''_i$ give appreciable contribution to \refb{euboundary}, \refb{esboundary} and
\refb{etboundary}. 

As a simple example we can consider the domain $s<-4$, $t<-4$ and $-4<u<-2$. In this case the divergences in
the integral \refb{edefiu} will come only from the region near $\sigma=0$. Therefore we can take the limit in which 
$\wt\RR_s$ shrinks to $\sigma=1$ and $\wt \RR_t$ recedes to infinity, making $\BB_s$ and $\BB_t$ vanish. To
determine $\BB_u$ we expand the bulk integrand in a power series in $\sigma$ to get
\ben
-{1\over 2\pi i} \, |\sigma|^{-u/2-4} |\sigma -1|^{-s/2 - 4} &=& -{1\over 2\pi i} \, |\sigma|^{-u/2-4}
\bigg\{1 + \left({s\over 4}+2\right) (\sigma +\bar\sigma) + \left({s\over 4}+2\right)^2 \sigma\bar\sigma
\nonumber \\
&& + {1\over 2}\left({s\over 4}+2\right) \left({s\over 4}+3\right)  (\sigma^2+\bar\sigma^2)+ \cdots
\bigg\}\, .
\een 
This gives, from \refb{euboundary}, 
\ben\label{ebunum}
\BB_u &=& {1\over 2\pi i} \, \int_{\p\wt\RR_u} \, d\sigma |\sigma|^{-u/2-4} \bigg[ -{4\over u+ 4} \bar \sigma 
-  \left({s\over 4}+2\right) {4\over u+ 4} \bar \sigma \sigma 
-  \left({s\over 4}+2\right) {4\over u} \bar\sigma^2 \nonumber \\
&& \hskip -.1in -  \left({s\over 4}+2\right)^2 {4\over u} \sigma \bar\sigma^2  
-{1\over 2}\left({s\over 4}+2\right) \left({s\over 4}+3\right)  \left({4\over u+ 4}
\sigma^2\bar\sigma + {4\over u-4} \bar\sigma^3
\right)
+ \cdots
\bigg]\, .
\een
Since \refb{eprelim} has been shown to be independent of the choice of contours
$\p\wt\RR_u$, $\p\wt\RR_s$ and $\p\wt\RR_t$, we can take $\wt\RR_u$ to be
a rectangular region around the origin. As the size of $\wt\RR_u$ shrinks, contribution to $\BB_u$ from
the first term inside the square bracket increases and the other terms decrease
for $-4<u<-2$.
By taking the contour to be sufficiently small, we can ignore the contribution from all terms other than the
first term in the square bracket. Adding this to the bulk integral given by the first term in \refb{eprelim}, with
$\wt\RR$ given by the complement of $\wt\RR_u$ in the whole complex plane, we can evaluate the full amplitude.
With relatively little effort we can reduce the error in computation
-- measured by comparing the result to the exact expression given in \refb{evir} -- 
to less than $.1\%$. 
The error comes from the higher order terms
inside the square bracket in \refb{ebunum}. We can further reduce the error either by including
these terms or by shrinking the size of $\wt\RR_u$, at the same time making sure that the error in the numerical
evaluation of the integrals are not significant.\footnote{For smaller size of $\wt\RR_u$, both the bulk and the boundary
integrals receive large contributions which cancel in the sum. Therefore if we take the size of $\wt\RR_u$ to be
very small, numerical errors will increase.
}

We shall now give an alternate expression that does not require infinite sum.  This requires taking the limit
in which the regions $\wt\RR_s$, $\wt\RR_t$ and $\wt\RR_u$ shrink to points 1, $\infty$ and 0 
respectively. 
In this limit only the terms with negative $\alpha_i+\beta_i$, $\alpha'_i+\beta'_i$ and 
$\alpha''_i+\beta''_i$
survive in the boundary terms.  We now use
\be \label{edivint}
{1\over \beta_i} \int_{\p\wt\RR_u} d\sigma \, \sigma^{\alpha_i-1} \, \bar\sigma^{\beta_i}
=  \int_{\wt\RR^c_u} d^2\sigma \, \sigma^{\alpha_i-1} \, \bar\sigma^{\beta_i-1}\, , \quad \hbox{for $\alpha_i+\beta_i<0$}\, ,
\ee
where $\wt\RR^c_u$ is the complement of the region $\wt\RR_u$. In the limit when $\wt\RR_u$ shrinks to a point
$\wt\RR^c_u$ will expand into the whole complex plane, but we do not want to take this limit yet since
the integral \refb{edivint} will diverge in this limit. Using \refb{edivint} and
analogous results for the $s$ and $t$-channel boundary terms, 
we can express \refb{eprelim} as
\ben \label{eampfin}
&& \int d^2 \sigma  \bigg[-{1\over 2\pi i} \, |\sigma|^{-u/2-4} |\sigma -1|^{-s/2 - 4} 
- \sum_{i,j\atop \alpha_i+\beta_i<0} C_{i} (\sigma -1)^{\alpha_i-1} 
 (\bar \sigma -1)^{\beta_i-1} \nonumber \\ &&
-\sum_{i,j\atop \alpha'_i+\beta'_i<0} C'_{i} \sigma ^{\alpha'_i-1} \bar \sigma ^{\beta'_i-1} 
- \sum_{i,j\atop \alpha''_i+\beta''_i<0} C''_{i} \sigma ^{-\alpha''_i-1} \bar \sigma ^{-\beta''_i-1}
\bigg]\, .
\een
The integral now runs over the full complex $\sigma$-plane, giving 
a finite expression for the amplitude since all the possible divergences near 0, 1 and
$\infty$ have been subtracted. In a related case, this form of the amplitude was used in \cite{1811.00032}.

\subsection{Comments} \label{s5.3}

We end this section with a few comments on our result.
\begin{enumerate}
\item 
One could ask whether the amplitudes defined via \refb{eprelim} or \refb{eampfin} agree with the ones defined 
by analytic continuation. Again string field theory can be invoked 
to prove the equality of these different procedures.
Each tree
level Feynman diagram constructed from string field theory is manifestly an analytic function of external
momenta except for poles when
an on-shell particle propagates in the intermediate state. Therefore if we begin in a region of the external momenta
where there are no divergences in the world-sheet description and the string field theory and world-sheet
descriptions coincide, and then analytically continue the result to the region where the direct world-sheet 
description gives divergent results, the result of analytic continuation will coincide with the one calculated from
string field theory \i.e.\ eqs.\refb{eprelim} and \refb{eampfin}.
\item Witten gave an operational procedure for performing the world-sheet integrals
that coincides with the analytic continuation of the world-sheet 
description\cite{1307.5124}.  
In spirit this corresponds to a different representation of the propagator \refb{e1}. Instead of 
\refb{e3.5}, we use \refb{e2alt}.
Using the relation between $\sigma\equiv y_4$ and $q$
described in \refb{emap1}, \refb{e2.18} and \refb{e2.23}, we can convert the resulting contribution from $s$, $t$ and
$u$ channel diagrams to appropriate contours in the complexified $(\sigma_R, \sigma_I)$ space where $\sigma_R$ and
$\sigma_I$ are real and imaginary parts of $\sigma$. Since \refb{e2alt} gives the correct propagator for 
$L_0+\bar L_0\ne 0$, this prescription is guaranteed to coincide with eqs.\refb{eprelim} and \refb{eampfin}.
\end{enumerate}

\sectiono{Higher point amplitudes} \label{shigher}

We shall now generalize the procedure described above to higher point amplitudes of generic vertex operators
for generic compactification. 
As in \S\ref{s5} we shall decompose each propagator into two parts according to \refb{e3.5} and include the
first part inside the square bracket as part of bulk integration while the second term inside the square bracket
will provide a boundary term at $|q|=e^{-\Lambda}$.
The main
strategy will be to begin with the boundary terms that are implied by string field theory 
and then convert them to a general coordinate system in the moduli space. 
Our goal will be to arrive at an algorithm to compute the boundary terms in a general coordinate system without
explicitly knowing the relation between the variables $q$ and the general coordinates  used to
parametrize the moduli space.

\begin{figure}
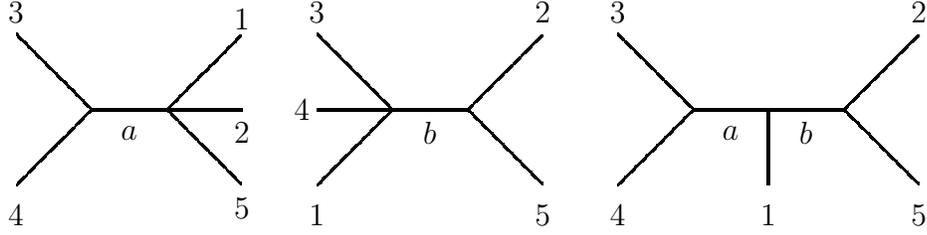


\begin{center}

\figgraph

\end{center}

\vskip -1in

\caption{Examples of Feynman diagrams and the labelling of their boundaries in our convention. \label{fig1}}

\end{figure}

We shall begin by introducing a convenient way of labelling the propagators of a Feynman diagram.
For a tree diagram, cutting a propagator divides the external states into two disconnected sets, with
each set containing at least two external states.
We associate to each propagator an index $s$ that specifies this division. 
Therefore $s$ can take $N$ possible values, where $N$ is 
the total number of ways in which the set of $(n+3)$ external states can be divided into two
sets, with each set containing at least two external states.
In a given Feynman diagram
different propagators carry different labels, but two propagators from two different Feynman diagrams
will carry the same label if cutting them leads to the same division of external particles into two sets.
This can be illustrated using Fig.~\ref{fig1} --  the left internal propagator in the third diagram carries the
same label as the propagator of the first diagram, while the right internal propagator in the third diagram carries the
same label as the propagator of the second diagram. We shall denote by $q_s$ the variable $q$ used to represent
the propagator carrying label $s$ as in \refb{e3.5}.

Next we shall introduce 
some subspaces of the moduli space on which we have to evaluate the
integrals. We denote by
$\CC^{(0)}$ the codimension 0 subspace of the moduli space of $(n+3)$-punctured sphere 
on which the bulk integration is performed. This includes contribution from the 
elementary $(n+3)$-point vertex, as well as all Feynman diagrams with propagators where from
each propagator we include only the first term inside the square bracket in \refb{e3.5}.
In the variables $\{q_s\}$ this corresponds to restricting $|q_s|\ge e^{-\Lambda}$ for each $s$.
We also denote by
$-\CC^{(1)}_s$ the boundary of $\CC^{(0)}$ corresponding to setting, in all Feynman diagrams
carrying a propagator with label $s$,
\be
 |q_s|=e^{-\Lambda}\equiv \eps, \quad |q_r|\ge\eps \ \hbox{for}\ r\ne s\, .
\ee 
The $-$ sign in front of $\CC^{(1)}$ corresponds to the fact that the normal to $\CC^{(1)}$ is taken to be along
the direction of increasing $|q_s|$, pointing into $\CC^{(0)}$.
The region inside $\CC^{(0)}$ corresponds to $|q_s|>\eps$. Even though for numerical evaluation it may
be better to work with small $\eps$, all our formulae\ will be valid for finite $\eps\le 1$.

We now denote by $\CC^{(2)}_{s_1s_2}$ the codimension 2
subspace of the moduli space $\CC^{(0)}$ associated with the intersection 
$\CC^{(1)}_{s_1}\cap \CC^{(1)}_{s_2}$.
This corresponds to setting, in all Feynman diagrams that contain propagators carrying labels $s_1$ and $s_2$,
\be
|q_{s_1}|=\eps, \quad |q_{s_2}|=\eps, \quad |q_s|\ge \eps \, \,  \hbox{for $s\ne s_1, s_2$} \, .
\ee
We can regard $\CC^{(2)}_{s_1s_2}$ as an oriented subspace that is anti-symmetric under the exchange
of $s_1$ and $s_2$ since this exchanges their transverse coordinates $|q_{s_1}|$ and $|q_{s_2}|$.
Generalizing this we define
$\CC^{(k)}_{s_1\cdots s_k}$  to be the codimension $k$ subspace obtained by setting 
\be
|q_{s}|=\eps \quad \hbox{for $s=s_1,\cdots, s_k$}, \quad |q_s|\ge \eps \, \,  \hbox{for $s\ne s_1, s_2,\cdots, s_k$} \, ,
\ee 
in all Feynman diagrams that contain propagators with labels $s_1,\cdots s_k$.
It is easy to verify the relations
\be \label{ecintersec}
C^{(k)}_{s_1\cdots s_k} = C^{(1)}_{s_1}\cap C^{(1)}_{s_2}\cap \cdots \cap C^{(1)}_{s_k}\, ,
\ee
and
\be \label{ecbound}
\p\,  \CC^{(k)}_{s_1\cdots s_k} = - \sum_{s} \CC^{(k+1)}_{s_1\cdots s_k s}\, ,
\ee
where the sum over $s$ runs over all labels for which  $\CC^{(k+1)}_{s_1\cdots s_k s}$ exists. As
before, $\CC^{(k)}_{s_1\cdots s_k}$ is antisymmetric under the exchange of the $s_\ell$'s. The minus
sign in \refb{ecbound} defines the orientation of the $\CC^{(k+1)}$'s for given orientation of the 
$\CC^{(k)}$'s.

Let $\II^{(0)}$ denote the bulk integrand that we need to integrate over the moduli space. 
The bulk contribution $\int_{\CC^{(0)}}\II^{(0)}$ comes from the sum of all Feynman diagrams,
where in each Feynman diagram we take 
the product of the interaction vertices with the first terms  inside the square bracket in
\refb{e3.5} in all the propagators. Different Feynman diagrams cover different regions of $\CC^{(0)}$.
Now it follows from the definition of $\CC^{(k)}_{s_1\cdots s_k}$ given above that if a Feynman diagram
covers part of the region 
in $\CC^{(0)}$ near $\CC^{(k)}_{s_1\cdots s_k}$, then it must contain the propagators carrying
labels $\{s_1,\cdots, s_k\}$ (and possibly other propagators), 
and we have $|q_s|\simeq e^{-\Lambda}$ for $s=s_1,\cdots, s_k$. 
Examination of the first term inside the square bracket in
\refb{e3.5} shows that in a region of $\CC^{(0)}$ near $\CC^{(k)}_{s_1\cdots s_k}$, 
$\II^{(0)}$ has the form
\be\label{egener1}
\II^{(0)} = \prod_{\ell=1}^k \left\{ dq_{s_\ell} \wedge d\bar q_{s_\ell} \right\}\wedge
d^{2(n-k)} m_{(s_1,\cdots, s_k)}  \sum_{i} C_i (m_{(s_1,\cdots, s_k)})
\prod_{\ell=1}^k q_{s_\ell}^{-1 + \gamma^{(s_\ell)}_i} \bar q_{s_\ell}^{-1 + \delta^{(s_\ell)}_i}\, ,
\ee
where $m_{(s_1,\cdots, s_k)}$ denote the coordinates of the $2(n-k)$ dimensional moduli space of the
punctured spheres into which the original sphere degenerates when we set
$q_{s_1}=\cdots = q_{s_k}=0$. 
In a Feynman diagram of string field theory, these moduli come from the moduli integration appearing in the
definition of the interaction vertices, as well as the parameters $q$ associated with the propagators other than
those carrying labels $s_1,\cdots,s_k$.
These, together with the phases of $q_{s_1},\cdots, q_{s_k}$ form the coordinates of
$\CC^{(k)}_{s_1\cdots s_k}$. 
$C_i(m_{(s_1,\cdots, s_k)})$ are 
some functions of these moduli given by the product of the vertices of the Feynman diagram and the
first term inside the square bracket in \refb{e3.5} for the propagators other than the ones carrying labels
$s_1,\cdots, s_k$.
$(\gamma^{(s_\ell)}_i, \delta^{(s_\ell)}_i)$ are the $(L_0,\bar L_0)$ quantum numbers of states that
can propagate in the propagator carrying the label $s_\ell$ for $1\le \ell\le k$.

Using the second term in \refb{e3.5} we can also write down the expression for 
the integrand for the boundary term 
$\int_{\CC^{(k)}_{s_1\cdots s_k}}\II^{(k)}_{s_1\cdots s_k}$, obtained by picking the second
term inside the square bracket in \refb{e3.5} for the propagators 
carrying labels $s_1,\cdots, s_k$ and the first term
inside the square bracket in \refb{e3.5} for all other propagators. The corresponding integrand
$\II^{(k)}_{s_1\cdots s_k}$ is given by the $(2n-k)$ form
\ben\label{ebou51}
\II^{(k)}_{s_1\cdots s_k} &=& (-1)^k dq_{s_k}\wedge \cdots \wedge dq_{s_1} \wedge 
d^{2(n-k)} m_{(s_1,\cdots, s_k)}  \nonumber \\
&& \sum_{i} C_i(m_{(s_1,\cdots, s_k)}) 
\prod_{\ell=1}^k \left\{(\delta^{(s_\ell)}_i)^{-1} (q_{s_\ell})^{-1+\gamma^{(s_\ell)}_i}
(\bar q_{s_\ell})^{\delta^{(s_\ell)}_i}
\right\}\, .
\een 
We can now express the
total contribution to the amplitude as:
\be \label{etotalI}
I = \sum_{k=0}^n  \sum_{\{s_1,\cdots ,s_k\}\atop s_1<s_2<\cdots < s_k}
\int_{\CC^{(k)}_{s_1\cdots s_k}} \II^{(k)}_{s_1\cdots s_k}\, .
\ee
The sum runs over all inequivalent combinations $\{s_1,\cdots, s_k\}$ for which $\CC^{(k)}_{s_1\cdots s_k}$
exists.

Using \refb{egener1} with $k$ replaced by $k+j$, one also finds that near $\CC^{(k+j)}_{s_1\cdots s_k r_1\cdots r_j}
\subset \CC^{(k)}_{s_1\cdots s_k}$,
\ben\label{ebou51new}
\II^{(k)}_{s_1\cdots s_k} &=& (-1)^k dq_{s_k}\wedge \cdots \wedge dq_{s_1} \wedge \left\{\prod_{a=1}^j
dq_{r_a}\wedge d\bar q_{r_a}\right\}  \wedge
d^{2(n-k-j)} m_{(s_1,\cdots, s_k,r_1,\cdots ,r_j)}  \nonumber \\
&& \hskip -.5in \sum_{i} C_i(m_{(s_1,\cdots, s_k,r_1,\cdots , r_j)}) 
\prod_{\ell=1}^k \left\{(\delta^{(s_\ell)}_i)^{-1} (q_{s_\ell})^{-1+\gamma^{(s_\ell)}_i}
(\bar q_{s_\ell})^{\delta^{(s_\ell)}_i} \right\}\left\{\prod_{b=1}^j (q_{r_b})^{-1+\gamma^{(r_{b})}_i}
(\bar q_{r_b})^{-1+\delta^{(r_{b})}_i} 
\right\}\, . \nonumber \\
\een 
Note that the $C_i$'s in \refb{ebou51} and \refb{ebou51new} are not the same, but the latter are obtained by
expanding the former near $\CC^{(k+j)}_{s_1\cdots s_k r_1\cdots r_j}$.
It is easy to verify from \refb{ebou51}, and \refb{ebou51new} with $j=1$, $k$ replaced 
by $(k-1)$, that in the neighborhood of $\CC^{(k)}_{s_1\cdots s_k}$:
\be \label{edikeq}
d \, \II^{(k)}_{s_1\cdots s_k} = \II^{(k-1)}_{s_1\cdots s_{k-1}} - \II^{(k-1)}_{s_1 \cdots  s_{k-2} s_k} + \cdots
+(-1)^{k-1} \II^{(k-1)}_{s_2 \cdots  s_k} \, .
\ee

Note that \refb{etotalI} and \refb{edikeq} are written in the coordinate free notation. Therefore we can use
these to give an expression for the amplitude in an arbitrary coordinate system in the moduli space without
referring to the coordinates $q_s,\bar q_s$. However we need to first check if these equations fix the amplitude
uniquely. First we see from \refb{edikeq} 
that for non-integer $\gamma^{(s)}_i, \delta^{(s)}_i$, there is an
ambiguity in determining $\II^{(k)}$ for given $\II^{(k-1)}$ of the form:
\be
\II^{(k)}_{s_1\cdots s_k}  \to \II^{(k)}_{s_1\cdots s_k}  + d\JJ^{(k)}_{s_1\cdots s_k} \, ,
\ee
where $\JJ^{(k)}_{s_1\cdots s_k}$ is an $(2 n-k-1)$ form defined near $\CC^{(k)}_{s_1\cdots s_k}$.
Due to \refb{edikeq} with $k$ replaced by $k+1$, 
this requires us to change $\II^{(k+1)}_{s_1\cdots s_k s_{k+1}}$ as
\be
\II^{(k+1)}_{s_1\cdots s_k s_{k+1}} \to \II^{(k+1)}_{s_1\cdots s_k s_{k+1}} + \sum_{\ell=1}^{k+1}
\JJ^{(k)}_{s_1\cdots s_{\ell-1}s_{\ell+1} s_{k+1}} (-1)^{k+1-\ell} \, .
\ee
Therefore the net ambiguity 
in the whole set $\{\II^{(k)}_{s_1\cdots s_k}\}$ for all $k$ and $\{s_1,\cdots, s_k\}$ satisfying
\refb{edikeq} takes the form
\be\label{eidxm1}
\Delta \II^{(k)}_{s_1\cdots s_k} = d\JJ^{(k)}_{s_1\cdots s_k}  + \sum_{\ell=1}^k \JJ^{(k-1)}_{s_1\cdots s_{\ell-1}
s_{\ell+1} \cdots s_k} (-1)^{k-\ell}\, .
\ee
We shall now compute its effect on the total integral $I$ defined in \refb{etotalI}. We get
\be \label{eidx0}
\Delta \, I = \sum_{k=0}^n  \sum_{\{s_1,\cdots ,s_k\}\atop s_1<s_2<\cdots < s_k}
\int_{\CC^{(k)}_{s_1\cdots s_k}} \left\{ d\JJ^{(k)}_{s_1\cdots s_k} + 
\sum_{\ell=1}^k \JJ^{(k-1)}_{s_1\cdots s_{\ell-1}
s_{\ell+1} \cdots s_k} (-1)^{k-\ell} \right\}\, .
\ee
Using \refb{ecbound} we get
\be \label{eidx1}
\int_{\CC^{(k)}_{s_1\cdots s_k}} d\JJ^{(k)}_{s_1\cdots s_k}  =
\int_{\p\CC^{(k)}_{s_1\cdots s_k}}  \JJ^{(k)}_{s_1\cdots s_k} =-\sum_{s\ne s_1,\cdots, s_k} \
\int_{\CC^{(k+1)}_{s_1\cdots s_k s}}  \JJ^{(k)}_{s_1\cdots s_k}\, .
\ee
On the other hand, replacing the summation variable $k$ by $k+1$  
we can express the second term in \refb{eidx0} as
\be \label{eidx2}
\sum_{k=0}^n \sum_{\{s_1,\cdots , s_{k+1}\}\atop s_1<s_2<\cdots < s_{k+1}} 
\sum_{\ell=1}^{k+1} \int_{\CC^{(k+1)}_{s_1\cdots s_k s_{k+1}}} \JJ^{(k)}_{s_1\cdots s_{\ell-1}
s_{\ell+1} \cdots s_{k+1}} (-1)^{k+1-\ell} = \sum_{k=0}^n
\sum_{\{s_1,\cdots , s_k\}\atop s_1<s_2<\cdots < s_k}
\sum_{s\ne s_1,\cdots , s_k} \int_{\CC^{(k+1)}_{s_1\cdots s_k s}} \JJ^{(k)}_{s_1\cdots s_k}
\ee
where in the last step we have rearranged the sum by relabelling $\{s_1,\cdots ,s_{\ell-1},
s_{\ell+1}, \cdots ,s_{k+1}\}$ as  $\{s_1,\cdots, s_k\}$ and $s_\ell$ as $s$, and used the antisymmetry
of  $\CC^{(k+1)}$ to write $(-1)^{k+1-\ell} \CC^{(k+1)}_{s_1\cdots s_k s_{k+1}}$ as $\CC^{(k+1)}_{s_1\cdots s_k s}$.
Substituting \refb{eidx1} and \refb{eidx2} into \refb{eidx0} we see that the two terms cancel and we get
\be
\Delta I = 0\, .
\ee
Therefore the ambiguity \refb{eidxm1}
in the determination of $\II^{(k)}$ by solving \refb{edikeq} does not affect the integral $I$ given in \refb{etotalI}.
This means that we can solve \refb{edikeq} to determine the boundary corrections directly by examining the form of the
bulk integrand in any given coordinate system, without having to know the relation 
between the chosen coordinates and the $q_s$'s near the boundaries. As  already emphasized before,
this requires us to work with generic external momenta so that the exponents $\gamma_i^{(s_\ell)}$,
$\delta_i^{(s_\ell)}$ are not integers. Otherwise the determination of $\II^{(1)}_s$ could have an additive 
ambiguity proportional to $q_s^{-1} dq_s d^{2(n-1)} m_{(s)} C(m_{(s)})$ that is not an exact differential 
$d\JJ^{(1)}_s$. Similar ambiguities could be present in the other $\II^{(k)}_{s_1\cdots s_k}$'s as well.

Now the expression for the amplitude given in \refb{etotalI} depends not only on the differential forms 
$\II^{(k)}_{s_1\cdots s_k}$ but also on the boundaries $\CC^{(k)}_{s_1\cdots s_k}$ on which these forms are
to be integrated. Since these are given by $|q_{s_1}|=\cdots=|q_{s_k}|=\eps$, the shapes of these boundaries in
a generic coordinate system depend on the relation between the special coordinates $\{q_{s_\ell}\}$ 
induced from string field theory and the coordinates we are using (which could be taken to be the positions 
$\sigma_1,\cdots,\sigma_n$ of $n$ of the vertex operators keeping three of them at fixed positions). We shall
now demonstrate that \refb{etotalI} is actually invariant under arbitrary deformations of $\CC^{(1)}_s$'s and
consequent deformation of the $\CC^{(k)}_{s_1\cdots s_k}$'s
following from \refb{ecintersec}.  
For this instead of considering the most general deformation let us consider the infinitesimal 
deformation $\delta_s$ that moves the boundary $\CC^{(1)}_s$ to a new position $\CC^{(1)\prime}_s$.
Clearly a generic infinitesimal deformation can be regarded 
as a linear combination of the $\delta_s$'s. Under such a deformation we have two kinds of effects:
\begin{enumerate}
\item The manifold $\CC^{(k+1)}_{s_1\cdots s_k s}$, being a subspace of $\CC^{(1)}_s$, 
gets shifted to a new 
manifold $\CC^{\prime(k+1)}_{s_1\cdots s_k s}$.
\item
The manifold $\CC^{(k)}_{s_1\cdots s_k}$ for $s\ne s_1,\cdots s_k$, having a boundary 
$-\CC^{(k)}_{s_1\cdots s_k s}$ inside  $\CC^{(1)}_s$, gets extended by an amount
$\delta_s \, \CC^{(k)}_{s_1\cdots s_k}$.
\end{enumerate}
Let us consider the subspace $\delta_s \CC^{(k)}_{s_1\cdots s_k}$. Its boundaries are given as follows.
$-\CC^{(k+1)}_{s_1\cdots s_k s}$ was a boundary of the original $\CC^{(k)}_{s_1\cdots s_k}$ which has now
moved to $-\CC^{\prime(k+1)}_{s_1\cdots s_k s}$. Therefore both $-\CC^{\prime(k+1)}_{s_1\cdots s_k s}$ and
$\CC^{(k+1)}_{s_1\cdots s_k s}$ form boundaries of $\delta_s \CC^{(k)}_{s_1\cdots s_k}$. Besides this
$\delta_s\CC^{(k)}_{s_1\cdots s_k}$ share the extensions of the boundaries $-\CC^{(k)}_{s_1\cdots s_kr}$ for $r\ne s$
of $\CC^{(k)}_{s_1\cdots s_k}$ which have been extended by $\delta_s\CC^{(k)}_{s_1\cdots s_kr}$. Therefore
we have the relation
\be \label{eshift1}
\p \delta_s \CC^{(k)}_{s_1\cdots s_k} = 
-\CC^{\prime(k+1)}_{s_1\cdots s_k s} + \CC^{(k+1)}_{s_1\cdots s_k s} - \sum_{r\ne s, s_1,\cdots ,s_k}
\delta_s \CC^{(k+1)}_{s_1\cdots s_k r}
\ee
The net change in $I$ under such a deformation is given by
\ben \label{eshift2}
\delta_s I &=& \sum_{k=0}^n  \sum_{\{s_1,\cdots ,s_k\}\atop s_1<s_2<\cdots < s_k, s_1,\cdots, s_k\ne s}
\int_{\delta_s\CC^{(k)}_{s_1\cdots s_k}} \II^{(k)}_{s_1\cdots s_k} \nonumber \\ &&
+ \sum_{k=0}^n  \sum_{\{s_1,\cdots ,s_k\}\atop s_1<s_2<\cdots < s_k,  s_1,\cdots, s_k\ne s}
\left[\int_{\CC^{\prime (k)}_{s_1\cdots s_k s}} \II^{(k+1)}_{s_1\cdots s_k s} - 
\int_{\CC^{(k)}_{s_1\cdots s_k s}} \II^{(k+1)}_{s_1\cdots s_k s}\right]\, .
\een
Using \refb{eshift1} we can express \refb{eshift2} as
\ben \label{eshift3}
\delta_s I &=& \sum_{k=0}^n  \sum_{\{s_1,\cdots ,s_k\}\atop s_1<s_2<\cdots < s_k,  s_1,\cdots, s_k\ne s}
\int_{\delta_s\CC^{(k)}_{s_1\cdots s_k}} \II^{(k)}_{s_1\cdots s_k} \nonumber \\ &&
\hskip -.3in - \sum_{k=0}^n  \sum_{\{s_1,\cdots ,s_k\}\atop s_1<s_2<\cdots < s_k,  s_1,\cdots, s_k\ne s}
\left[\int_{\p \delta_s \CC^{(k)}_{s_1\cdots s_k}} \II^{(k+1)}_{s_1\cdots s_k s} +
\sum_{r\ne s, s_1,\cdots, s_k}
\int_{\delta_s\CC^{(k+1)}_{s_1\cdots s_k r}} \II^{(k+1)}_{s_1\cdots s_k s}\right]\, .
\een
Using \refb{edikeq} the first term inside the square bracket in the second line can be manipulated as
\be\label{eshift4}
\int_{\p \delta_s \CC^{(k)}_{s_1\cdots s_k}} \II^{(k+1)}_{s_1\cdots s_k s} 
= \int_{\delta_s \CC^{(k)}_{s_1\cdots s_k}} d\, \II^{(k+1)}_{s_1\cdots s_k s} 
=\int_{\delta_s \CC^{(k)}_{s_1\cdots s_k}} \left[\II^{(k)}_{s_1\cdots s_k} 
+ \sum_{\ell=1}^k  (-1)^{k+1-\ell} 
\II^{(k)}_{s_1\cdots s_{\ell-1} s_{\ell+1} \cdots s_k s} \right]\, .
\ee
Substituting this into \refb{eshift3} we get
\ben \label{eshift5}
\delta_s I &=& \sum_{k=0}^n  \sum_{\{s_1,\cdots ,s_k\}\atop s_1<s_2<\cdots < s_k, s_1,\cdots, s_k\ne s}
\int_{\delta_s\CC^{(k)}_{s_1\cdots s_k}} \sum_{\ell=1}^k (-1)^{k-\ell} \II^{(k)}_{s_1\cdots s_{\ell-1} s_{\ell+1} \cdots s_k s}
\nonumber \\ &&
- \sum_{k=0}^n  \sum_{\{s_1,\cdots ,s_k\}\atop s_1<s_2<\cdots < s_k, s_1,\cdots, s_k\ne s}
\sum_{r\ne s, s_1,\cdots, s_k}
\int_{\delta_s\CC^{(k+1)}_{s_1\cdots s_k r}} \II^{(k+1)}_{s_1\cdots s_k s}\, .
\een
By making a $k\to (k+1)$ shift in the first term and 
relabelling $s_\ell$ as $r$, $\{s_1,\cdots s_{\ell-1}, s_{\ell+1},\cdots , s_{k+1}\}$ as
$\{s_1,\cdots , s_k\}$, we see that the two terms in \refb{eshift5} 
cancel and we have
\be \label{eshift6}
\delta_s I = 0\, .
\ee
This shows that \refb{etotalI} is invariant under the deformations of the subspaces $\CC^{(1)}_s$.

Using these results we arrive at the following algorithm for evaluating an
$(n+3)$-point amplitude: 
\begin{enumerate}
\item We can take three of the vertex operators at fixed locations
$y_{n+1}$, $y_{n+2}$ and $y_{n+3}$ and take the locations $\sigma_1,\cdots \sigma_n$ of the other vertex
operators as coordinates of the moduli space.
\item 
We now list all possible ways of dividing the set $1,\cdots ,(n+3)$ into two subsets, with each subset containing
at least two particles. Associated with each such decomposition, labelled by $s$, there is a possible degeneration
in which the original sphere degenerates into two spheres, each carrying external punctures belonging to one
of the two subsets. For every such degeneration, we can construct, in terms original moduli $\{\sigma_i\}$, a new
complex
parameter $\xtau_s$ that vanishes at the degeneration and a set of $(n-1)$ complex moduli $m_{(s)}$ 
labelling the moduli of the spheres left after degeneration. The change of variables from $\{\sigma_i\}$ to
$\xtau_s$, $m_{(s)}$ may not be globally defined over the whole moduli space spanned by $m_{(s)}$ and
we typically will have to use different $\xtau_s$ in different patches. 

\item
Multiple degenerations will correspond to several of the $\xtau_s$'s vanishing simultaneously. Near a degeneration
where $\xtau_{s_1},\cdots, \xtau_{s_k}$ vanish, we can construct, in terms of the $\{\sigma_i\}$'s, 
a set of $(n-k)$ complex 
coordinates $m_{(s_1,\cdots , s_k)}$ which remain finite at the degeneration, labelling the moduli
of the spheres into which the original sphere degenerates. We can use $\xtau_{s_1},\cdots, \xtau_{s_k}$ and
$m_{(s_1,\cdots , s_k)}$ to parametrize the moduli space near such degenerations.
\item We now denote by $\DD_s$ an open tubular neighborhood 
of $\xtau_s=0$ and define $\CC^{(0)}$ to be the region
of the moduli space that excludes $\DD_s$ for all $s$. We also denote by $\CC^{(1)}_s$
the intersection $\p\DD_s\cap \CC^{(0)}$, forming a component of the boundary of $\CC^{(0)}$,
and by
$\CC^{(k)}_{s_1\cdots s_k}$ the codimension $k$ 
intersection $\CC^{(1)}_{s_1}\cap \CC^{(1)}_{s_2}\cap\cdots\cap \CC^{(1)}_{s_k}$.

\item Once we know the relation between the coordinates $\sigma_1,\cdots, \sigma_n$ and $\{\xtau_s\}$ and
$\{m_{(s_1,\cdots ,s_k)}\}$, we can determine $\II^{(k)}_{s_1\cdots s_k}$ as follows. $\II^{(0)}$ is the integrand of the
original amplitude given by correlation functions of vertex operators in the CFT. $\II^{(0)}$ is originally expressed
in terms of $\sigma_i$'s, but given the known relation between the $\sigma_i$'s and $\xtau_s$, $m_{(s)}$ we can
expand it near $\CC^{(1)}_s$ in a power series in $\xtau_s$ of the form:
\be \label{epow1}
\II^{(0)} = d\xtau_s\wedge d\bar\xtau_s\wedge d^{2(n-1)} m_{(s)}\,
 \sum_i K^{(s)}_i(m_{(s)}) \, \xtau_s^{-1+\alpha^{(s)}_i} \, \bar\xtau_s^{-1+\beta^{(s)}_i}\, ,
\ee
for appropriate constants $\alpha_i^{(s)}, \beta_i^{(s)}$ 
and functions $K^{(s)}_i(m_{(s)})$. Given this we can find 
$\II^{(1)}_s$ satisfying \refb{edikeq} as
\be \label{epow2}
\II^{(1)}_s = - d\xtau_s\wedge  d^{2(n-1)} m_{(s)}\,
 \sum_i (\beta^{(s)}_i)^{-1}\, K^{(s)}_i(m_{(s)}) \, \xtau_s^{-1+\alpha^{(s)}_i} \, 
 \bar\xtau_s^{\beta^{(s)}_i}\, ,
\ee
up to addition of exact forms which, as we have argued before,
does not affect the final result. We now expand 
$\II^{(1)}_{s_1}$ and $\II^{(2)}_{s_2}$ near $\CC^{(2)}_{s_1s_2}$ as a power series in 
$\xtau_{s_1}$, $\xtau_{s_2}$ and their complex conjugates, with the coefficients of expansion 
given as functions of $m_{(s_1,s_2)}$. This can be done by using the known relations between the
coordinate systems $\{ \xtau_{s_1},  \bar\xtau_{s_1}, m_{(s_1)}\}$, 
$\{ \xtau_{s_1},  \bar\xtau_{s_1}, \xtau_{s_2},  \bar\xtau_{s_2},m_{(s_1,s_2)}\}$ and
$\{ \xtau_{s_2},  \bar\xtau_{s_2}, m_{(s_2)}\}$, each of which can be related to the original
coordinates $\{\sigma_1,\cdots, \sigma_n\}$. Given $\II^{(1)}_{s_1}$ and $\II^{(1)}_{s_2}$, both
expressed in the same coordinate system, we now look for $\II^{(2)}_{s_1s_2}$ satisfying
\be
d\, \II^{(2)}_{s_1s_2} = \II^{(1)}_{s_1}-\II^{(1)}_{s_2}\, ,
\ee
with $\II^{(2)}_{s_1s_2}$ having an expansion in powers of $\xtau_{s_1}$, $\xtau_{s_2}$ and their
complex conjugates, with the coefficients of expansion given as functions of the remaining moduli
$m_{(s_1,s_2)}$. 
The
existence of such solutions is guaranteed by our general argument.
Proceeding this way we can construct all the $\II^{(k)}_{s_1\cdots s_k}$'s using \refb{edikeq}.
\item Once all the $\II^{(k)}_{s_1\cdots s_k}$'s and $\CC^{(k)}_{s_1\cdots s_k}$'s have been constructed,
we can compute the amplitude using \refb{etotalI}.
\item As already mentioned, the relation between the coordinate system 
$\{\xtau_s\}$, $\{m_{(s_1,\cdots, s_k)}\}$ and the global coordinates $\{\sigma_1,\cdots,\sigma_n\}$ may
differ in different patches of the moduli space. In particular 
the natural choice of the coordinates 
$\xtau_{s_1},\cdots, \xtau_{s_k}$
near $\CC^{(k)}_{s_1\cdots s_k}$ may differ from the natural choice of the $\xtau_s$'s 
near the $\CC^{(k-1)}$'s that are used to describe the
$\II^{(k-1)}$'s. 
Since the relations between these coordinates and $\{\sigma_i\}$'s and the
old coordinates and $\{\sigma_i\}$'s are known, we can always express the $\II^{(k-1)}$'s determined in the
previous step in the new coordinate system and proceed as above. Example of such changes in coordinate system will
be described in appendix \ref{sb}, where we have  described possible choice of coordinates $\{\xtau_s\}$,
$m_{(s_1,\cdots ,s_k)}$  on a five punctured sphere.
\end{enumerate}

\sectiono{Amplitudes in superstring theory} \label{ssuper}

In superstring theory the computation of amplitudes suffers from divergences similar to the ones 
described above for bosonic string theory -- appearing from propagation of internal states carrying
negative $L_0+\bar L_0$ eigenvalue. However there is another source of divergences in superstring
theory -- from the wrong choice of PCOs. Consider for example a pair of integrated $-1$ picture vertex operators
$e^{-\phi}\psi^\mu \bar\p X^\nu(z,\bar z)$ and $e^{-\phi} \psi^\rho \bar\p X^\sigma(w,\bar w)$. The rule that follows from
superstring field theory is that when we bring them close, we must also bring a PCO close to them so that
their product generates a $-1$ picture state. This can be achieved {\it e.g.} by converting the second
vertex operator to a zero picture vertex operator $\p X^\rho \bar\p X^\sigma(w,\bar w)$. The leading singularity
in the collision of the two vertex operators now is proportional to 
$(\bar z-\bar w)^{-2} e^{-\phi} \psi^\mu
\p X^\rho \, \eta^{\nu\sigma}$. If instead we had taken the product of the two $-1$ picture vertex operators, then
we would have gotten a stronger leading singularity proportional to $(z-w)^{-2}(\bar z-\bar w)^{-2} 
e^{-2\phi} \eta^{\mu\rho}\eta^{\nu\sigma}$. A similar situation would arise if we had converted both vertex
operators to zero picture before bringing them close to each other.

Now often it is convenient to take the PCO's to coincide with some vertex operators before computing the
correlation function and use the same arrangement everywhere in the moduli space. 
According to the discussion in the preceding paragraph this would give the wrong integrand
$\II^{(0)}$ near many degenerations, {\it e.g.} when
two or more zero picture vertex operators come together their product would carry a net
picture number 0 instead of $-1$. Therefore
this would be the wrong starting point for implementing the procedure described in \S\ref{shigher}.
However one can show that the error that one makes by taking the wrong choice of PCO locations adds 
to $\II^{(0)}$ an exact differential in $\CC^{(0)}$. We shall now argue that the addition of such exact differentials
does not affect the amplitude $I$ defined in \refb{etotalI}. For this let us suppose that we change $\II^{(0)}$
to $\II^{(0)}+ d \LL^{(1)}$ where $\LL^{(1)}$ is a globally defined $2n-1$ form on $\CC^{(0)}$. Then the equation
$d\II^{(1)}_s = \II^{(0)}$ may be solved by shifting $\II^{(1)}_s$ to $\II^{(1)}_s + \LL^{(1)}$. The next equation
\be
d\, \II^{(2)}_{s_1s_2} = \II^{(1)}_{s_1} - \II^{(1)}_{s_2}\, ,
\ee
is not affected by this change since the extra term $\LL^{(1)}$ in $\II^{(1)}_{s_1}$ and $\II^{(1)}_{s_2}$ cancel.
Therefore $\II^{(k)}_{s_1\cdots s_k}$ for $k\ge 2$ remain unchanged.

We can now study the effect of the change in $\II^{(0)}$ and $\II^{(1)}$ on $I$ defined in \refb{etotalI}. We have
\be \label{echan1}
\int_{\CC^{(0)}} \II^{(0)} \to \int_{\CC^{(0)}} \II^{(0)} + \int_{\CC^{(0)}} d\, \LL^{(1)}
=  \int_{\CC^{(0)}} \II^{(0)} +\int_{\p \CC^{(0)}} \LL^{(1)} 
= \int_{\CC^{(0)}} \II^{(0)} - \sum_s \int_{\CC^{(1)}_s} \LL^{(1)}\, ,
\ee
and 
\be \label{echan2}
\sum_s \int_{\CC^{(1)}_s} \II_s^{(1)} \to \sum_s \int_{\CC^{(1)}_s} \II_s^{(1)} 
+ \sum_s \int_{\CC^{(1)}_s} \LL^{(1)}\, .
\ee
Therefore the net change in the sum of the left hand sides of \refb{echan1} and \refb{echan2} vanishes,
showing that the expression corrects itself even if we evaluate the integrand with the wrong choice of
PCO locations. However this procedure works only for generic external momenta for which the
exponents $\gamma^{(s)}_i,\delta^{(s)}_i$ appearing in \refb{egener1} are not integers.

\sectiono{Mass shift under marginal deformation} \label{s6}

We shall now consider a different situation in tree level string theory where short distance divergences appear
in the world-sheet of the string, and show how string field theory removes the divergence.
The case that we shall describe involves marginal deformation of the matter CFT that is used in describing
the target space geometry. Since the deformed background is also a conformal field theory, one can formulate
string theory around the deformed background as well. The spectrum and S-matrix of the new theory is in
principle computable from the data in the original CFT before the deformation using conformal perturbation theory.
This however requires ultraviolet regularization on the world-sheet, since the naive computation requires 
integrating correlation functions of marginal operators and other operators over the locations of the marginal
operators and they diverge when the marginal operators collide with each other or with other operators.
Our goal will be to show how in string field theory we can carry out the computation without encountering any
divergence.

Although the procedure we shall describe is valid for any marginal deformation, in order to get concrete results
we shall focus on a particular class of examples where 
the target space includes a compact circle and the marginal deformation corresponds to deforming the radius
of the compact circle. 
We shall denote
the world-sheet scalar field corresponding to the compact circle by $Y$. 
In string field theory the marginal deformation that
changes the radius of the circle  can be represented as the effect
of switching on a background string field solving the classical equations of motion. We shall use string field theory
to compute the shift in the masses of various states under this deformation to second order in the deformation
parameter and compare the result to known results. In doing this we shall make use of the general  strategy
described in \cite{1411.7478} for solving classical equations of motion of string field theory.
During this analysis we shall see that while in the intermediate stages of calculation the results depend on
the choice of local coordinates used to define the theory, the final result is independent of this choice.

\subsection{Bosonic string theory}

The equations of motion of bosonic string field theory is given by \refb{eeom}:
\be \label{eeomrep}
Q_B|\Psi\rangle + \sum_{N=2}^\infty {1\over N!} \, [\Psi^N] = 0\, .
\ee
We shall solve this equation in a power series in the parameter $\mu$:
\be
|\Psi^{\cl} \rangle =\sum_{n\ge 1} \mu^n |\Psi_n^{\cl}\rangle\, ,
\ee 
by starting with a solution
\be \label{epsi1cl}
\mu\, |\Psi^{\cl}_1\rangle =\mu\, c_1 \bar c_1 V(0)|0\rangle, \quad V = \p Y \bar\p Y\, .
\ee
It is clear that this solves \refb{eeomrep} to order $\mu$ since $|\Psi^{\cl}_1\rangle$ is annihilated by $Q_B$.
The second order correction $\mu^2 |\Psi_2^{\cl}\rangle$ 
to the solution can be expressed as\cite{1411.7478}
\be \label{e5.3x}
|\Psi^{\cl}_2\rangle = -{1\over 2} {b_0^+\over L_0^+} (1-P) [\Psi^{\cl}_1 \Psi^{\cl}_1 ] + |\psi^{\cl}_2\rangle
\ee
where $P$ is the projection operator into the $L_0^+=0$ states and $|\psi^{\cl}_2\rangle\in P\HH$ satisfies:
\be \label{eabove}
Q_B|\psi^{\cl}_2\rangle = -{1\over 2}\, P [\Psi^{\cl}_1 \Psi^{\cl}_1 ]\, .
\ee
It is easy to see however that for the choice given in \refb{epsi1cl}, 
the right hand side of the above equation vanishes. For this we can take
the inner product of this with $\langle \phi|c_0^-$ for
any ghost number 2 state $|\phi\rangle$ in $\HH$ with $L_0^+=0$
and identify this with a 3-point function $\{\phi \Psi^{\cl}_1 \Psi^{\cl}_1\}$
using \refb{edefsquare}. Since \refb{epsi1cl} is invariant under $Y\to -Y$, we can restrict to states invariant
under this transformation. 
Also we can ignore states involving excitations in the parts 
of the matter CFT other than the one
involving $Y$, since one point functions of the corresponding vertex operators vanish.
The relevant states $|\phi\rangle$ are
\be \label{elist}
c_1 c_{-1} |0\rangle, \quad \bar c_1 \bar c_{-1}|0\rangle, \quad
c_1 \bar c_1 V(0)|0\rangle
\, .
\ee
For the first two states in \refb{elist},  $\{\phi \Psi^{\cl}_1 \Psi^{\cl}_1\}$ vanishes since in the
three point function the total ghost number has to add to 3 separately in the left and the
right sector. For the last state
in \refb{elist},  $\{\phi \Psi^{\cl}_1 \Psi^{\cl}_1\}$ 
vanishes since the three point function of three $\p Y\bar \p Y$'s vanishes
due to separate $Y\to -Y$ symmetry in the left and the right sector.\footnote{For a general marginal deformation,
vanishing of the three point function is a requirement for exact marginality.}
Therefore we can take
\be \label{epsi2cl}
|\psi^{\cl}_2\rangle = 0\, .
\ee
Since the states listed in \refb{elist} are all BRST invariant, we
could include an arbitrary linear combination of these states in the definition of 
$|\psi_2^{\cl}\rangle$. This will correspond
to a redefinition of  $\mu$ or the string coupling constant, or a gauge transformation.

Even though we shall use the solution to order $\mu^2$, it is instructive to examine how the solution can be
extended to order $\mu^3$. We can solve \refb{eeomrep} by taking\cite{1411.7478}
\be
|\Psi^{\cl}_3\rangle = -{b_0^+\over L_0^+} (1-P) \left({1\over 6} [\Psi^{\cl}_1 \Psi^{\cl}_1 \Psi^{\cl}_1] 
+  [ \Psi^{\cl}_1 \Psi^{\cl}_2]  \right)
+ |\psi^{\cl}_3\rangle 
\ee
where $|\psi_3^{\cl}\rangle\in P\HH$ satisfies
\be\label{efail}
Q_B|\psi_3^{\cl}\rangle =-{1\over 6} P [\Psi^{\cl}_1 \Psi^{\cl}_1 \Psi^{\cl}_1] 
-P  [ \Psi^{\cl}_1 \Psi^{\cl}_2]\, .
\ee
Possible obstruction to extending the solution to order $\mu^3$ can arise from the failure to solve
\refb{efail}.
Using \refb{e5.3x} and \refb{epsi2cl}, we can 
express this as
\be
Q_B|\psi_3^{\cl}\rangle = -P \left({1\over 6} [\Psi^{\cl}_1 \Psi^{\cl}_1 \Psi^{\cl}_1] 
-{1\over 2}  [ \Psi^{\cl}_1 b_0^+ (L_0^+)^{-1} (1-P) [\Psi^{\cl}_1\Psi_1^{\cl}]]  \right)\, .
\ee
A consistency condition for the existence of a solution to this equation is the vanishing of the inner product
of both sides with $\langle \phi|c_0^-$ for the three states $|\phi\rangle$ listed in \refb{elist}. The inner product of the
left hand side with all the states vanish since all of these states are BRST invariant. The inner product of the right 
hand side with the first two states in \refb{elist} 
vanish due to the vanishing of the correlation functions of odd number of
$\p Y$'s and / or odd number of $\bar\p Y$'s. Since the last state in \refb{elist}
is given by $|\Psi_1^{\cl}\rangle$,
we need to check the vanishing of
\ben\label{e7.11}
A&\equiv& \langle \Psi_1^{\cl}| c_0^- P \left( [\Psi^{\cl}_1 \Psi^{\cl}_1 \Psi^{\cl}_1] 
-3 [ \Psi^{\cl}_1 b_0^+ (L_0^+)^{-1} (1-P) [\Psi^{\cl}_1\Psi_1^{\cl}]]  \right)
\nonumber \\
&=& \{\Psi^{\cl}_1 \Psi^{\cl}_1 \Psi^{\cl}_1\Psi^{\cl}_1\} -3\{\Psi^{\cl}_1 \Psi^{\cl}_1
b_0^+ (L_0^+)^{-1} (1-P) [\Psi^{\cl}_1\Psi_1^{\cl}]\} \, .
\een
Formally this can be regarded as the four point function of four $\Psi_1^{\cl}$ states, with the first term representing
the contribution from the elementary vertex and the second term representing the contribution from the s, t and u
channel diagrams all of which are equal. 
However unlike the four point function which has divergences from collision
of vertex operators, \refb{e7.11} is manifestly finite. 
To evaluate it we shall choose a local coordinate system of
the type described in \S\ref{s2} with large $\lambda$. We emphasize however that the choice of large $\lambda$ is a matter of
convenience, but is in no way necessary.

Since the external momenta all vanish and are not generic, we can have  ambiguities in applying the procedure of
\S\ref{s5},\ref{shigher}. Therefore we shall evaluate \refb{e7.11} directly. First we have
\ben \label{e4psicl}
\{\Psi^{\cl}_1 \Psi^{\cl}_1 \Psi^{\cl}_1\Psi^{\cl}_1\} 
&=& {1\over 2\pi i}\int_{\RR} d\sigma\wedge d\bar\sigma \, \langle \Psi^{\cl}_1(0) \Psi^{\cl}_1(1) \Psi^{\cl}_1(\infty) \,
V(\sigma)\rangle 
\nonumber \\
&=& -{1\over 32\pi i} \, \int_\RR  d\sigma\wedge d\bar\sigma 
\,  \left[\sigma^{-2}+ (\sigma-1)^{-2}+1\right] \left[\bar\sigma^{-2}
+ (\bar\sigma-1)^{-2}+1\right] \nonumber\\
&=& {1\over 32\pi i} \, \int_\RR  d\sigma\wedge d\bar\sigma  
\, \left[\sigma^{-2}+ (\sigma-1)^{-2}+1\right]  \p_{\bar\sigma} \left[\bar\sigma^{-1}
+ (\bar\sigma-1)^{-1}-\bar\sigma\right] \nonumber\\
&=& {1 \over 32\pi i} \, \int_{\p\RR_s+\p\RR_u+\p\RR_t}  d\sigma 
\, \left[\sigma^{-2}+ (\sigma-1)^{-2}+1\right] \left[\bar\sigma^{-1}
+ (\bar\sigma-1)^{-1}-\bar\sigma\right]\, ,\nonumber \\
\een
where we have used $\p\RR=-\p\RR_s-\p\RR_u-\p\RR_t$.
Since all the external states are identical, the contribution from $\p\RR_u$, $\p\RR_s$ and $\p\RR_t$ are 
identical. So we shall calculate the contribution from one of the boundaries $\p\RR_u$ around $\sigma=0$
and multiply the result by 3.
Now we see from \refb{e3.13x} and \refb{e2.18} that in the large $\lambda$ limit,  $\p\RR_u$ given by 
$|q|=1$ corresponds approximately to a circle around $\sigma=0$ of radius of order $\lambda^{-2}$. If we drop terms
proportional to negative powers of $\lambda$, which will drop out when we take the $\lambda\to\infty$ limit, then
the only contribution to \refb{e4psicl} comes from the $\sigma^{-2}\bar\sigma^{-1}$
term in the integrand. This leads to
\be\label{esample}
\{\Psi^{\cl}_1 \Psi^{\cl}_1 \Psi^{\cl}_1\Psi^{\cl}_1\}  = {3 \over 32\pi i} \int_{\p\RR_u} d\sigma\, \sigma^{-2}
\bar\sigma^{-1}\, .
\ee
Since the $\p\RR_u$ is not strictly a circle in the $\sigma$-plane, it
will be more convenient to express the boundary term in the $q$ plane via the relation \refb{e2.18}. 
This gives
\be \label{ei1bbb}
\{\Psi^{\cl}_1 \Psi^{\cl}_1 \Psi^{\cl}_1\Psi^{\cl}_1\}   =
 - {3 \over 32\pi i}  \int_{|\xi|=1/\lambda^2}d\xi\, {g'(\xi)\over (1-g(\xi))^2} {1\over 1 - \overline{g(\xi)}}
\, ,
\ee
where $\xi=q/\lambda^2$. Since the integral has to be evaluated at $|\xi|=1/\lambda^2$, we can expand 
$g(\xi)$ in a power series in $\xi$ for large $\lambda$.  Using  \refb{e3.13x}
we have
\be \label{ei1b}
g(\xi) = 1 - {\xi\over h_3'(1)^2} + \OO(\xi^2), \quad {g'(\xi)\over (1-g(\xi))^2} = {\p\over \p\xi}\left({1\over 1- g(\xi)}\right)
=-{ h_3'(1)^2\over \xi^2} + \OO(1)\, .
\ee
Substituting this into \refb{ei1bbb} we get
\be \label{ei1finbb}
\{\Psi^{\cl}_1 \Psi^{\cl}_1 \Psi^{\cl}_1\Psi^{\cl}_1\}  \simeq  {3 \over 16} {\lambda^4 \, |h_3'(1)|^4} \, ,
\ee
where $\simeq$ means that we have dropped terms involving inverse powers of $\lambda$.

The second term on the right hand side of \refb{e7.11} is given by
\be \label{esample2}
-3\{\Psi^{\cl}_1 \Psi^{\cl}_1
b_0^+ (L_0^+)^{-1} [\Psi^{\cl}_1\Psi_1^{\cl}]\}
= 
-3 \sum_{p,q} \left\{\Psi^{\cl}_1\Psi_1^{\cl} \xi_p\right \} \left\langle \xi^c_p|
b_0^+ (L_0^+)^{-1} (1-P) \delta_{L_0^-} b_0^- |\xi_q^c\right\rangle
\{\xi_q\Psi^{\cl}_1 \Psi^{\cl}_1\}\, ,
\ee
where $|\xi_p\rangle$ denote a complete basis of ghost number 2
states in $\HH$ and $|\xi_p^c\rangle$ is a conjugate
basis of ghost number 4 states in $c_0^-\HH$ satisfying
\be\label{ecomplete}
\langle \xi^c_p |\xi_q\rangle = \delta_{pq} = \langle \xi_q |\xi^c_p\rangle,
\qquad \sum_p |\xi_p\rangle \langle \xi_p^c| = \sum_p |\xi_p^c\rangle \langle \xi_p|
={\bf 1}\, .
\ee
In arriving at the right hand side of \refb{esample2}, we have replaced
$[\Psi^{\cl}_1\Psi_1^{\cl}]$ by $b_0^-c_0^- [\Psi^{\cl}_1\Psi_1^{\cl}]$ using the fact that
$[\Psi^{\cl}_1\Psi_1^{\cl}]\in\HH$, and then inserted a sum over complete set of states
between $b_0^-$ and $c_0^-$.
Now from \refb{e2.2} we see that if $\xi_p$ has $L_0^+$ eigenvalue $\Delta_p$, then for dimension 0
primaries $A$, $B$, the $\{AB\xi_p\}$ carries
a factor of $|dw_3/dz|_{z=z_3}^{-\Delta_p}=
(\lambda\, |h'_3(1)|)^{-\Delta_p}$. Since $\Psi_1^{\cl}$ is a dimension 0 primary, it follows 
that for large $\lambda$ we can restrict the sum over $p,q$ in \refb{esample2} to states with
$\Delta_p\le 0$. Now the contributions from the $\Delta_p=0$ states are already removed by the projection
operator $(1-P)$, therefore we have to focus on states with $\Delta_p< 0$. The only such state is the ground
state with $\Delta_p=-2$, giving
\be 
|\xi_p\rangle =|\xi_q\rangle = c_1\bar c_1|0\rangle,
\quad |\xi_p^c\rangle =|\xi_q^c\rangle = - c_0 \bar c_0 c_1\bar c_1|0\rangle\, .
\ee
Therefore we have
\ben\label{esample3}
&& \left\{\Psi^{\cl}_1\Psi_1^{\cl} \xi_p\right \} =  - {1\over 4} 
(\lambda\, |h'_3(1)|)^{2}, \quad \{\xi_q\Psi^{\cl}_1 \Psi^{\cl}_1\} = \{\Psi^{\cl}_1 \Psi^{\cl}_1\xi_q\}
= - {1\over 4} (\lambda\, |h'_3(1)|)^{2}, \nonumber \\ 
&&
\left\langle \xi^c_p\left|
b_0^+ (L_0^+)^{-1} (1-P) \delta_{L_0^-} b_0^- \right|\xi_q^c\right\rangle = 1.
\een
This gives
\be \label{ei4nonbb}
-3\{\Psi^{\cl}_1 \Psi^{\cl}_1
b_0^+ (L_0^+)^{-1} [\Psi^{\cl}_1\Psi_1^{\cl}]\}
\simeq  -{3 \over 16} (\lambda\, |h'_3(1)|)^{4} \, .
\ee

\refb{ei4nonbb} exactly cancels \refb{ei1finbb}. Therefore \refb{e7.11} gives 
\be
A=0\, .
\ee
This in turn shows that there is no obstruction to extending the solution to order $\mu^3$. This is equivalent to
proving the vanishing of the $\beta$-function for the marginal deformation $V$ to order $\mu^3$ without having 
to introduce ultraviolet cut-off at any step. 

Earlier explicit analysis of 
marginal deformations in closed string field theory\cite{9201040} examined the
solution to second order and therefore did not encounter dependence on the local coordinate $\lambda\, h_3(z)$
in the intermediate steps of the calculation. As stated below \refb{e7.11}, and noted in \cite{9201040},
at order $\lambda^3$ the existence of the
solution formally requires the vanishing of the four point function of marginal operators. However the corresponding
integrand diverges when vertex operators collide and one has to regularize it. 
In contrast, string field theory gives finite result by treating $(L_0^+)^{-1}$ correctly.

We now turn to the computation of the spectrum of the deformed theory to order $\mu^2$.
For this we need to study the fluctuation of the string field
around the new background. Let us define:
\be 
|\Phi\rangle = |\Psi\rangle - |\Psi^{\cl}\rangle\, .
\ee
Then the equation of motion \refb{eeomrep}
to linear order in $\Phi$ is given by:
\be \label{elin}
Q_B|\Phi\rangle + \sum_{N=1}^\infty {1\over N!}[(\Psi^{\cl})^N\Phi] = 0\, .
\ee
Let us for definiteness consider the tachyon state carrying momentum $n/R$ along the compact direction and
$k$ along the non-compact directions. Starting with the leading order solution to \refb{elin}, we shall compute
systematic corrections to the solution following the procedure described in \cite{1411.7478}. We shall denote by
$|\Phi_\ell\rangle$ the full solution to order $\mu^\ell$.
To order $\mu^0$ the solution to \refb{elin} takes the form:
\be \label{edefk0}
|\Phi_0\rangle = |\phi_0\rangle, \quad |\phi_0\rangle=
c\bar c e^{i k_{(0)}.X}  e^{i n Y / R} , \quad k_{(0)}^2 = -m^2 - {n^2\over R^2}=4 - {n^2\over R^2}\, .
\ee

For subsequent analysis we need to introduce a projection operator $\PP$ that projects 
to states which carry momentum $k=k_{(0)}+\OO(\mu)$ along
non-compact directions, $n/R$ along the compact direction and have $L_0^+=\OO(\mu)$. 
It is easy to verify that in this case only vertex operators invariant under $\PP$ are those of the form
$c\bar c e^{i k.X}  e^{i n Y / R}$. We normalize $|\Phi_\ell\rangle$ such that
\be \label{edefphin}
|\phi_\ell\rangle \equiv \PP |\Phi_\ell\rangle = c_1\bar c_1 \, e^{i k_{(\ell)}.X}(0) 
e^{i n Y/R}(0)|0\rangle, \quad 
k_{(\ell)} = k_{(0)}+\OO(\mu)\, .
\ee
As discussed in \cite{1411.7478}, since $k_{(\ell)}$ is expected to 
be different from $k_{(0)}$  due to a change in the mass,  it is
not convenient to use the solution $|\Phi_0\rangle$ and correct it to obtain 
$|\Phi_\ell\rangle$. 
Instead we begin with 
$|\phi_\ell\rangle$ given in \refb{edefphin} as the seed solution and correct it to order $\mu^\ell$ to determine
$|\Phi_\ell\rangle$ and $k_{(\ell)}$.

We begin with the
ansatz for
the solution $|\Phi_1\rangle$:
\be  \label{edefk1}
|\Phi_1\rangle = |\phi_1\rangle +\OO(\mu), \qquad \phi_1
=c\bar c e^{i k_{(1)}.X}  e^{i n Y / R} , \qquad  k_{(1)}^2 = k_{(0)}^2 + a_1\mu\, ,
\ee
where the constant $a_1$ will be determined shortly. Taking \refb{edefk1} as the leading order
solution, we can substitute this into the right hand side of \refb{elin} to get a solution to order $\mu$:
\be
|\Phi_1\rangle = - \mu\, {b_0^+\over L_0^+} (1-\PP) \, [\Psi^{\cl}_1 \phi_1] + |\phi_1\rangle \, ,  
\ee
provided $|\phi_1\rangle$ satisfies:
\be \label{eqbford}
Q_B |\phi_1\rangle = - \mu\, \PP  [\Psi^{\cl}_1 \phi_1] +\OO(\mu^2)\, .
\ee
If we define
\be \label{e7.30x}
\tilde\phi_1=c\bar c e^{-i k_{(1)}.X}  e^{-i n Y / R}\, ,
\ee
satisfying the normalization
\be\label{esignphi11}
\langle \tilde \phi_1|c_0\bar c_0 | \phi_1\rangle = -1\, ,
\ee
and take the inner product of $\langle\tilde\phi_1|c_0^-$ with \refb{eqbford}, we get
\be  \label{ek1k1pre}
\langle \tilde\phi_1| c_0^- Q_B |\phi_1\rangle = - \mu \, \langle \tilde\phi_1| c_0^- |[\Psi^{\cl}_1 \phi_1]\rangle
= - \mu\, \{ \tilde\phi_1\Psi^{\cl}_1 \phi_1\}
= - \mu\, \langle \tilde \phi_1| c \bar c \p Y \bar \p Y (1) 
|\phi_1\rangle +\OO(\mu^2)\, ,
\ee
where in the last step we have used the fact that $\Psi^{\cl}_1$ is a dimension 0 primary and $\phi_1$ and $\tilde\phi_1$
are primaries of dimension of order $\mu$. Only the $(c_0L_0+\bar c_0\bar L_0)$ term
in $Q_B$ contributes to the left hand side.
This gives
\be\label{ek1k1}
-{1\over 4} \left(k_{(1)}^2 - 4 + n^2 R^{-2}
\right) = - {n^2\over 4\, R^2} \, \mu  \quad \Rightarrow \quad k_{(1)}^2 = 4 - n^2 \, R^{-2} \left(1-\mu\right)\, .
\ee

We now turn to 
order $\mu^2$ computation. We begin with the ansatz
\be 
|\Phi_2\rangle = |\phi_2\rangle+\OO(\mu)\, , 
\ee
with $|\phi_2\rangle$ as defined in \refb{edefphin}, 
and solve the equations of motion \refb{elin} iteratively to get successive order solutions:
\be
|\Phi_2\rangle = - \mu \, {b_0^+\over L_0^+} (1-\PP) \, [\Psi^{\cl}_1 \phi_2] + |\phi_2\rangle+\OO(\mu^2)\, ,
\ee
and
\ben
&& \hskip -.3in |\Phi_2\rangle = -  {b_0^+\over L_0^+} (1-\PP) \, \left(\mu \, [\Psi^{\cl}_1 \phi_2] - \mu^2 [\Psi^{\cl}_1
{b_0^+\over L_0^+}(1-\PP) [\Psi^{\cl}_1\phi_2]]+ \mu^2
[\Psi^{\cl}_2 \phi_2] + {\mu^2\over 2} [\Psi^{\cl}_1\Psi^{\cl}_1 \phi_2]\right)
+ |\phi_2\rangle\nonumber \\ && \hskip 1in + \OO(\mu^3) \, ,
\een
provided $|\phi_2\rangle$ satisfies, to order $\mu^2$,
\ben \label{e7.38xx}
Q_B |\phi_2\rangle &=&  - \PP \left(\mu\,
[\Psi^{\cl}_1 \phi_2] - \mu^2 \left[\Psi^{\cl}_1 {b_0^+\over L_0^+} (1-\PP) \,  [\Psi^{\cl}_1 \phi_2] \right] \right. \nonumber \\ &&
\left. -{\mu^2\over 2}\left[ \left({b_0^+\over L_0^+} (1-P) [\Psi^{\cl}_1 \Psi^{\cl}_1]\right) \phi_2 \right] 
+ {\mu^2\over 2} [\Psi^{\cl}_1\Psi^{\cl}_1 \phi_2]\right)
\, . %\nonumber \\
\een
In the second line we have used the expression for $|\Psi_2^{\cl}\rangle$ given in \refb{e5.3x},
\refb{epsi2cl}.
Defining
\be \label{edefk2a}
\tilde\phi_2 =  c\bar c e^{-i k_{(2)}.X}  e^{-i n Y / R}\, ,
\ee
and
taking the inner product of $\langle \tilde \phi_2| c_0^-$ with \refb{e7.38xx}, using the analog of
\refb{esignphi11} with $\phi_1$, $\tilde\phi_1$ replaced by $\phi_2$, $\tilde\phi_2$, we get
\ben \label{esecondorder}
\hskip -.5 in -{1\over 4} \left(k_{(2)}^2 - 4 + n^2 R^{-2}
\right) &=& -\mu\, \{\tilde \phi_2 \Psi_1^{\cl} \phi_2\}
+ \mu^2 \left\{ \tilde \phi_2 \Psi^{\cl}_1 {b_0^+\over L_0^+} (1-\PP) \,  [\Psi^{\cl}_1 \phi_2] \right\} \nonumber \\
&& +{\mu^2\over 2}\left\{\tilde \phi_2 \phi_2 \left({b_0^+\over L_0^+} (1-P) [\Psi^{\cl}_1 \Psi^{\cl}_1]\right)  \right\}
-{\mu^2\over 2} \left\{\tilde \phi_2  \phi_2 \Psi^{\cl}_1\Psi^{\cl}_1\right\}\, .
\een

We shall now evaluate the different terms appearing on the right hand side of \refb{esecondorder}. First of all
we have:
\ben\label{ek2k1}
\mu\, \{\tilde \phi_2 \Psi_1^{\cl} \phi_2\} &=& {\mu\over 4}  n^2 R^{-2} \, \left( {\lambda |h_3'(1)|}\right)^{-k_{(2)}^2 + 4 - n^2 R^{-2}} \nonumber \\
&=& {\mu\over 4}  n^2 R^{-2}  \left\{ 1- \left(k_{(2)}^2 - 4 + n^2 R^{-2}\right) \ln (\lambda|h_3'(1)|)\right\} + \OO(\mu^3)\, ,
\een
where the $\left( {\lambda |h_3'(1)|}\right)^{-k_{(2)}^2 + 4 - n^2 R^{-2}}$ factor arises from the fact that the states
$\phi_2$, $\tilde\phi_2$ have $L_0^+$ eigenvalues $\Delta=(k_{(2)}^2 - 4 + n^2 R^{-2})/2$ and therefore 
$\{\tilde \phi_2 \Psi_1^{\cl} \phi_2\}$ will carry factors of $|dw_3/dz|_{z=z_i}^{-\Delta}$ for $i=1,3$.
Using \refb{e2.1}, \refb{e2.2} and \refb{e2.2b} we get $(dw_1/dz)_{z=z_1}=(dw_3/dz)|_{z=z_3}=\lambda h_3'(1)$.
To evaluate the right hand side of \refb{ek2k1} to order $\mu^2$, we can replace $k_{(2)}$ by $k_{(1)}$
at the cost of making an error of order $\mu^2$. Using \refb{ek1k1} we get
\be \label{efirstfin}
-\mu\, \{\tilde \phi_2 \Psi_1^{\cl} \phi_2\} = -{\mu\over 4}  n^2 R^{-2} + {\mu^2\over 4} n^4 R^{-4}\,
\ln (\lambda|h_3'(1)|) +\OO(\mu^3)\, .
\ee

The evaluation of the last three terms on the right hand side of \refb{esecondorder} can be simplified by noting
that since all these terms
already have explicit factors of $\mu^2$, 
we can replace the momenta $k_{(2)}$ by $k_{(0)}$ given in \refb{edefk0}. 
This makes all the external states on-shell. 

Let us begin with the evaluation of the last term in \refb{esecondorder}. This has the same expression as
the on-shell four point function of the states $\phi$, $\tilde\phi$, $\Psi^{\cl}_1$ and $\Psi^{\cl}_1$, except that
the integration over the moduli runs over the region $\RR$ introduced in \S\ref{slocal} 
instead of the whole complex plane.
The integrand for the on-shell amplitude is given by
\ben \label{e4integrand}
\II &\equiv& {1\over 2\pi i} \langle c\bar c \p Y \bar\p Y (0) \, c\bar c e^{i k_{(0)}.X}  e^{i n Y / R}(\infty) \, 
c\bar c e^{-i k_{(0)}.X}  e^{-i n Y / R}(1)  \, \p Y \bar\p Y(\sigma)
\rangle  \nonumber \\
&=& - {1\over 2\pi i} \left\{ {1\over 2\sigma^2} - {n^2\over 4 R^2} {1\over \sigma -1}\right\}
\left\{ {1\over 2\bar\sigma^2} - {n^2\over 4 R^2} {1\over \bar\sigma -1}\right\}\, .
\een
Therefore we have
\be \label{e5furpoint}
-{\mu^2\over 2} \, \left\{\tilde \phi_2  \phi_2 \Psi^{\cl}_1\Psi^{\cl}_1\right\}=-{\mu^2\over 2} \, 
\int_{\RR} d^2\sigma \, \II \equiv I_1+I_2+I_3\, ,
\ee
where 
\be 
I_1 = {\mu^2\over 16\pi i} \, \int_\RR d^2 \sigma \, \sigma^{-2} \bar\sigma^{-2}\, ,
\ee
\be
I_2 = -{\mu^2\over 32\pi i} \, n^2 R^{-2} \, \int_\RR d^2 \sigma \left\{ \sigma^{-2} (\bar\sigma-1)^{-1}
+ \bar \sigma^{-2} (\sigma-1)^{-1}\right\}\, ,
\ee
and
\be
I_3 = {\mu^2\over 64\pi i} \, n^4 R^{-4} \, \int_\RR d^2 \sigma \,  (\sigma-1)^{-1} (\bar\sigma-1)^{-1}
\, .
\ee

While the analysis can be carried out for any choice of local coordinate system encoded in the functions
$\lambda\, h_i(z)$ introduced in \S\ref{slocal}, we shall simplify our analysis by taking $\lambda$ to be large.
In this case it follows from \refb{emap1},
\refb{e2.18} and
\refb{e2.23} that the excluded regions $|q|\le 1$, denoted by $\RR_s$, $\RR_u$ and $\RR_t$ in \S\ref{slocal},
correspond to small regions around 1, 0 and a region outside a large radius
respectively. Therefore for large $\lambda$
we can ignore the excluded regions $\RR_s$, $\RR_t$ and/or $\RR_u$
as long as the integrands do not encounter any divergence.\footnote{Note that we are not taking the large $\lambda$
limit of individual terms, but dropping terms with inverse powers of $\lambda$ in anticipation of the fact that
eventually we shall take the $\lambda\to\infty$ limit after adding all the terms. In this limit the terms with
inverse powers of $\lambda$ will drop out. We are however perfectly entitled to keep $\lambda$ finite and add
up the contribution from all the terms.} 
For example in $I_1$ we can ignore the excluded
regions $\RR_s$ and $\RR_t$ since the integrand does not have any divergence from 1 and $\infty$, but
cannot ignore the excluded region $\RR_u$ around the origin. 
On the other hand in $I_3$ we can ignore the excluded region $\RR_u$ around the origin, but cannot ignore
$\RR_s$ and $\RR_t$.

Let us begin with the evaluation of $I_1$. Since the only excluded region in this expression is a small region
$\RR_u$ around 0, we can express this as
\be \label{ei1a}
I_1 = -{\mu^2 \over 16\pi i}\, \int_\RR d\sigma\wedge d\bar \sigma \, {\p\over \p\bar\sigma} \left(\sigma^{-2}
\bar\sigma^{-1}\right) =-{\mu^2 \over 16\pi i}\int_{\p \RR_u} \, d\sigma \, \sigma^{-2} \bar\sigma^{-1}\, ,
\ee
where we have used $\p \RR=-\p\RR_u$. 
This has the same structure as \refb{esample} and can be evaluated using identical procedure, leading to the
analog of \refb{ei1finbb}:
\be \label{ei1fin}
I_1 \simeq  -{\mu^2 \over 8} {\lambda^4 \, |h_3'(1)|^4} \, ,
\ee
where $\simeq$ denotes equality up to terms containing inverse powers of $\lambda$.

$I_2$ can be evaluated by expressing this as
\be
I_2 = {\mu^2\over 32\pi i} \, n^2 R^{-2} \, \int_\RR d^2 \sigma \left\{ {\p\over\p\sigma} \left(\sigma^{-1}
(\bar\sigma-1)^{-1} \right)
+ {\p\over \p\bar\sigma} \left(\bar \sigma^{-1}(\sigma-1)^{-1}\right) \right\}\, .
\ee
We can evaluate this using integration by parts, picking up boundary contributions from 
$\p\RR$. The only boundary that contributes is the boundary $-\p\RR_s$ around $\sigma=1$.
This gives
\be
I_2 = {\mu^2\over 8} \, n^2 R^{-2} \, .
\ee

$I_3$ can be evaluated by noting that the integral has logarithmic divergence both near 1 and $\infty$, and
therefore we need to use eqs.\refb{emap1} and \refb{e2.23} to determine the cut-off on $\sigma$ integral
by identifying the $|q|=1$ curves. Using \refb{e3.13x} we see that
in the $\sigma=y_4$ plane these curves are at
\be
|\sigma-1| \simeq |h'_3(1)|^{-2} \, \lambda^{-2}\, ,
\ee
and 
\be
|\sigma| \simeq \lambda^2 \, |h'_3(1)|^2\, ,
\ee
respectively. Since the integrals are at most logarithmically divergent we do not need to know the corrections
to these curves. This gives
\be 
I_3 \simeq {\mu^2\over 64\pi i} \, n^4 R^{-4} \, \int_{|h'_3(1)|^{-2}\lambda^{-2} \le |\sigma-1|\le \lambda^2 |h'_3(1)|^2} 
d^2 \sigma \,  |\sigma-1|^{-2}=-{\mu^2 \over 4}  \, n^4 R^{-4} \, \ln (\lambda\, |h'_3(1)|)
\, .
\ee

Using the values of $I_1,I_2,I_3$ determined above, we finally get from \refb{e5furpoint}
\be\label{e6.43}
-{\mu^2\over 2} \, \left\{\tilde \phi_2  \phi_2 \Psi^{\cl}_1\Psi^{\cl}_1\right\} = I_1+I_2+I_3
\simeq -{\mu^2\over 8} \, {\lambda^4\, |h'_3(1)|^4} + {\mu^2\over 8} \, {n^2\over R^2} 
-{\mu^2 \over 4}  \, n^4 R^{-4} \, \ln \left(\lambda\, |h'_3(1)|\right)\, .
\ee

Let us now turn to the contribution from the third term on the right hand side of \refb{esecondorder}.
Evaluation of this proceeds exactly as that of \refb{esample2} with the only difference that the first two
factors of $\Psi_1^{\cl}$ are replaced by $\tilde \phi_2 \phi_2$. The 
first equation in \refb{esample3} is replaced by
\be  
\left\{\tilde \phi_2 \phi_2 \xi_p\right \} = -(\lambda\, |h'_3(1)|)^{2}\, .
\ee
This gives the analog of \refb{ei4nonbb}:
\be \label{ei4non}
I_4\equiv {\mu^2\over 2}\left\{\tilde \phi_2 \phi_2 \left({b_0^+\over L_0^+} (1-P) \delta_{L_0^-}
b_0^- c_0^- [\Psi^{\cl}_1 \Psi^{\cl}_1]\right)  \right\} \simeq   {\mu^2\over 8} (\lambda\, |h'_3(1)|)^{4} \, .
\ee

The contribution from the second term on the right hand side of \refb{esecondorder} 
vanishes in the large $\lambda$ limit due to the
following reason. As already argued before, for evaluation of this term we can set $k_{(2)}=k_{(0)}$. 
In this case momentum conservation forces the intermediate state to carry a factor of
$e^{ik_{(0)}.X + i n  Y/R}$ which has $L_0^+=2$. Therefore the lowest $L_0^+$ eigenvalue
state propagating in this channel is $c\bar c e^{ik_{(0)}.X + i n Y/R}$ with $L_0^+=0$. The contribution from
this state is projected out by the $(1-\PP)$ operator insertion. Therefore the contribution comes only from states
with positive $L_0^+$ eigenvalue, 
leading to terms with negative powers of $\lambda$. Such terms will vanish in the large
$\lambda$ limit.
Therefore 
\be \label{ei5vanish}
\left\{ \tilde \phi_2 \Psi^{\cl}_1 {b_0^+\over L_0^+} (1-\PP) \,  [\Psi^{\cl}_1 \phi_2] \right\} \simeq 0\, .
\ee

We now have from \refb{esecondorder},  \refb{efirstfin}, \refb{e6.43}, \refb{ei4non} and \refb{ei5vanish}, 
\be - {1\over 4} \left(k_{(2)}^2 - 4 + n^2 R^{-2}
\right)  \simeq -\mu\, \{\tilde \phi_2 \Psi_1^{\cl} \phi_2\}  + I_1+I_2+I_3+I_4 \nonumber \\
= -\left({\mu\over 4} - {\mu^2\over 8}\right) n^2 R^{-2} \, .
\ee
This gives
\be\label{ebosfin}
k_{(2)}^2 = 4 - n^2 R^{-2} \left(1 -\mu + {\mu^2\over 2}\right)\, .
\ee
Therefore the mass$^2$ of the state to order $\mu^2$ is given by
\be
m^2 = - k_{(2)}^2 = -4 + n^2 R^{-2} \left(1 -\mu + {\mu^2\over 2}\right)\, .
\ee
This is consistent with the expectation that the marginal deformation induces a deformation of the radius $R$
of the compact direction. Note that we have arrived at this result without encountering any divergence from
collision of the pair of marginal operators or of the marginal operator with the vertex operator of the tachyon. The 
$\mu$ dependent terms can be regarded as an expansion of $e^{-\mu}$ but we do not have a compelling reason to
believe that this pattern will continue to hold at higher order. 

\subsection{Heterotic string theory} \label{shetb}

We shall now repeat the analysis of the previous section for heterotic string theory. The marginal
deformation corresponds to switching on NS sector string field.
For definiteness we shall take the states, whose mass shift we compute, also to be in the NS sector, but the
generalization to Ramond sector states is straightforward.
Since in the 
NS sector the structure of heterotic string field theory is identical to that of bosonic string field theory, 
most of the analysis takes identical form. For this reason, we shall mention only the differences.

The first difference is in the form of the leading order classical solution describing the shifted vacuum. 
We take this to be of the form:
\be 
\mu\, |\Psi_1^{\cl}\rangle = \mu\, c_1 \bar c_1 e^{-\phi}(0) V(0) |0\rangle, \qquad 
V(z,\bar z) = -2 \chi(z) \bar\p Y(\bar z)\, ,
\ee
where we have used the notation of \S\ref{s2} for various fields.
The form of \refb{e5.3x} and \refb{eabove} remain unchanged. The possible list of 
$L_0^+=0$ states in $\HH$ which have the correct total ghost and picture 
numbers and $(Y\to -Y, \chi\to -\chi)$ symmetry
for having non-zero inner product with
$c_0^-[\Psi^{\cl}_1\Psi^{\cl}_1]$ is
\be \label{elisthet}
c_1 \eta_{-1}|0\rangle, \quad 
\bar c_1 \bar c_{-1} c_1\, \xi_{-1}\, e^{-2\phi}(0)|0\rangle, \quad 
c_1 \bar c_1 e^{-\phi}(0) V(0)|0\rangle
\, .
\ee
However all of these actually have vanishing inner product with $c_0^-[\Psi^{\cl}_1\Psi^{\cl}_1]$ due to 
separate ghost
charge conservation in holomorphic and anti-holomorphic sectors 
and / or the $\chi\to -\chi$ symmetry. Therefore we can take $|\psi_2^{\cl}\rangle =0$ as
in bosonic string theory. Showing that the solution extends to order $\mu^3$ is also straightforward, but we
shall not describe it here since we only use the solution to order $\mu^2$.

Analysis of the fluctuations around the deformed vacuum also proceeds as in the case of bosonic string theory.
We replace the tachyon carrying spatial momentum $k$ and internal momentum $n/R$ by a massless field
with the same momenta. The ansatz for $\phi_\ell$ and $\tilde \phi_\ell$, replacing 
\refb{edefphin}, \refb{e7.30x}  and \refb{edefk2a} takes the
form:
\ben 
\phi_\ell=-2\, c\bar c e^{-\phi} \psi^1 \bar \p X^2 \, e^{i k_{(\ell)}.X}  e^{i n Y / R} , \quad
\tilde\phi_\ell=-2\, c\bar c e^{-\phi} \psi^1 \bar \p X^2 \, e^{-i k_{(\ell)}.X}  
e^{-i n Y / R}, 
\nonumber \\ \quad k_{(\ell)}^2=-n^2 R^{-2}+\OO(\mu), \qquad  \ell
=0,1,2\, , \hskip 2in
\een
satisfying the normalization condition:
\be\label{esignphi2het}
\langle \tilde \phi_\ell|c_0\bar c_0 | \phi_\ell\rangle = -1\, .
\ee
$k_{(\ell)}$'s have vanishing spatial components along 1 and 2 directions.
The rest of the analysis proceeds as in bosonic string theory. 
To first order in $\mu$ we get the analog of
\refb{ek1k1pre}:
\be  \label{ek1k1prehet}
\langle \tilde\phi_1| c_0^- Q_B |\phi_1\rangle 
= - \mu\, \{ \tilde\phi_1\Psi^{\cl}_1 \phi_1\}
= 2\,  \mu\, \langle \tilde \phi_1| \XX(p) \, c \bar c \, e^{-\phi} \, \chi\,  \bar \p Y (1) 
|\phi_1\rangle+\OO(\mu^2)\, ,
\ee
where $p$ is the PCO location on the NS-NS-NS interaction vertex chosen according to the rules
discussed in \S\ref{s4}.
Explicit evaluation gives a result independent of $p$:
\be\label{ek1k1het}
-{1\over 4} \left(k_{(1)}^2 + n^2 R^{-2}\right)  = - \mu \{ \tilde\phi_1\Psi^{\cl}_1 \phi_1\}
= - {n^2\over 4\, R^2} \, \mu  \quad \Rightarrow \quad k_{(1)}^2 = - n^2 \, R^{-2} \left(1-\mu\right)\, .
\ee

At order $\mu^2$ we have the analog of \refb{esecondorder}:
\ben \label{esecondorderhet}
\hskip -.3in -{1\over 4} \left(k_{(2)}^2  + n^2 R^{-2}
\right)  &=& -\mu \, \{\tilde \phi_2 \Psi_1^{\cl} \phi_2\}
+\mu^2 \left\{ \tilde \phi_2 \Psi^{\cl}_1 {b_0^+\over L_0^+} (1-\PP) \,  [\Psi^{\cl}_1 \phi_2] \right] \nonumber \\
&& +{\mu^2\over 2}\left\{\tilde \phi_2 \phi_2 \left({b_0^+\over L_0^+} (1-P) [\Psi^{\cl}_1 \Psi^{\cl}_1]\right)  \right\}
-{\mu^2\over 2} \left\{\tilde \phi_2  \phi_2 \Psi^{\cl}_1\Psi^{\cl}_1\right\}\, .
\een
The first term on the right hand side can be evaluated exactly as in bosonic string theory, leading to the
analog of \refb{efirstfin}:
\be \label{efirstfinhet}
-\mu\, \{\tilde \phi_2 \Psi_1^{\cl} \phi_2\} = -{\mu\over 4}  n^2 R^{-2} + {\mu^2\over 4} n^4 R^{-4}\,
\ln (\lambda\, |h'_3(1)|) +\OO(\mu^3)\, .
\ee
In this case the contribution from the second and the third terms on the right hand side of 
\refb{esecondorderhet} carry only negative powers of $\lambda$ since there are no states with $L_0^+<0$
and the contributions from the $L_0^+=0$ states are removed by the projection operators 
$(1-P)$ and $(1-\PP)$. Therefore we are left to evaluate the last term. 

As in the bosonic string theory,
while evaluating $\left\{\tilde \phi_2  \phi_2 \Psi^{\cl}_1\Psi^{\cl}_1\right\}$ we can take
the external states to be on-shell. The
integrand depends on the choice of PCO locations but under a change in PCO locations the integrand changes
by a total derivative. Therefore we can adjust them at will in the interior of $\RR$, as  long as on
$\p\RR$ they coincide
with the PCO locations in $\p\RR_s$, $\p\RR_t$ and $\p\RR_u$, fixed by the arrangements described in
\S\ref{s4}. Our strategy will be to take the PCO locations to coincide with the locations of the vertex operators
$\Psi_1^{\cl}$ in the interior of $\RR$, and at the boundary $\p\RR$ make them jump to the values they take
inside $\RR_s$, $\RR_u$ and $\RR_t$. The effect of this jump can be computed via vertical 
integration\cite{1408.0571,1504.00609}. If we had chosen a different PCO assignment inside $\RR$, the bulk
integrand will change by a total derivative. However the result of vertical integration will also change, precisely
cancelling this effect.

The effect of taking the PCO
locations to the locations of $\Psi^{\cl}_1$ is to convert the $\Psi^{\cl}_1$ inserted at 0 to an unintegrated
0 picture vertex operator given in \refb{eunint0}
\be 
\lim_{z\to 0} \XX(z) \Psi^{\cl}_1(0) = c_1 \bar c_1 \, \p Y \bar\p Y(0)|0\rangle
+{1\over 2} \eta \bar c e^{\phi} \chi \bar \p Y(0)|0\rangle\, ,
\ee
and the $\Psi^{\cl}_1$ inserted at $\sigma$ to an integrated zero picture vertex operator
\be
-\p Y \bar \p Y(\sigma)\, .
\ee
Using these we get
\be \label{efourthhet}
-{\mu^2\over 2} \left\{\tilde \phi_2  \phi_2 \Psi^{\cl}_1\Psi^{\cl}_1\right\} =
-{\mu^2\over 2} \int_\RR \, \II + \BBB_{v}\, ,
\ee
where
\ben \label{eiihet}
\II &=& 4\, {1\over 2\pi i} d^2 \sigma \, \Big\langle 
c\bar c e^{-\phi} \psi^1 \bar \p X^2 \, e^{-i k_{(0)}.X}  e^{-i n Y / R}(1)
\p Y \bar \p Y (\sigma) \nonumber \\ && 
\left(c\bar c \p Y \bar \p Y (0)
+{1\over 2} \eta \bar c e^{\phi} \chi \bar \p Y(0)\right)
 c\bar c e^{-\phi} \psi^1 \bar \p X^2 \, e^{i k_{(0)}.X}  e^{i n Y / R}(\infty)
\Big\rangle\, ,
\een
and $\BBB_v$ is the result of vertical integration that moves the PCOs from their
locations at 0 and $\sigma $ to the required values on $\p\RR$ so that they coincide with the positions of the
PCOs on $\p\RR_s$, $\p\RR_t$ and $\p\RR_u$ given in \S\ref{s4}. If we denote by $W_1$ and $W_2$ 
the final locations of the PCOs, and follow the convention that we first move the PCO at 0 to $W_1$ and
then move the PCO at $\sigma $ to $W_2$, then 
this has the effect of replacing inside the correlation function\cite{1408.0571,1504.00609} the factor
\be 
\XX(0)\, \XX(\sigma) \, d\sigma\wedge d\bar\sigma \, \left(-\ointop_\sigma b(w)dw\right)
 \left(-\ointop_\sigma \bar b(\bar w)d\bar w\right)\, ,
 \ee
by
\ben \label{epcomove}
&& \left(d\sigma \, \ointop_\sigma b(w)dw +
d\bar\sigma \ointop_\sigma \bar b(\bar w)d\bar w 
\right) \{(\xi(0) - \xi(W_1)) \XX(\sigma ) +  \XX(W_1)  (\xi(\sigma ) - \xi(W_2))\} \nonumber \\
&& - \{\xi(\sigma ) - \xi(W_2)\}  d\sigma \, {\p W_1\over \p\sigma}\, \p\xi(W_1)\, .
\een
Only the terms involving $\ointop_\sigma \bar b(\bar w)d\bar w$ survive after imposing ghost charge conservation in
the anti-holomorphic sector.
This gives
\ben \label{epco1}
\BBB_v &=& 2\, {\mu^2\over 2\pi i} \, \int_{\p\RR} d\bar \sigma  \, \Big\langle 
c\bar c e^{-\phi} \psi^1 \bar \p X^2 \, e^{-i k_{(0)}.X}  e^{-i n Y / R}(1) 
c\bar c e^{-\phi} \psi^1 \bar \p X^2 \, e^{i k_{(0)}.X}  e^{i n Y / R}(\infty) \nonumber \\ &&
\Big[-2 \{\xi(0) - \xi(W_1)\} \left\{ c \p Y + {1\over 2} \eta e^{\phi} \chi \right\} 
\bar \p Y (\sigma ) \,  c\bar c \, e^{-\phi}\chi \bar \p Y (0)
\nonumber \\ &&
+ 4 \,  \{\xi(\sigma ) -\xi(W_2)\} \, \XX(W_1) 
\, c \, e^{-\phi}\chi \bar \p Y (\sigma )  \, c\bar c \, e^{-\phi}\chi  \bar \p Y (0)
\Big]
\Big\rangle\, .
\een

First let us compute the contribution from the first term on the right hand side of \refb{efourthhet}.
The $\phi$-charge conservation (or equivalently $\xi$-$\eta$ charge conservation) tells us that the term
proportional to $\eta$ 
in the second line of \refb{eiihet}
does not contribute. Evaluation of the rest of the correlator gives
\be
\II = - {1\over 2\pi i} \left\{ {1\over 2\sigma^2} - {n^2\over 4 R^2} {1\over \sigma -1}\right\}
\left\{ {1\over 2\bar\sigma^2} - {n^2\over 4 R^2} {1\over \bar\sigma -1}\right\}\,.
\ee
This is identical to the integrand in \refb{e4integrand}. Therefore we get, using \refb{e6.43}
\be \label{e6.84}
-{\mu^2\over 2} \int_\RR \II  \simeq 
-{\mu^2\over 8} \, {\lambda^4\, |h'_3(1)|^4} + {\mu^2\over 8} \, {n^2\over R^2} 
-{\mu^2 \over 4}  \, n^4 R^{-4} \, \ln (\lambda\, |h'_3(1)|)\, .
\ee

Next we turn to the analysis of $\BBB_v$ given in \refb{epco1}. Using the relation 
\be
\p\RR = -\p\RR_s -\p\RR_t-\p\RR_u
\ee
we can express $\BBB_v$ as
\be 
\BBB_v = \BBB_s + \BBB_t + \BBB_u\, ,
\ee
where
\ben \label{epco3}
\BBB_{s,t,u} &=& -2\, {\mu^2\over 2\pi i} \, \int_{\p\RR_{s,t,u}} d\bar \sigma  \, \Big\langle 
c\bar c e^{-\phi} \psi^1 \bar \p X^2 \, e^{-i k_{(0)}.X}  e^{-i n Y / R}(1) 
c\bar c e^{-\phi} \psi^1 \bar \p X^2 \, e^{i k_{(0)}.X}  e^{i n Y / R}(\infty) \nonumber \\ &&
\Big[-2 \{\xi(0) - \xi(W_1)\} \left\{ c \p Y(\sigma) + {1\over 2} \eta e^{\phi} \chi (\sigma)\right\} 
\bar \p Y (\sigma ) \,  c\bar c \, e^{-\phi}\chi \bar \p Y (0)
\nonumber \\ &&
+ 4 \,  \{\xi(\sigma ) -\xi(W_2)\} \, \XX(W_1) 
\, c \, e^{-\phi}\chi \bar \p Y (\sigma )  \, c\bar c \, e^{-\phi}\chi  \bar \p Y (0)
\Big]
\Big\rangle\, .
\nonumber \\
\een
We shall furthermore express $\BBB_u$ as
\be
\BBB_u = \BBB_u' +\BBB_u''\, ,
\ee
where
\ben
\BBB_u' &=&-2\, {\mu^2\over 2\pi i} \, \int_{\p\RR_{u}} d\bar \sigma  \, \Big\langle 
c\bar c e^{-\phi} \psi^1 \bar \p X^2 \, e^{-i k_{(0)}.X}  e^{-i n Y / R}(1) 
c\bar c e^{-\phi} \psi^1 \bar \p X^2 \, e^{i k_{(0)}.X}  e^{i n Y / R}(\infty) \nonumber \\ &&
\Big[-2 \{\xi(0) - \xi(1)\} \left\{ c \p Y + {1\over 2} \eta e^{\phi} \chi \right\} 
\bar \p Y (\sigma ) \,  c\bar c \, e^{-\phi}\chi \bar \p Y (0)
\Big]
\Big\rangle\, ,
\een
and
\ben \label{edefbbbupp}
\BBB_u''&=&-2\, {\mu^2\over 2\pi i} \, \int_{\p\RR_{u}} d\bar \sigma  \, \Big\langle 
c\bar c e^{-\phi} \psi^1 \bar \p X^2 \, e^{-i k_{(0)}.X}  e^{-i n Y / R}(1) 
c\bar c e^{-\phi} \psi^1 \bar \p X^2 \, e^{i k_{(0)}.X}  e^{i n Y / R}(\infty) \nonumber \\ &&
\Big[-2 \{\xi(1) - \xi(W_1)\} \left\{ c \p Y + {1\over 2} \eta e^{\phi} \chi \right\} 
\bar \p Y (\sigma ) \,  c\bar c \, e^{-\phi}\chi \bar \p Y (0)
\nonumber \\ &&
+ 4 \,  \{\xi(\sigma ) -\xi(W_2)\} \, \XX(W_1) 
\, c \, e^{-\phi}\chi \bar \p Y (\sigma )  \, c\bar c \, e^{-\phi}\chi  \bar \p Y (0)
\Big]
\Big\rangle\,.
\een
We have shown in appendix \ref{sa} 
that in the large $\lambda$ limit, $\BBB_s$, $\BBB_t$ and $\BBB_u''$ vanish so that
\be \label{ebvbu}
\BBB_v=\BBB_u'\, .
\ee
Intuitively the vanishing of $\BBB_s$, $\BBB_t$ and $\BBB_u''$ may be understood as follows.
Let us start with $\BBB_s$. For large $\lambda$ the integration region is near a degeneration where
the points 0 and $\infty$ are on one sphere and the points $\sigma$ and $1$ are on another
sphere, with the two spheres connected by a narrow neck. In this case the dominant contribution
comes from nearly on-shell states propagating along the neck. In the initial configuration the PCO
at 0 lies on the first sphere and the PCO at $\sigma$ lies on the second sphere. In the final configuration
one of the PCOs ($W_1$) lies on
the first sphere and the other PCO ($W_2$) lies on the second sphere. $\BBB_s$ describes the
effect of moving the first PCO from 0 to $W_1$ and moving the second PCO from $\sigma$ to $W_2$.
Each of the PCOs remains on its own sphere. Since for on-shell three point function, moving the PCO
on the sphere does not have any effect, we expect $\BBB_s$ to vanish in the large $\lambda$ limit.
Similar argument can be given for $\BBB_t$. This does not apply to $\BBB_u$ since the initial position
of the two PCOs, at 0 and $\sigma$, lie on the same sphere while the final arrangements $W_1$ and
$W_2$ must lie on different spheres. We analyze this as a combination of two moves: first move one
of the PCO's from 0 to 1, and call this contribution $\BBB_u'$, and then move the pair of PCO's 
at $1$ and $\sigma$, which are now on different spheres, to $W_1$ and $W_2$. The latter contribution,
called $\BBB_u''$ vanishes in the large $\lambda$ limit due to the same arguments as for $\BBB_s$ and
$\BBB_t$.

We shall now analyze $\BBB_u'$.  Due to $\xi$-$\eta$ charge conservation, only the term proportional
to $\eta$ inside the curly bracket contributes. This gives:
\ben
\BBB_u'&=& 2\, {\mu^2\over 2\pi i} \, \int_{\p\RR_{u}} d\bar \sigma  \, \Big\langle 
c\bar c e^{-\phi} \psi^1 \bar \p X^2 \, e^{-i k_{(0)}.X}  e^{-i n Y / R}(1) 
c\bar c e^{-\phi} \psi^1 \bar \p X^2 \, e^{i k_{(0)}.X}  e^{i n Y / R}(\infty) \nonumber \\ &&
\{\xi(0) - \xi(1)\} \eta e^{\phi} \chi 
\bar \p Y (\sigma ) \,  c\bar c \, e^{-\phi}\chi \bar \p Y (0)
\Big\rangle \nonumber \\
&=& -{\mu^2\over 16\pi i} \int_{\p \RR_u} d\bar\sigma \, \sigma^{-1} \, \left( {1\over \bar\sigma^{2}} - {n^2\over 2 R^2} 
{1\over \bar\sigma-1}\right)\, .
\een
For large $\lambda$, $\p\RR_u$ is a  small contour around the origin. In this case the contribution from the term
proportional to $(\bar\sigma-1)^{-1}$ is suppressed by inverse power of $\lambda$ and can be neglected. The remaining
term has the same structure as the right hand side of
\refb{ei1a} with  the roles of $\sigma$ and $\bar\sigma$ interchanged.
This generates an extra minus sign, leading to \refb{ei1fin} multiplied by $-1$:
\be \label{ebsp}
\BBB_u' \simeq  {\mu^2 \over 8} {\lambda^4\, |h'_3(1)|^4} \, .
\ee
Substituting \refb{e6.84}, \refb{ebvbu} and \refb{ebsp} into \refb{efourthhet} we get
\be \label{efourthhetnew}
-{\mu^2\over 2} \left\{\tilde \phi_2  \phi_2 \Psi^{\cl}_1\Psi^{\cl}_1\right\} =
{\mu^2\over 8} \, {n^2\over R^2} 
-{\mu^2 \over 4}  \, {n^4 \over R^{4}} \, \ln (\lambda\, |h'_3(1)|)\, .
\ee

Using \refb{efirstfinhet}, 
\refb{efourthhetnew}, and the fact that the contribution from the second and third
terms on the right hand side of
\refb{esecondorderhet} vanishes, we get
\be 
-{1\over 4} \left(k_{(2)}^2 + n^2 R^{-2}\right)= -\left({\mu\over 4} - {\mu^2\over 8}\right) n^2 R^{-2} \, .
\ee
This gives
\be \label{ehetfin}
k_{(2)}^2 = - n^2 R^{-2} \left(1 -\mu + {\mu^2\over 2}\right)\, .
\ee

Although the $\mu$ dependent corrections have 
the same form as \refb{ebosfin}, there is a subtle difference in the analysis.
In the case of bosonic string theory the potential short distance divergence, reflected in the 
term $I_1$  in \refb{ei1fin} that grows as $\lambda^4$ in the large $\lambda$ limit, can be identified to the
contribution from the intermediate tachyon state.
The naive divergence in the world-sheet integral
over $\sigma$ can be traced to the wrong treatment of the tachyon propagator as in \refb{e2}. In contrast
the potential short distance divergence in the heterotic string theory, reflected in the term proportional
to $\lambda^4$ in \refb{e6.84}, can be traced to the wrong assignment of PCO locations, and is 
cancelled by the boundary term $\BBB_u'$ that corrects the PCO location via vertical integration.

\sectiono{Comments of higher genus amplitudes} \label{sdiss}

As we have emphasized earlier, there is a subtle difference between the analysis in \S\ref{s5},\ref{shigher},\ref{ssuper} 
and that in \S\ref{s6}.  The analysis in sections \ref{s5}, \ref{shigher} and \ref{ssuper} was 
carried out in a manner that is manifestly independent of the string field theory data --
choice of local coordinate system and PCO
locations. On the other hand the analysis in \S\ref{s6} required, in the intermediate steps, use of the local
coordinate system, {\it e.g.} the function $h_3(z)$ in \refb{efirstfin} and $g_4(z)$ in \refb{ei1b}, 
although at the end the dependence cancelled. There is an intrinsic reason for this difference. The on-shell
amplitudes discussed in \S\ref{s5},\ref{shigher},\ref{ssuper} are expected to be genuinely independent of the
choice of string field theory data, and so it is not surprising that the analysis can be made independent
of these choices. The result of \S\ref{s6}, describing the effect of marginal deformation, is not expected to be 
manifestly independent of the string field theory data although the dependence on these data is
expected to be removable by a redefinition of the deformation parameter $\mu$. This is due to the fact that a change
in the string field theory 
data causes a redefinition of the string fields including the one corresponding to the marginal operator, 
and such field redefinitions will induce a redefinition of the deformation parameter. Therefore the 
total independence of the final
results \refb{ebosfin}, \refb{ehetfin} on the string field theory data 
is accidental, and we expect that in general there will be such dependence of the result on these data.

It is natural to ask if the analysis can be generalized to give a systematic procedure for computing higher genus
amplitudes that gives manifestly finite results, and yet minimizes the dependence on the choice of string field
theory data. To this end note that for higher genus amplitudes, the dependence on the string field theory data
of the kind presented in \S\ref{s6} will always be present even in the absence of marginal deformations. This
is due to the fact that at higher genus, 
under quantum corrections there will be mass and wave-function renormalizations of
all the external states. Now since different choices of string field theory data  lead to string field theories that are
related by field redefinition, there will be two 
effects\cite{AMS,1209.5461,1411.7478}.
First the amplitudes computed in different string field theories will differ due to different wave-function renormalizations
of external states. Second the definition of the moduli fields, {\it e.g.} the string coupling constant encoded in the
dilaton, will differ in different string field theories, causing a change in the amplitude. These will lead to ambiguities
in the final result that depend on the string field theory data. However such ambiguities can be absorbed into
a finite renormalization of external states and the values of the moduli. Therefore, as in this paper,
one could proceed with the computation
assuming the existence of a consistent set of string field theory data  without making a particular choice and then
verify at the end that the final result depends on these data only through the normalization of the external states
and definition of the moduli fields.

We shall end this section by describing the different kinds of degenerations that we need to deal with for higher
genus amplitudes:
\begin{enumerate}
\item The first type of degeneration  is separating type degeneration with generic momentum
flowing across the degenerating punctures. 
These correspond to the original Riemann surface degenerating into a pair
of Riemann surfaces, each of which carries two or more external punctures. These
can be treated in the same way as in \S\ref{shigher},\ref{ssuper} and do not introduce any ambiguity in the final result.
\item The second type of degenerations, analyzed extensively in \cite{1209.5461}, 
involve separating type degenerations with special momentum flowing across
the degenerating punctures. Examples of these involve degenerations where the original Riemann surface degenerates
into a pair of Riemann surfaces, one of which carries either no puncture or one puncture. The first one represents tadpole 
type diagrams with zero momentum flowing across the degenerating puncture 
while the second one represents mass and wave-function
renormalization diagrams with on-shell momentum flowing across
the degenerating puncture. Near these degenerations 
the expansion of the integrand in powers of the variables $\xtau,\bar\xtau$ 
that vanish at the degeneration involves integer exponents
and we cannot apply the general trick of \S\ref{shigher},\ref{ssuper} to remove these divergences. There may be
genuine divergences of the form $\int d^2\xtau/|\xtau|^2$ signaling the presence of massless tadpoles and
renormalization of physical masses. There may also be
ambiguities in determining the $\II^{(k)}$'s of the form $d\xtau/\xtau$ encoding redefinitions of massless
moduli fields and/or external states. We need to use
the analog of the procedure described in \S\ref{s6} to address these cases. The general procedure based on
string field theory can be found in \cite{1411.7478}. In simple cases,  one can follow the procedure described in  
\S\ref{s6} to
minimize the
dependence on the explicit knowledge of string field theory, reproducing the 
results in \cite{1311.1257,1401.7014,1404.6254}.
\item The final category of degenerations involves non-separating type degenerations -- degenerations where one
pinches the handle of a Riemann surface but the Riemann surface still remains connected after degeneration. This
case lies in between the two cases described earlier, in that the momentum flowing across the degenerating puncture is
a loop momentum that needs to be integrated. For most of the range of integration over momenta 
the momentum is generic, but on
codimension one subspaces of the loop momentum space
the $L_0+\bar L_0$ eigenvalue corresponding to that momentum may vanish, 
causing the integrand to diverge. In four or less dimensions the momentum integrals themselves are divergent
reflecting the presence of infrared divergences in the theory and the procedure for getting finite result is complicated
-- requiring the same methods that are normally used in quantum field theories with massless fields. However in
higher than four dimensions there is no genuine divergence and one should be able to extract finite results. 
Nevertheless the answer is not free from ambiguity without additional input, since the $i\eps$ prescription in the
integration over the loop momenta is hidden in the prescription of how we deal with the pole in the
$1/(L_0+\bar L_0)$ factor.  This can be done following the procedure described in \cite{1307.5124}, where 
we replace the $(L_0+\bar L_0)^{-1}$ factor by the representation
\refb{e2alt}. This means that if the original integral has the form
\be 
\int dq\wedge d\bar q \, |q|^{-2} \, f(q,\bar q)=-2 i  \, \int 
ds \wedge d\theta\,  f(q,\bar q)\, , \quad q\equiv e^{-(s+i\theta)},
\ee
where we have suppressed the integration over the other moduli, we replace it by\footnote{Even though 
the Cutkosky rules are not manifest in the procedure described in
\cite{1307.5124}, this has been shown to be equivalent to the procedure of \cite{1604.01783} 
and therefore satisfies the
Cutkosky rules\cite{1610.00443}.}

\be \label{eXXin}
-2 i \int_0^{2\pi} d\theta \int_{s=0}^\Lambda ds \,  f(q,\bar q) - 2 i \int_0^{2\pi} d\theta \int_{s=\Lambda}^{\Lambda+i\infty}
ds \, e^{i \eps s}\, f(q,\bar q) \, .
\ee
The relation between the variables $q,\bar q$ and some predetermined moduli parameters on the punctured Riemann
surface can be determined if we know the string field theory data. However this is not necessary. Let us suppose that
$\xtau,\bar\xtau$ correspond to some other set of variables with the property that $\xtau$ vanishes linearly
with $q$ near the degeneration. Parametrizing $\xtau$ as $e^{-t-i\phi}$, we can conclude that if the original
integrand has the form
\be
d\xtau\wedge d\bar \xtau \, |\xtau|^{-2} \, F(\xtau,\bar \xtau)=-2 i  \, dt \, \wedge d\phi\,  
F(\xtau,\bar \xtau)\, , 
\ee
then we can restrict the range of $\xtau$ integration to $|\xtau|\ge e^{-\Lambda}$ and add to it a term
\be \label{eXXfin}
- 2 i \int_{t=\Lambda}^{\Lambda+i\infty}
dt \,  \int_0^{2\pi} d\phi \, e^{i \eps t}\, F(\xtau,\bar \xtau) \, .
\ee
The equivalence between these two procedures -- one in $q,\bar q$ variable and the other in the
$\xtau,\bar\xtau$ variable --  follows from the fact that near the degeneration $q$ and
$\xtau$ are related to each other 
in a one to one fashion. Therefore if we map the integration
contour used in \refb{eXXfin} to 
the $(s,\theta)$ plane, the resulting contour can be smoothly deformed to the contour in
\refb{eXXin} without passing through any singularity. In dimensions larger than four, and 
for generic momenta of external states, 
the integrand $F(\xtau,\bar\xtau)$ has power law suppression in inverse powers of 
$-\ln|\xtau|$ that makes 
the integral convergent from the large $\Imm(t)$ region even without the $e^{i\eps t}$ factor. 
Therefore we can compute the integral directly, without having to take limits of integrals.
This remains true even for multiple integrals of this kind
where more than one handle degenerates simultaneously.
An explicit example of this for one loop amplitude can be found in \cite{1607.06500} (see {\it e.g.} eq.(3.16)).

\end{enumerate}
 
\bigskip

{\bf Acknowledgement:}
I wish to thank Roji Pius and Barton Zwiebach for useful discussions and
Barton Zwiebach for his very useful comments on an earlier version of this manuscript.
This work was
supported in part by the J. C. Bose fellowship of 
the Department of Science and Technology, India and also by the Infosys Chair Professorship.

\appendix

\sectiono{Choice of local coordinates on a five punctured sphere} \label{sb}

\begin{figure}
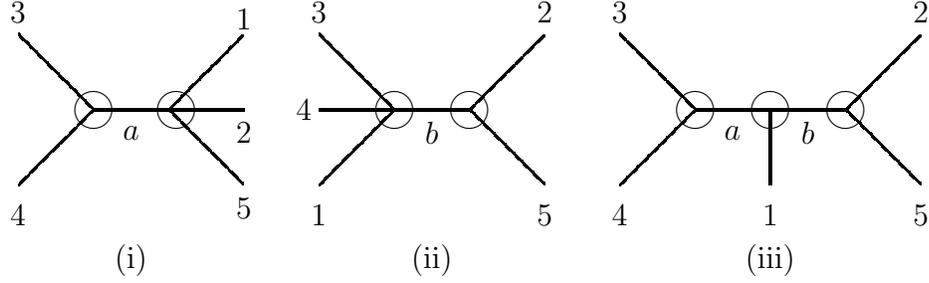


\begin{center}

\figgraphrep

\end{center}

\vskip -1in

\caption{Examples of boundaries $\CC^{(k)}_{s_1\cdots s_k}$. The left figure describes
$\CC^{(1)}_a$, the middle figure described $\CC^{(1)}_b$ and the right figure describes 
$\CC^{(2)}_{ab}$. \label{fig1rep}}

\end{figure}

In this appendix we shall describe, for the five punctured sphere, the relation between the global coordinates
$(\sigma_1,\sigma_2)$ and the coordinates $\xtau_{s_1},\cdots,\xtau_{s_k}, m_{(s_1,\cdots, s_k)}$ for some
choices of $\{s_1,\cdots s_k\}$. 

We begin by making a specific choice of global coordinates. This will be done by
fixing the puncture locations $y_1,\cdots y_5$ in the complex plane to be at
\be
y_1=\sigma_1, \quad y_2=\sigma_2, \quad y_3=1, \quad y_4=2, \quad y_5=0\, .
\ee
The boundaries we shall consider are shown in Fig.~\ref{fig1rep}. These are similar to the Feynman diagrams
shown in Fig.~\ref{fig1}, but Fig.~\ref{fig1rep} should be regarded as depicting regions {\it near} those
represented by the Feynman diagrams in Fig.~\ref{fig1} for $|q|\sim e^{-\Lambda}$
-- not necessarily the precise regions that follow from a
string field theory. For example Fig.~\ref{fig1rep}(i) represents region near the boundary $\CC^{(1)}_a$ 
where the original
sphere is near a degeneration into a four punctured sphere carrying the original punctures 1, 2 and 5 and a three
punctured sphere carrying the original punctures 3 and 4. Our goal will be to introduce the coordinates
$\xtau_a, m_{(a)}$ near this boundary in terms of the global coordinates $\sigma_1$, $\sigma_2$. This is
done as follows.
Let us take the three punctured sphere on the left, carrying global coordinate $z$, 
to have puncture 3 at $z=1$, 
puncture 4 at $z=2$ and the
sewing puncture at $z=0$. We also take the four punctured 
sphere on the right, carrying global coordinate $z'$,  to have puncture 1 at 
$z'=m_{(a)}$,
puncture 2 at $z'=2$, puncture 5 at $z'=1$ and the sewing puncture at $z'=0$. $m_{(a)}$ should keep a finite
distance away from 0, 1 and 2 so that the four punctured sphere is not close to degeneration -- as will be discussed
later,
for $m_{(a)}$ close to 0, 1 or 2, we need to
choose the coordinate systems differently. 
We now sew the two spheres
via the relation
\be
z z'=\hat\xtau_a\, .
\ee
We have used $\hat\xtau_a$ instead of $\xtau_a$ to take into account the fact that the correct
candidate for the coordinate $\xtau_a$ may have different form in different domains in the moduli space.
We shall see that while the $\hat\xtau_a$ appearing in the above equation is the correct choice of
$\xtau_a$ as long as $m_{(a)}$ is kept away from 0, 1 and 2, we need modifications when $m_{(a)}$
approaches any of these points.
In the $z$ coordinate the punctures are located at:
\be
3:\, z=1, \quad 4:\, z=2, \quad 5: z=\hat\xtau_a, \quad 1:\, z=\hat\xtau_a/m_{(a)}, \quad 2:\, z=\hat\xtau_a/2\, .
\ee
In order to bring the 5th puncture at 0 leaving the third and fourth punctures at 1 and 2 respectively, we make
a change of coordinates:
\be
y ={2 (z-\hat\xtau_a)\over \hat\xtau_a(z-3)+2}\, .
\ee
The location of the punctures 1 and 2 in the $y$ plane are now given by, respectively,
\be \label{egl1}
\sigma_1= {2 (1-m_{(a)})\, \hat\xtau_a \over \hat\xtau_a(\hat\xtau_a-3 m_{(a)}) +2\, m_{(a)} }, \quad
\sigma_2 = -{2\, \hat\xtau_a\over \hat\xtau_a(\hat\xtau_a-6) +4 }\, .
\ee
Since we shall use this formula only
for small $\hat\xtau_a$, we shall replace \refb{egl1} by a simpler equation:
\be \label{egl1new}
\sigma_1= {(1-m_{(a)})\, \hat\xtau_a \over m_{(a)} }, \qquad
\sigma_2 = -{\hat\xtau_a\over 2 }\, .
\ee
We shall take \refb{egl1new} as the definitions of the coordinates $\{\hat\xtau_a, m_{(a)}\}$ 
in terms of
the global coordinates $\sigma_1,\sigma_2$ of the moduli space near $\CC^{(1)}_a$. These can be taken
to represent the coordinates $(\xtau_a, m_{(a)})$ in the notation of \S\ref{shigher}
when $m_{(a)}$ is not close to
0, 1 or 2.

Let us now turn to the region near the boundary $\CC^{(1)}_b$ represented by Fig.~\ref{fig1rep}(ii). 
We take the four punctured sphere on the left,
carrying global coordinate $z$, to have puncture 3 at $z=1$, puncture 4 at $z=2$, puncture 1
at $z=m_{(b)}$ and the
sewing puncture at $z=0$. 
$m_{(b)}$ needs to keep finite distance away from 0, 1 and 2 
so that this sphere is not close to degeneration.
We also take the three punctured sphere on the right, carrying global coordinate $z'$,  to have 
puncture 2 at $z'=2$, puncture 5 at $z'=1$ and the sewing puncture at $z'=0$. 
We now sew the two Riemann
surfaces via the relation
\be
z z'=\hat\xtau_b\, .
\ee
In the $z$ coordinate the punctures are located at:
\be
3:\, z=1, \quad 4:\, z=2, \quad 5: z=\hat\xtau_b, \quad 1:\, z=m_{(b)}, \quad 2:\, z=\hat\xtau_b/2\, .
\ee
In order to bring the 5th puncture to 0 leaving the third and fourth punctures at 1 and 2 respectively, we make
a change of coordinates:
\be
y ={2 (z-\hat\xtau_b)\over \hat\xtau_b(z-3)+2}\, .
\ee
The location of the punctures 1 and 2 in the $y$ plane are now given by, respectively,
\be \label{egl2}
\sigma_1= {2 (m_{(b)}-\hat\xtau_b)\over \hat\xtau_b(m_{(b)}-3)+2}, \quad
\sigma_2 = -{2\, \hat\xtau_b\over \hat\xtau_b(\hat\xtau_b-6) +4 }\, .
\ee
Again, since we shall use this coordinate system for small $\hat\xtau_b$, we replace this by a simpler set
of equations:
\be \label{egl2new}
\sigma_1= m_{(b)}, \quad
\sigma_2 = -{\hat\xtau_b\over 2 }\, .
\ee
\refb{egl2new} defines the coordinates $\hat\xtau_b$, $m_{(b)}$ appropriate near $\CC_{(b)}$
in terms of the global coordinates $\sigma_1,\sigma_2$ of the moduli space.
These can be taken
to represent the coordinates $(\xtau_b, m_{(b)})$ in the notation of \S\ref{shigher} 
when $m_{(b)}$ is not
close to 0, 1 or 2. 

Next we turn to the region near $\CC^{(2)}_{ab}$ represented by the diagram \ref{fig1rep}(iii).
Keeping in mind that the variables $\hat\xtau_a$ and $\hat\xtau_b$ introduced earlier may not exactly match with
the variables suitable for parametrizing the region near $\CC^{(2)}_{ab}$, we shall denote the new
parameters by $\tilde\xtau_a$ and $\tilde\xtau_b$. 
We take the left sphere 
carrying global coordinate $z$ to have puncture 3 at $z=1$, puncture 4 at $z=2$ and the
sewing puncture at $z=0$,  the middle sphere carrying global coordinate $z'$  to have puncture 1 at 
$z'=1$, the left sewing puncture at $z'=0$ and the right sewing puncture at $z'=\infty$ and the right sphere
carrying global coordinate $z''$ to have the sewing puncture at $z''=0$,  puncture 2 at $z''=2$ and 
puncture 5 at $z''=1$. 
We now sew the three spheres
 via the relation
\be
z z'=\tilde\xtau_a, \quad z''/z' = \tilde\xtau_b\, .
\ee
In the $z$ coordinate the punctures are located at:
\be
3:\, z=1, \quad 4:\, z=2, \quad 5: z= \tilde\xtau_a \, \tilde\xtau_b, 
\quad 1:\, z=\tilde\xtau_a, \quad 2:\, z=\tilde\xtau_a\, \tilde\xtau_b/2 \, .
\ee
We now introduce new coordinate 
\be
y ={2 (z-\tilde\xtau_a \tilde\xtau_b)\over \tilde\xtau_a \tilde\xtau_b (z-3)+2}\, ,
\ee
so that we have
\be
y_3=1, \quad y_4=2, \quad y_5=0\, .
\ee
The locations of the punctures 1 and 2 in the $y$ plane are now given by, respectively,
\be \label{egl3}
\sigma_1= {2 (1-\tilde\xtau_b)\, \tilde\xtau_a \over \tilde\xtau_a \tilde\xtau_b (\tilde\xtau_a-3) +2}, \quad
\sigma_2 = -{2\, \tilde\xtau_a \tilde\xtau_b \over \tilde\xtau_a \tilde\xtau_b(\tilde\xtau_a\tilde\xtau_b-6) +4 }\, .
\ee
Again since we shall be using this change of coordinates for small $\tilde\xtau_a,\tilde\xtau_b$, we shall replace
this by
\be  \label{egl3new}
\sigma_1= \tilde\xtau_a, \quad
\sigma_2 = -{\tilde\xtau_a \tilde\xtau_b \over 2 }\, .
\ee
\refb{egl3new} gives the definition of the coordinates $\tilde\xtau_a$, $\tilde\xtau_b$ appropriate 
near $\CC^{(2)}_{(ab)}$ in terms of the global coordinates $\sigma_1,\sigma_2$ of the moduli space.
Comparing \refb{egl1new}, \refb{egl2new} and \refb{egl3new} we can find the relations between the
coordinate systems near $\CC^{(1)}_a$, $\CC^{(2)}_{ab}$ and $\CC^{(1)}_b$:
\be\label{ectrs}
\hat\xtau_a = \tilde\xtau_a \tilde\xtau_b, \quad m_{(a)} = {\tilde\xtau_b\over 1+\tilde\xtau_b}, 
\quad \hat\xtau_b = \tilde\xtau_a \tilde\xtau_b,
\quad m_{(b)} = \tilde\xtau_a\, .
\ee

We now recall that for Fig.~\ref{fig1rep}(i), identification of 
the coordinate system $(\hat\xtau_a, m_{(a)})$ with the coordinates $(\xtau_a, m_{(a)})$ introduced
in \S\ref{shigher}
breaks down for $m_{(a)}$
close to 0 since the right sphere degenerates in this limit. 
By examining the choice of coordinates of the punctures
on the original sphere one can see that this degeneration is precisely the one 
depicted in Fig.\refb{fig1rep}(iii). 
Therefore in this region we can identify $(\tilde\xtau_a, m_{(a)})$, instead of
$(\hat\xtau_a, m_{(a)})$,  
with the coordinates $(\xtau_a, m_{(a)})$ introduced in \S\ref{shigher}.
Similar modifications must also be made when $m_{(a)}$ approaches 1 and 2 
by analyzing good coordinate systems
near other degenerations. The choice of the coordinate system $(\xtau_b, m_{(b)})$ needs to be similarly
modified when $m_{(b)}$ approaches 0, 1 and 2. For example  when $m_{(b)}$ approaches 0 we can use 
$(\tilde\xtau_b, m_{(b)})$ to label coordinates near $\CC^{(1)}_{(b)}$.

Using this coordinate system we can also define the boundaries $\CC^{(1)}_a$ and $\CC^{(1)}_b$. For
example when $m_{(a)}$ is finite distance away from 0, 1 and 2, we can use $|\hat\xtau_a|=\eps$ for
defining $\CC^{(1)}_a$, but when $m_{(a)}$ is close to 0, we use $|\tilde\xtau_a|=\eps$ as the definition
of $\CC^{(1)}_a$. 
Similarly we can define $\CC^{(1)}_b$ to be given by $|\hat\xtau_b|=\eps$ when $m_{(b)}$ is away from
0, 1 and 2 but $|\tilde\xtau_b|=\eps$ when $m_{(b)}$ is close to zero.
These can be formally stated as follows. 
Let $H(x)$ be a smooth function of a complex variable $x$ that approaches 1
for large $|x|$ and zero for small $|x|$, {\it e.g.}
\be
H(x)\equiv {|x|^2\over |x|^2 + \eta^2}\, ,
\ee
where $\eta$ is a  fixed number. Then we define:
\be
\CC^{(1)}_a: |\hat\xtau_a|=\eps H(m_{(a)}) + \eps |m_{(a)}| (1 - H(m_{(a)})),
\qquad \CC^{(1)}_b: |\hat\xtau_b|=\eps H(m_{(b)}) + \eps |m_{(b)}| (1 - H(m_{(b)}))\, .
\ee
When $m_{(a)}$ ($m_{(b)}$) is close to 1 or 2, the definitions of $\CC^{(1)}_a$ ($\CC^{(2)}_b$)
need to be further modified along the same line. $\CC^{(1)}_{ab}$ is simply the intersection of
these two subspaces, given approximately by $|\tilde\xtau_a|\simeq |\tilde\xtau_b|\simeq \eps$
for small $\eps$. Note however that we do not need to take the $\eps\to 0$ limit since \refb{etotalI} gives
the correct result even when the $\CC^{(1)}_s$'s have finite size.

Given the original integrand $\II^{(0)}$ for the five point function, computed from correlation
functions of vertex operators in the conformal field theory, we 
can now construct the differential forms $\II^{(1)}_a$, $\II^{(1)}_b$ and $\II^{(2)}_{ab}$ as follows.
We first consider expansions of $\II^{(0)}$ near $\CC^{(1)}_a$ and $\CC^{(1)}_b$: 
\ben
\II^{(0)} &=& d\hat\xtau_a\wedge d\bar {\hat\xtau}_a\wedge 
dm_{(a)} \wedge d\bar m_{(a)} \sum_i C_i(m_{(a)}) \hat\xtau_a^{-1+\alpha_i} 
\bar{\hat\xtau}_a^{-1+\beta_i} \quad \hbox{near \ $\CC^{(1)}_a$}
\nonumber \\
&=&  d\hat\xtau_b\wedge d\bar {\hat\xtau}_b\wedge 
dm_{(b)} \wedge d\bar m_{(b)} \sum_i \wt C_i(m_{(b)}) \hat\xtau_b^{-1+\tilde\alpha_i} 
\bar{\hat\xtau}_b^{-1+\tilde\beta_i} 
\quad \hbox{near \ $\CC^{(1)}_b$}\, .
\een
We now obtain $\II^{(1)}_a$ by solving the $d\, \II^{(1)}_a=\II^{(0)}$ near $\CC^{(1)}_a$. A solution is
\ben\label{ekx1}
\II^{(1)}_a &=&  -d\hat\xtau_a \wedge 
dm_{(a)}\wedge d\bar m_{(a)}  \sum_i (\beta_i)^{-1} \,
C_i(m_{(a)}) \hat\xtau_a^{-1+\alpha_i} \bar{\hat\xtau}_a^{\beta_i}  \nonumber \\
&=& - d\tilde\xtau_a \wedge d\tilde \xtau_b \wedge d\bar{\tilde\xtau}_b  \,
|1+\tilde \xtau_b|^{-4} 
\tilde\xtau_b
\sum_i (\beta_i)^{-1} \,
C_i\left({\tilde\xtau_b\over 1+\tilde\xtau_b}\right) (\tilde\xtau_a\tilde\xtau_b)^{-1+\alpha_i} 
\left(\bar{\tilde \xtau}_a \bar{\tilde\xtau}_b\right)^{\beta_i} \, ,\nonumber \\
\een
where in the second line we have displayed its behavior in the coordinate system
appropriate near $\CC^{(2)}_{ab}$ using the
coordinate  transformations \refb{ectrs}. Similarly we have
\ben \label{ekx2}
\II^{(1)}_b &=& -d\hat\xtau_b\wedge 
dm_{(b)} \wedge d\bar m_{(b)} \sum_i  (\tilde\beta_i)^{-1} \wt C_i(m_{(b)}) 
\hat\xtau_b^{-1+\tilde\alpha_i} \bar{\hat\xtau}_b^{\tilde\beta_i} \nonumber \\
&=& -d\tilde \xtau_b\wedge d\tilde \xtau_a \wedge d\bar{\tilde\xtau}_a \, \tilde\xtau_a
\sum_i  (\tilde\beta_i)^{-1}  \wt C_i(\tilde\xtau_a)
(\tilde\xtau_a\tilde\xtau_b)^{-1+\tilde\alpha_i} \left(\bar{\tilde \xtau}_a 
\bar{\tilde\xtau}_b\right)^{\tilde\beta_i} \, .
\een
We can now expand $C_i$ and $\wt C_i$ in the second lines of \refb{ekx1} and \refb{ekx2} in power series
expansion in $\tilde\xtau_b$ and $\tilde\xtau_a$ respectively to find the expressions for $\II^{(1)}_a$ and
$\II^{(1)}_b$ near $\CC^{(2)}_{ab}$. $\II^{(2)}_{ab}$ is then obtained by solving
the equation:
\be
d\II^{(2)}_{ab} = \II^{(1)}_a - \II^{(1)}_b\, .
\ee

\sectiono{Analysis of $\BBB_u''$, $\BBB_s$ and $\BBB_t$} \label{sa}

In this appendix we shall show that $\BBB_s$, $\BBB_t$ and $\BBB_u''$, defined in \refb{epco3}, 
\refb{edefbbbupp}, vanish
in the large $\lambda$ limit.  We begin with $\BBB_s$. Keeping only the part of the correlator in
\refb{epco3} that does not vanish by ghost charge conservation, we have
\ben \label{eapp3}
\BBB_s &=& -2\, {\mu^2\over 2\pi i} \, \int_{\p\RR_{s}} d\bar \sigma  \, \Big\langle 
c\bar c e^{-\phi} \psi^1 \bar \p X^2 \, e^{-i k_{(0)}.X}  e^{-i n Y / R}(1) 
c\bar c e^{-\phi} \psi^1 \bar \p X^2 \, e^{i k_{(0)}.X}  e^{i n Y / R}(\infty) \nonumber \\ && \hskip -.6in
\Big[-\{\xi(0) - \xi(W_1)\} \eta e^{\phi} \chi (\sigma)
\bar \p Y (\sigma ) \,  c\bar c \, e^{-\phi}\chi \bar \p Y (0)
\nonumber \\ &&  \hskip -.6in
-  \{\xi(\sigma ) -\xi(W_2)\} \, \Big\{ \p \eta \, e^{2\phi} \, b(W_1)
+ \p\left(\eta \, e^{2\phi} \, b(W_1)\right)\Big\}
\, c \, e^{-\phi}\chi \bar \p Y (\sigma )  \, c\bar c \, e^{-\phi}\chi  \bar \p Y (0)
\Big]
\Big\rangle\, .
%\nonumber \\
\een
For large $\lambda$, $\p\RR_s$ represents a contour around 1 of linear size of order $\lambda^{-2}$. 
Also $W_1$ is a point at finite distance away from 1 and 
$W_2$ is a point within distance $\lambda^{-2}$ of 1.
We shall now
estimate the integrand for $\sigma\in\p\RR_s$ and show that the integral is suppressed in the large $\lambda$ limit.
The integration measure $d\bar\sigma$ contributes a factor of $\lambda^{-2}$ since the contour has linear size
$\lambda^{-2}$. Therefore the integrand must grow as $\lambda^2$ for getting a finite contribution. For the term in the 
second line of \refb{eapp3} the $b,c,\bar b, \bar c$ correlators give finite contribution, the $\phi$ correlator
gives a contribution of order $(\sigma-1)\sim\lambda^{-2}$, 
the $\xi,\eta$ correlator gives finite contribution and the
matter correlator gives a contribution of order $(\sigma-1)^{-1}\sim
\lambda^2$ from the $Y$ contribution. Therefore the correlator
grows as order $\lambda^0$ and its contribution to the integral vanishes for large $\lambda$. For the term in
the third line of \refb{eapp3} the $b,c,\bar b, \bar c$ correlators give a contribution of order $(\sigma-1)\sim
\lambda^{-2}$ from the $c$-$c$ operator product, 
the $\phi$ correlator
gives a contribution of order $(\sigma-1)^{-1}\sim
\lambda^2$,  the
matter correlator gives a contribution of order $(\sigma-1)^{-1}\sim
\lambda^2$ from the $Y$ contribution and the
$\xi,\eta$ correlator gives a contribution of order $(\sigma-W_2)\sim
\lambda^{-2}$ due to the fact that the points  $\sigma$ and
$W_2$ where $\xi$ is inserted are within a distance $\lambda^{-2}$ of each other. This again makes the
integrand is of order $\lambda^0$ in the large $\lambda$ limit, making the integral vanish in this limit. This shows
that $\BBB_s$ vanishes.

A similar analysis can be carried out for $\BBB_t$ which has the same expression as
\refb{eapp3} except that the integration contour $\p\RR_t$ lies in the large $\sigma$ region
($\sigma\sim\lambda^2$), $W_1$ lies at a finite point and $W_2$ also becomes large of order $\lambda^2$. However this
case is related to that for $\BBB_s$ via a $z\to z/(z-1)$ transformation accompanied by a reversal in sign of
$k_{(0)}$ and $n$. Therefore vanishing of $\BBB_s$ in the large $\lambda$ limit also implies vanishing of
$\BBB_t$.

Let us now turn to $\BBB_u''$ which, according to \refb{edefbbbupp}, is given by
\ben \label{eapp4}
\BBB_u'' &=& -2\, {\mu^2\over 2\pi i} \, \int_{\p\RR_{u}} d\bar \sigma  \, \Big\langle 
c\bar c e^{-\phi} \psi^1 \bar \p X^2 \, e^{-i k_{(0)}.X}  e^{-i n Y / R}(1) 
c\bar c e^{-\phi} \psi^1 \bar \p X^2 \, e^{i k_{(0)}.X}  e^{i n Y / R}(\infty) \nonumber \\ &&
\hskip -.6in \Big[-\{\xi(1) - \xi(W_1)\} \eta e^{\phi} \chi (\sigma)
\bar \p Y (\sigma ) \,  c\bar c \, e^{-\phi}\chi \bar \p Y (0)
\nonumber \\ &&\hskip -.6in 
-  \{\xi(\sigma ) -\xi(W_2)\} \, \Big\{ \p \eta \, e^{2\phi} \, b(W_1)
+ \p\left(\eta \, e^{2\phi} \, b(W_1)\right)\Big\}
\, c \, e^{-\phi}\chi \bar \p Y (\sigma )  \, c\bar c \, e^{-\phi}\chi  \bar \p Y (0)
\Big]
\Big\rangle\, .
%\nonumber \\
\een
In this case the integration contour $\p\RR_u$ encloses 0 and, according to \refb{e3.13x},
\refb{e2.18}, represents approximately a circle of radius $\propto\lambda^{-2}$ 
around the origin, with corrections of
order $\lambda^{-4}$.
The point $W_1$ is at finite distance away from 0 and the point $W_2$ is within a distance of order 
$\lambda^{-2}$ of 0. Now it follows from the analysis in \S\ref{slocal}, \ref{s4} that $W_1$ and $W_2$ are
holomorphic functions of $q$ and hence of $\sigma$. Evaluating the correlator in \refb{eapp4} we find that 
$\BBB_u''$ has the form:
\be
\BBB_u''= \int_{\p\RR_u} d\bar \sigma \, f(\sigma) \left( {1\over \bar\sigma^2} - {n^2\over 2 R^2} {1\over \bar\sigma-1}
\right)\, ,
\ee
for some holomorphic function $f(\sigma)$ with a regular Taylor series expansion around $\sigma=0$.
It follows from this that for large $\lambda$, $\BBB_u''$ is suppressed by inverse powers of $\lambda$.

\end{document}